\documentstyle[seceq,mbf,epsf,wick,wrapft]{ptptex}
\def\kdelta{\delta_{k_1{+}k_2}}

\def\trho{\tilde\rho}
\def\tr{\mathop{\rm Tr}}

\def\aa{x_u}
\def\bb{x_\propto}
\def\cc{x_\infty}
\def\dd{x_\Omega}
\def\ee{x_4}

\def\xu{x_u}
\def\xa{x_\propto}

\def\xc{x_{\rm c}}

\def\m1{{-1}}

\newcommand{\nn}{\nonumber\\}
\newcommand{\QB}{Q_{\rm B}}
\newcommand{\tQB}{{\tilde Q}_{\rm B}}
\newcommand{\half}{\frac{1}{2}}
\newcommand{\ttau}{{\tilde{\tau}}}
\newcommand{\abs}[1]{\left| #1 \right|}
\newcommand{\bra}[1]{\left< #1 \right|}
\newcommand{\ket}[1]{\left| #1 \right>}
\newcommand{\sbra}[1]{\,\langle#1|\,}
\newcommand{\sket}[1]{\,|#1\rangle\,}
\newcommand{\bidx}[1]{{\textstyle {\atop #1}}\!\!}%{}_{#1}\hspace{-0.1em}}
\def\o{{\rm o}}
\def\c{{\rm c}}

\def\bP#1{{(b_0^-{\cal P})}^{(#1^\c)}}

\newcommand{\Vtho}[1]{\,\langle V_3^{\rm o}(#1)|\,}
\newcommand{\Vfo}[1]{\,\langle V_4^{\rm o}(#1)|\,}
\newcommand{\Vthc}[1]{\,\langle V_3^{\rm c}(#1)|\,}
\newcommand{\UU}[1]{\,\langle U(#1)|\,}
\newcommand{\Valp}[1]{\,\langle V_\propto(#1)|\,}
\newcommand{\Vinf}[1]{\,\langle V_\infty(#1)|\,}
\newcommand{\Uomg}[1]{\,\langle U_\Omega(#1)|\,}
\newcommand{\vfo}[1]{\,\langle v_4^{\rm o}(#1)|\,}
\newcommand{\vtho}[1]{\,\langle v_3^{\rm o}(#1)|\,}
\newcommand{\vthc}[1]{\,\langle v_3^{\rm c}(#1)|\,}
\newcommand{\uv}[1]{\,\langle u(#1)|\,}
\newcommand{\valp}[1]{\,\langle v_\propto(#1)|\,}
\newcommand{\vinf}[1]{\,\langle v_\infty(#1)|\,}
\newcommand{\uomg}[1]{\,\langle u_\Omega(#1)|\,}

\def\calO{{\cal O}}
\def\calA{{\cal A}}

\def\calG{{\cal G}}
\def\VEV#1{\left<#1\right>}

\def\of{{\ket{\Psi}}}
\def\cf{{\ket{\Phi}}}

\def\Idx<#1,#2,#3>{{\mbf #1}_{#2:#3}}
\def\Idxc<#1,#2,#3>{{\mbf #1}^\c_{#2:#3}}
\def\rhoz(#1){\rho_0^{(#1)}}
\def\zz(#1){z_0^{(#1)}}
\def\hzz(#1){\hat{z}_0^{(#1)}}
\def\tzz(#1){\tilde{z}_0^{(#1)}}
\def\shalf{{\textstyle\frac{1}{2}}}
\def\v#1#2(#3#4){\vhide#1#2(#3#4\nil\nil\nil\nil)}
\def\vhide#1#2(#3#4#5#6#7#8){{\xdef\tmpone{#1}\xdef\tmptwo{#2}\xdef\open{o}
\xdef\closed{c}\xdef\oMg{\Omega}\xdef\aLp{\propto}\xdef\iNf{\infty}
\if 3\tmpone
 \ifx \open\tmptwo \vtho{#3,#4,#5} \else
 \ifx \closed\tmptwo \vthc{#3^\c,#4^\c,#5^\c} \else\fi\fi \else
\if 4\tmpone \vfo{#3,#4,#5,#6;#7} \else
\ifx \oMg\tmpone \uomg{#3,#4,#5^\c;#6} \else
\ifx \aLp\tmpone \valp{#3,#4;#5,#6} \else
\ifx \iNf\tmpone \vinf{#3^\c,#4^\c;#5} \else
\if 2\tmpone \uv{#3,#4^\c} \else
\fi\fi\fi\fi\fi\fi}}
\def\V#1#2(#3#4){\VVhide#1#2(#3#4\nil\nil\nil)}
\def\VVhide#1#2(#3#4#5#6#7){{\xdef\tmpone{#1}\xdef\tmptwo{#2}\xdef\open{o}%
\xdef\closed{c}\xdef\oMg{\Omega}\xdef\aLp{\propto}\xdef\iNf{\infty}%
\if3\tmpone%
\ifx\open\tmptwo \Vtho{#3,#4,#5} \else
\ifx\closed\tmptwo \Vthc{#3^\c,#4^\c,#5^\c} \else\fi\fi\else
\if4\tmpone \Vfo{#3,#4,#5,#6} \else
\ifx\oMg\tmpone \Uomg{#3,#4,#5^\c} \else
\ifx\aLp\tmpone \Valp{#3,#4} \else
\ifx\iNf\tmpone \Vinf{#3^\c,#4^\c} \else
\if2\tmpone \UU{#3,#4^\c} \else
\fi\fi\fi\fi\fi\fi}}
\def\VN#1(#2){\left\langle V_{(#1)}(#2)\right|}
\def\braR(#1#2){\left\langle R(#1,#2)\right|}
\def\braRo(#1#2){\left\langle R^\o(#1,#2)\right|}
\def\braRc(#1#2){\left\langle R^\c(#1^\c\!,#2^\c)\right|}
\def\ketR(#1#2){\left|R(#1,#2)\right\rangle}
\def\ketRo(#1#2){\left|R^\o(#1,#2)\right\rangle}
\def\ketRc(#1#2){\left|R^\c(#1^\c\!,#2^\c)\right\rangle}
\def\vfour(#1#2#3#4){{\sbra{v_4^{\o\,(\alpha_{#1},\alpha_{#2},\alpha_{#3},\alpha_{#4})}%
(#1,#2,#3,#4;\sigma_0)}}}
\def\sdot{\!\cdot\!}
\def\1loop{{\hbox{\scriptsize 1-loop}}}
\preprintnumber[3cm]{%<-- [..]: optional width of preprint # column.
KUNS-1575\\ hep-th/9905043\\
May, 1999}
\title{%        %You can use \\ for explicit line-break
One-Loop Tachyon Amplitude\\
in Unoriented Open-Closed String Field Theory
}
\author{%       %Use \sc for the family name
Tsuguhiko {\sc Asakawa},$^{1,}$\footnote{JSPS Research Fellow. 
E-mail address: asakawa@gauge.scphys.kyoto-u.ac.jp}
Taichiro {\sc Kugo}$^{1,}$\footnote{E-mail address:
kugo@gauge.scphys.kyoto-u.ac.jp}
and Tomohiko {\sc Takahashi}$^{2,}$\footnote{
JSPS Research Fellow. E-mail address: tomo@hep-th.phys.s.u-tokyo.ac.jp}
}

\inst{%         %Affiliation, neglected when [addenda] or [errata]
$^1$Department of Physics, Kyoto University, Kyoto 606-8502
\\
$^2$Department of Physics, University of Tokyo, Tokyo 113-0033
}

\recdate{%      %Editorial Office will fill in this.
\today
}

\abst{%       %this abstract is neglected when [addenda] or [errata]
We calculate one-loop 2-point tachyon amplitudes in unoriented open-closed
string field theory, and determine all the coupling constants of the 
interaction vertices in the theory. It is shown that the planar, 
nonplanar and nonorientable amplitudes are all reproduced correctly, and 
nontrivial consistencies between the determined coupling constants are 
observed. The necessity for the gauge group to be $SO(2^{13}{=}8192)$ 
is also reconfirmed. }

\begin{document}
\maketitle

\section{Introduction}

In previous two papers,\cite{rf:KugoTaka,rf:AKT2} which we refer to as I 
and II, we constructed a 
string field theory (SFT) for an unoriented open-closed string mixed 
system and proved the BRS/gauge invariance of the action at the `tree 
level': namely, we have classified the terms in the BRS transform of the
action into two categories, `tree terms' and `loop terms' and have shown 
that all the `tree terms' indeed cancel themselves. And for the other 
`loop terms', we have identified which anomalous one-loop diagrams they 
are expected to cancel.

It was left for a future work to show that those loop diagrams are 
indeed anomalous and the `loop terms' really cancel them. However, the 
task to show this BRS invariance for general one-loop diagrams is 
technically rather hard and we have not yet completed it. Here in this
paper, we address ourselves to an easier problem to compute the
one-loop 2-point (open-string) tachyon amplitudes in our SFT. These
one-loop amplitudes already suffer from a BRS anomaly and so the
present calculation gives partially an affirmative answer to the above
expected cancellation between the `loop terms' and the one-loop
anomaly. Moreover with this computation we shall confirm that the
one-loop tachyon amplitudes are correctly reproduced in our SFT and
can determine all the remaining coupling constants left undetermined
in the previous work; in particular, we show that the gauge group
must be
$SO(2^{13}{=}8192)$.%\cite{rf:DougGrin,rf:Weinberg,rf:ItoyamaMoxhay,rf:Ohta} 

The action of the present system, containing seven interaction terms, 
is given by
\begin{eqnarray}
S &=& -{1\over2}\bra{\Psi}\tQB^\o\Pi\ket{\Psi}
-{1\over2}\bra{\Phi}\tQB^\c(b_0^-{\cal P}\Pi)\ket{\Phi} \nn
&&+\frac{g}{3}\V3o(123)\of_{321} 
+\ee \frac{g^2}{4}\V4(1234)\of_{4321}      
+\bb \frac{g^2}{2}\V\propto(12)\of_{21} \nn
&&+\xc \frac{g^2}{3!}\V3c(123)\cf_{321}
     +\cc \frac{g^2}{2}\V\infty(12)\cf_{21} \nn
&&+\aa g \UU{1,2^\c}\cf_{2}\of_{1}
     +\dd \frac{g^2}{2}\V\Omega(123)\cf_3\of_{21}\,,
\label{eq:action}
\end{eqnarray}
where $\ee$, $\bb$, $\xc$, $\cc$, $\aa$ and $\dd$ are coupling constants
(relative to the open 3-string coupling constant $g$). 
For notation and conventions, we follow our previous papers I and II. 
The open and closed string fields are denoted by $\ket{\Psi}$ and 
$\ket{\Phi}$, respectively, both of which are Grassmann {\it odd}. The 
multiple products of string fields are denoted for brevity as 
%\begin{equation}
$\of_{n\cdots21}\equiv\ket{\Psi}_n\cdots\ket{\Psi}_2\ket{\Psi}_1$.
%\end{equation}
The BRS charges $\tQB$ with tilde here, introduced in I, are given by the 
usual BRS charges $\QB$ plus counterterms for the `zero intercept' 
proportional to the squared string length parameter $\alpha^2$:
\begin{equation}
\tQB^\o = \QB^\o + \lambda_\o g^2 \alpha^2 c_0\,, \qquad 
\tQB^\c = \QB^\c + \lambda_\c g^2 \alpha^2 c_0^+\,.
\end{equation}
The ghost zero-modes for the closed string are defined by
$c_0^+\equiv(c_0+\bar c_0)/2,\  c_0^- \equiv c_0-\bar c_0$, and 
$b_0^+\equiv b_0+\bar b_0, \ b_0^-\equiv(b_0-\bar b_0)/2$. 
%\begin{eqnarray}
%&&c_0^+\equiv(c_0+\bar c_0)/2\,,\qquad \quad c_0^- \equiv c_0-\bar c_0\,, \nn
%&&b_0^+\equiv b_0+\bar b_0\,, \qquad \quad b_0^-\equiv(b_0-\bar b_0)/2\,. 
%\end{eqnarray}

The string fields are always accompanied by the unoriented projection 
operator $\Pi$, which is given by using the twist operator
$\Omega$ in the form $\Pi= (1+\Omega)/2$, where $\Omega$ for the 
open string case means also taking transposition of the matrix index. 
The closed string is further accompanied by the projection operator 
${\cal P}\equiv\int^{2\pi}_0(d\theta/2\pi)\exp i\theta(L_0-\bar{L}_0)$, 
projecting out the $L_0-\bar L_0=0$ modes, 
%${\cal P}$, projecting out the $L_0-\bar L_0=0$ modes
%\begin{equation}
%{\cal P}\equiv\int^{2\pi}_0
%\frac{d\theta}{2\pi}\exp i\theta(L_0-\bar{L}_0),
%\end{equation}
and the corresponding anti-ghost zero-mode factor 
$b_0^-=(b_0-\bar b_0)/2$. 

Among the seven vertices, the open 3-string vertex $V_3^\o$, 
open-closed transition vertex $U$ and 
open-string self-intersection vertex $V_\propto$ are relevant in this paper 
and have the following form:\cite{rf:AKT2}
\begin{eqnarray}
&& \Vtho{1,2,3} = \vtho{1,2,3} \prod_{r=1,2,3} \Pi^{(r)}\,, \nn
&&
\bra{U(1,2^\c)}=\bra{u(1,2^\c)}\bP{2}
\prod_{r=1,2^\c}\Pi^{(r)}\,, \nn
&&\bra{V_\propto(1,2)}=\int_{0\leq\sigma_1\leq\sigma_2\leq\pi\alpha_1}
\hspace{-1.5em}d\sigma_1 d\sigma_2\sbra{v_\propto(1,2;\sigma_1,\sigma_2)}
b_{\sigma_1}b_{\sigma_2}\!
\prod_{r=1,2}\!\Pi^{(r)}\,.
\end{eqnarray}
The vertices here denoted by lower case letters are those constructed 
by the procedure of LeClair, Peskin and Preitschopf (LPP).\cite{rf:LPP}
The $b_{\sigma_r}$ $(r=1,2)$ are anti-ghost 
factors\cite{rf:AGMV,rf:GM,rf:KugoSuehiro} 
corresponding to the moduli $\sigma_r$ representing the two interaction 
points in the intersection vertex $v_\propto$. Those LPP vertices generally 
have the structure 
\begin{equation}
\int\Bigl(\prod_r{d^dp_r\over(2\pi)^d}\Bigr)(2\pi)^d\delta^d(\sum_rp_r)
\bidx{1}\bra{p_1}\bidx{2}\bra{p_2}\cdots 
\bidx{n}\bra{p_n}e^{E(\alpha)}
\label{eq:LPPzero}
\end{equation}
where $\bidx{r}\bra{p_r}$ is the Fock bra vacuum of string $r$ with 
momentum eigenvalue $p_r$ and the exponent $E(\alpha)$ is a quadratic form of 
string oscillators with Neumann coefficients determined by the way of 
gluing the participating strings. A point to be noted here is that 
since the gluing way depends on the set of the string lengths $\alpha_r$ 
and, in our $\alpha=p^+$ HIKKO type theory,\cite{rf:kugozwie} 
the string length $\alpha_r$ is 
identified with the + component of string momentum $p_r^\mu$; i.e., 
$\alpha_r=2p_r^+$ for open string and $\alpha_r=p_r^+$ for closed string, the 
exponent function $E(\alpha)$ has a nontrivial dependence on $\alpha_r$s.
Note that $p^\pm\equiv(p^0\pm p^{d-1})/\sqrt2$ and $p^2=-2p^+p^-+{\mib p}^2$.
Therefore the integration over $d^dp=dp^+dp^-d^{d-2}{\mib p}$ in 
Eq.~(\ref{eq:LPPzero}) is quite nontrivial.

In the previous papers, we have shown that the theory is BRS 
(and hence gauge) invariant at `tree level' if the coupling constants 
satisfy the relations
\begin{eqnarray}
 && \lambda_\c=2\lambda_\o= -\lim_{\epsilon\rightarrow0}
      \frac{32n\aa^2}{\epsilon^2}, \label{eq:rel-1}\\
 && x_\infty=nx_u^2=-4\pi ix_\propto, \label{eq:rel-2}\\
 && x_4=1 \label{eq:rel-3},\\
 && x_u=x_\Omega,%\label{eq:rel-4}, 
\quad  x_\c=8\pi ix_\Omega\,,\label{eq:rel-5}
\end{eqnarray}
where the sign of $x_\propto$ has been changed from the previous papers I and
II since we change the sign convention for the $V_\propto$ vertex in this 
paper by the reason as will be made clear in \S5. These relations 
(\ref{eq:rel-1}) -- (\ref{eq:rel-5}) leave only two parameters free, 
e.g., $x_u$ and $n$, or $x_u$ and $x_\propto$. We shall determine all the 
three parameters $x_u$, $x_\propto$ and $n$, thus giving a nontrivial 
consistency check of the theory.

The rest of this paper is organized as follows. 
First in \S2, we show which diagrams contribute to the on-shell tachyon 
2-point amplitude at one-loop level by using Feynman rule in the present
SFT. To evaluate the amplitudes of those diagrams explicitly we need 
to conformally map those diagrams into the torus plane and to compute 
the conformal field theory (CFT) correlation functions on the torus. 
We present in \S3 such conformal mapping for each diagram explicitly, 
and calculate in \S4 the Jacobians for the changes of moduli parameters 
associated with the mappings. In \S5 we first give some discussions on 
the `generalized gluing and resmoothing theorem' 
(GGRT)\cite{rf:LPP,rf:LPP2} to fix the normalizations of CFT correlation
functions on the torus, and then evaluate the necessary correlation 
functions explicitly. Gathering those results in \S\S3 - 5, we finally 
evaluate the tachyon amplitude explicitly in \S6, where the coupling 
constants $x_u$ and $x_\propto$ and $n$ of the gauge group $SO(n)$ are also 
determined. Finally in \S7, we conclude and present some discussions on 
the relations between the BRS anomaly in the present SFT and the Lorentz
invariance anomaly in the light-cone gauge 
SFT.\cite{rf:ST1,rf:ST2,rf:KikkawaSawada} 

For the variables and functions appearing in the one-loop 
amplitudes, we use the same notations as much as possible as those in 
Chapter 8 of the textbook by Green, Schwarz and Witten (GSW).\cite{rf:GSW} 
We cite in Appendix some modular transformation relations between such 
one-loop functions which will be used in the text.

\section{One-loop 2-point tachyon amplitude: preliminaries}

\begin{figure}[tb]
   \epsfxsize= 10cm   %or \epsfysize= HEIGHT cm
   \centerline{\epsfbox{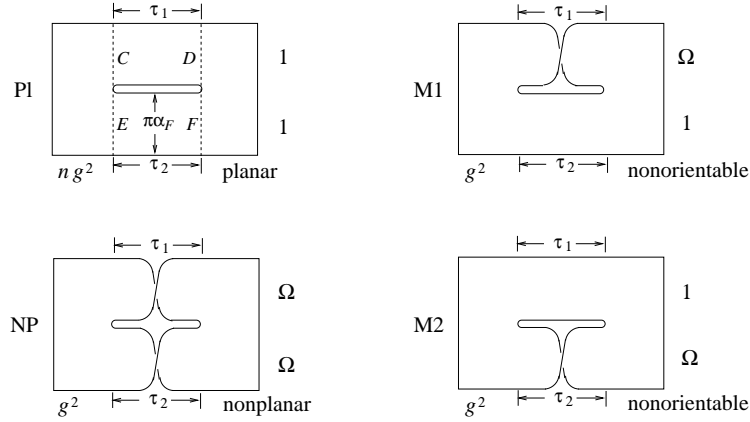}}
 \caption{Four diagrams contributing to the one-loop 2-point amplitude. 
For tachyon external states, actually, only the 
configurations with $\tau_1=\tau_2$ contribute as we shall see later.}
 \label{fig:loop}
\end{figure}

The one-loop amplitude obtained by using open 3-string vertex 
$V_3^\o$ twice contributes to the effective action $\Gamma_\1loop$ at one 
loop as
\begin{eqnarray}
 \Gamma_\1loop &=& i^{-1}{1\over2!}\,2\sdot3\sdot3\,
   \left(g\over3\right)^2\! \V3o(CE1)\of_1
   \wick{23}{<1\of_E <2\of_C {\Vtho{D,2,F}}
%   \wick{23}{<1\of_E <2\of_C {\V3o(D2F)}
   >1\of_F {\of_2} >2\of_D} \nn
 &=& -ig^2 \V3o(CE1)\V3o(D2F)
   \wick{2}{<1\of_C >1\of_D}\wick{2}{<1\of_E >1\of_F}
   \of_2 \of_1 \nn
 &=& ig^2 \int_{0}^{\infty}d\tau_1 d\tau_2
    \sbra{\tilde V(1,2;\tau_1,\tau_2)}
   \,b_{\tau_1}b_{\tau_2}\of_2 \of_1 ,
\label{eq:effaction}
\end{eqnarray}
where\footnote{Note that the factor $i^{-L}$ is multiplied to the 
$L$-loop level effective action in the present Feynman rule where the 
factors $i^{-1}$ and $i$ are omitted from the propagators and the 
vertices, respectively.}  
the open string propagator is given by
\begin{eqnarray}
 \wick{2}{<1\of_C >1\of_D} 
 &\equiv& \left({b_0\over L_0}\right)^{(C)}\ketRo(CD)
 = \left({b_0\over L_0}\right)^{(D)}\ketRo(CD)  \nn
 &=& \int_{0}^{\infty}dT_1 b_0^{(D)}e^{-L_0^{(D)} T_1}\ketRo(CD)
 = \int_{0}^{\infty}d\tau_1 b_{\tau_1}e^{-L_0 \tau_1}\ketRo(CD)
\hspace{3em}
\end{eqnarray}
and the glued vertex $\sbra{\tilde V(1,2;\tau_1,\tau_2)}$ is defined by 
\begin{equation}
\sbra{\tilde V(1,2;\tau_1,\tau_2)} \equiv 
   \V3o(CE1)\V3o(D2F)
   e^{-L_0 \tau_1}\ketRo(CD) e^{-L_0 \tau_2}\ketRo(EF) .
\end{equation}
Here $\tau_1=\alpha_DT_1$ is the time interval on the $\rho$ plane defined later
and $b_{\tau_1}=\alpha_D^{-1}b_0^{(D)}$ is the anti-ghost factor corresponding to 
the moduli $\tau_1$. 
Our SFT vertex $\V3o(123)$ contains the unoriented projection operators 
$\Pi^{(i)}=(1+\Omega^{(i)})/2$ $(i=1,2,3)$, as an effect of which 
the propagators of the two intermediate strings $D$ and $F$ 
are multiplied by the projection operators:
\begin{eqnarray}
&&\sbra{\tilde V(1,2;\tau_1,\tau_2)} = \v3o(CE1)\v3o(D2F) \nn
&& \qquad \qquad \times\Pi^{(D)}\Pi^{(F)}
   e^{-L_0 \tau_1}\ketRo(CD)e^{-L_0 \tau_2}\ketRo(EF) 
\prod_{r=1}^2\Pi^{(r)} .
\label{eq:v3v3glue}
\end{eqnarray}
This product of projection operators yields 
four terms, $\Pi^{(D)}\Pi^{(F)} = (\,1+\Omega^{(D)}+\Omega^{(F)}
 +\Omega^{(D)}\Omega^{(F)}\,)/4$, 
and accordingly the glued vertex $\sbra{\tilde V(1,2;\tau_1,\tau_2)}$ 
contains four different configurations as drawn in 
Fig.~\ref{fig:loop}, 
%\begin{wrapfigure}[6]{r}{6.6cm}
%\end{wrapfigure}
which we call planar (P), nonorientable (or M\"obius) (M1 and M2) 
and nonplanar (NP) diagrams, respectively. To those the following 
four LPP vertices correspond:
\begin{equation}
\sbra{\tilde V(1,\!2;\tau_1,\!\tau_2)}
 = {1\over4}\bigl\{n\bra{v_{\rm P}(\tau_1,\!\tau_2)}
+ \bra{v_{\rm M1}(\tau_1,\!\tau_2)}+\bra{v_{\rm M2}(\tau_1,\!\tau_2)}+
   \bra{v_{\rm NP}(\tau_1,\!\tau_2)}\bigr\}
\prod_{r=1}^2\!\Pi^{(r)}
\label{eq:LPPvertices}
\end{equation}
where the factor $n$ in front of $\bra{v_{\rm P}}$ has come from the inner 
endpoint loop carrying Chan-Paton index in the planar diagram.

We take the external open string states to be on-shell tachyon states
\begin{equation}
 \of_r = \sket{\varphi^0(k_r)}_r = 
 \big[c(w_r)\,e^{ik_r\cdot X^{(r)}(w_r)}\,\big]_{w_r=0}\ket{0}_r
 \qquad \left(k_r^2=2\right).
\end{equation}
Then, the one-loop effective action (\ref{eq:effaction}) with 
Eq.~(\ref{eq:LPPvertices}) substituted, yields the following one-loop 
2-point tachyon amplitudes 
\begin{eqnarray}
 &&\hspace*{-2em} 
\left\{\,{\cal A}_{\rm P}+{\cal A}_{\rm M1}
  +{\cal A}_{\rm M2}+{\cal A}_{\rm NP}\,\right\} \kdelta \nn
%\qquad \qquad \quad (\kdelta\equiv(2\pi)^d \delta^d(k_1+k_2))\nn
 &&= 2i\cdot {g^2\over4} \int_{0}^{\infty}d\tau_1 d\tau_2
   \bigl\{\,n\bra{v_{\rm P}(\tau_1,\tau_2)}
   + \bra{v_{\rm M1}(\tau_1,\tau_2)}+\bra{v_{\rm M2}(\tau_1,\tau_2)}  \nn
 &&\hspace{10em}
 + \bra{v_{\rm NP}(\tau_1,\tau_2)} \,\bigr\}
   \,b_{\tau_1}b_{\tau_2} 
%\Pi^{(2)}\of_2\Pi^{(1)}\of_1
%  \bigl\{\,n\bra{v_{\rm P}}+\bra{v_{\rm M1}}+\bra{v_{\rm M2}}+
%  \bra{v_{\rm NP}}\,\bigr\}b_{\tau_1}b_{\tau_2}
  \sket{\varphi^0(k_2)}_2 \sket{\varphi^0(k_1)}_1, 
\label{eq:Xamp}
\end{eqnarray}
with an abbreviation $\kdelta\equiv(2\pi)^d \delta^d(k_1+k_2)$, where
the individual amplitude is evaluated by the CFT on the corresponding 
torus $J=$ P, M1, M2, NP:
\begin{equation}
\bra{v_J(\tau_1,\tau_2)}
   \,b_{\tau_1}b_{\tau_2} 
  \sket{\varphi^0(k_2)}_2 \sket{\varphi^0(k_1)}_1
 = 
  \left<b_{\tau_1}b_{\tau_2}\,c(Z_2)e^{ik_2\cdot X(Z_2)}
c(Z_1)e^{ik_1\cdot X(Z_1)}\right>_J.
%  \left<\,e^{ik_2\cdot X(Z_2)}\,e^{ik_1\cdot X(Z_1)}\,\right>_X.
\label{eq:CFTfnX}
\end{equation}
Here we note two points.  First, the RHS is generally multiplied by a factor 
\begin{equation}
\prod_{r=1,2}\left({du\over dw_r}\right)_{u=Z_r}^{{k_r^2\over2}-1} 
\label{eq:ConfFactor}
\end{equation}
which is associated with the conformal mapping of the operators 
$c(w_r)e^{ik_r\cdot X(w_r)}$ from the unit disk $w_r$ to the 
torus $u$ plane and $Z_r=u(w_r{=}0)$ are the positions of punctures 
on the torus representing the external strings. But here the factor 
(\ref{eq:ConfFactor}) is 1 since the conformal weights $(k_r^2/2) -1$ 
are zero for the {\it on-shell} tachyons. Secondly, this 
Eq.~(\ref{eq:CFTfnX}) {\it determines} the CFT correlation functions on 
the RHS including their signs and weights. This constitutes the content 
of GGRT;\cite{rf:LPP2,rf:AKT1} namely, the loop level LPP vertex $\bra{v_J}$ with $J=$ P, NO 
(M1 and M2), NP are already {\it defined} above as glued vertices of the
two tree level LPP vertices $\bra{v_3^\o}$ by Eq.~(\ref{eq:v3v3glue}) 
with (\ref{eq:LPPvertices}). So these torus correlation functions are 
already fixed including their coefficients. We defer the explicit 
evaluation of these correlation functions until \S5.

\begin{figure}[tb]
   \epsfxsize= 11.5cm   %or \epsfysize= HEIGHT cm
   \centerline{\epsfbox{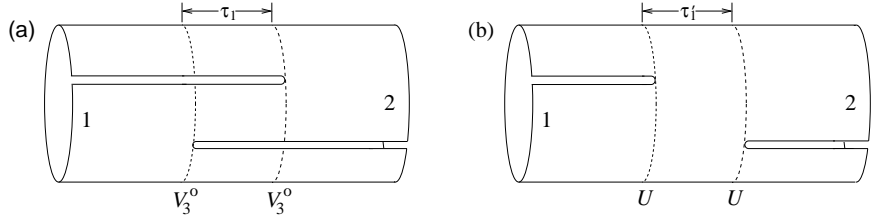}}
 \caption{(a) Loop diagram equivalent to the nonplanar diagram NP in 
Fig.~\protect\ref{fig:loop}. (b) A tree nonplanar diagram obtained 
by using the $U$ vertex twice.}  
 \label{fig:nonplanar}
\end{figure}

Here we first consider the nonplanar diagram NP, for which the two 
intermediate strings are both twisted. 
The amplitude $\calA_{\rm NP}$ corresponding to this diagram alone does 
not give the full nonplanar amplitude correctly. Indeed, this can 
easily be seen if we redraw the diagram NP in the form as depicted in 
the diagram (a) in Fig.~\ref{fig:nonplanar}. 
%\begin{wrapfigure}[6]{r}{6.6cm}
%\end{wrapfigure}
So it can cover only the $\tau_1>0$ part of the full nonplanar amplitude, 
and the remaining $0>\tau_1(=-\tau'_1)$ part is supplied by the `tree' diagram 
given by using the open-closed transition vertex $U$ twice as drawn in 
the diagram (b) in Fig.~\ref{fig:nonplanar}.\cite{rf:KK} 
The amplitude for this $UU$ diagram is given by 
\begin{eqnarray}
{\cal A}_{UU} \kdelta &=& {1\over2!}\cdot2\left(\xu g\right)^2 \UU{1,A^{\c}}
   \wick{2}{<1\cf_{A^{\c}} \sket{\varphi^0(k_1)}_1 \UU{2,B^{\c}}
   >1\cf_{B^{\c}}} \sket{\varphi^0(k_2)}_2 \nn
% &=& \xu^2 g^2 \uv{1,A^{\c}}\uv{2,B^{\c}}
%   \wick{2}{{\bP A} <1\cf_{A^{\c}} {\bP B} >1\cf_{B^{\c}}}
%   \of_2 \of_1 \nn
 &=& \xu^2 g^2 \int d\tau_1{d\sigma_1\over2\pi}
   \bra{v_{UU}(\tau_1,\sigma_1)}\,b_0\bar{b}_0\,
    \sket{\varphi^0(k_2)}_2\sket{\varphi^0(k_1)}_1,
\label{eq:UUamp}
\end{eqnarray}
where the Wick contraction gives the closed string propagator
\begin{eqnarray}
&& \wick{2}{{\bP A} <1\cf_{A^{\c}} {\bP B} >1\cf_{B^{\c}}}
 = \left({b_0\bar{b}_0\over L_0+\bar{L}_0}\right)^{(B^{\c})}
   {\cal P}^{(B^{\c})}\ketRc(AB) \nn
% &&\quad = \int dT_1 e^{-(L_0+\bar{L}_0)^{(B^{\c})}T_1}
%   \int{d\theta_{B^{\c}}\over2\pi}e^{-i\theta_{B^{\c}}(L_0-\bar{L}_0)^{(B^{\c})}}
%   \left(b_0\bar{b}_0\right)^{(B^{\c})}\ketRc(AB) \nn
 &&\quad  
 = \int d\tau_1{d\sigma_1\over2\pi}e^{-(L_0+\bar{L}_0)\tau_1}e^{-i\sigma_1(L_0-\bar{L}_0)}
   \,b_0\bar{b}_0\,\ketRc(AB)
\end{eqnarray}
and the following glued LPP vertex has been defined:
\begin{equation}
 \bra{v_{UU}(\tau_1,\sigma_1)}=\v2(1A)\v2(2B)
  e^{-(L_0+\bar{L}_0)\tau_1}e^{-i\sigma_1(L_0-\bar{L}_0)}
  \ketRc(AB).
\label{eq:UUvertex}
\end{equation}
Again the amplitude is evaluated by referring to the CFT on the torus:
\begin{equation}
   \bra{v_{UU}(\tau_1,\sigma_1)}\,b_0\bar{b}_0\,
   \sket{\varphi^0(k_2)}_2 \sket{\varphi^0(k_1)}_1
 = 
   \left\langle\,b_0\bar{b}_0\,c(Z_2)e^{ik_2\cdot X(Z_2)}c(Z_1)
   e^{ik_1\cdot X(Z_1)}\right>_{UU}.
\label{eq:CFTfnUU}
\end{equation} 
The amplitudes $\calA_{\rm NP}$ in Eq.~(\ref{eq:Xamp}) and $\calA_{UU}$ 
in Eq.~(\ref{eq:UUamp}) should smoothly connect with each other at $\tau_1=0$ 
to reproduce the correct nonplanar amplitude, and this condition will 
determine the coupling constant $x_u$, as we shall obtain later.

\begin{figure}[tb]
   \epsfxsize= 9cm   %or \epsfysize= HEIGHT cm
   \centerline{\epsfbox{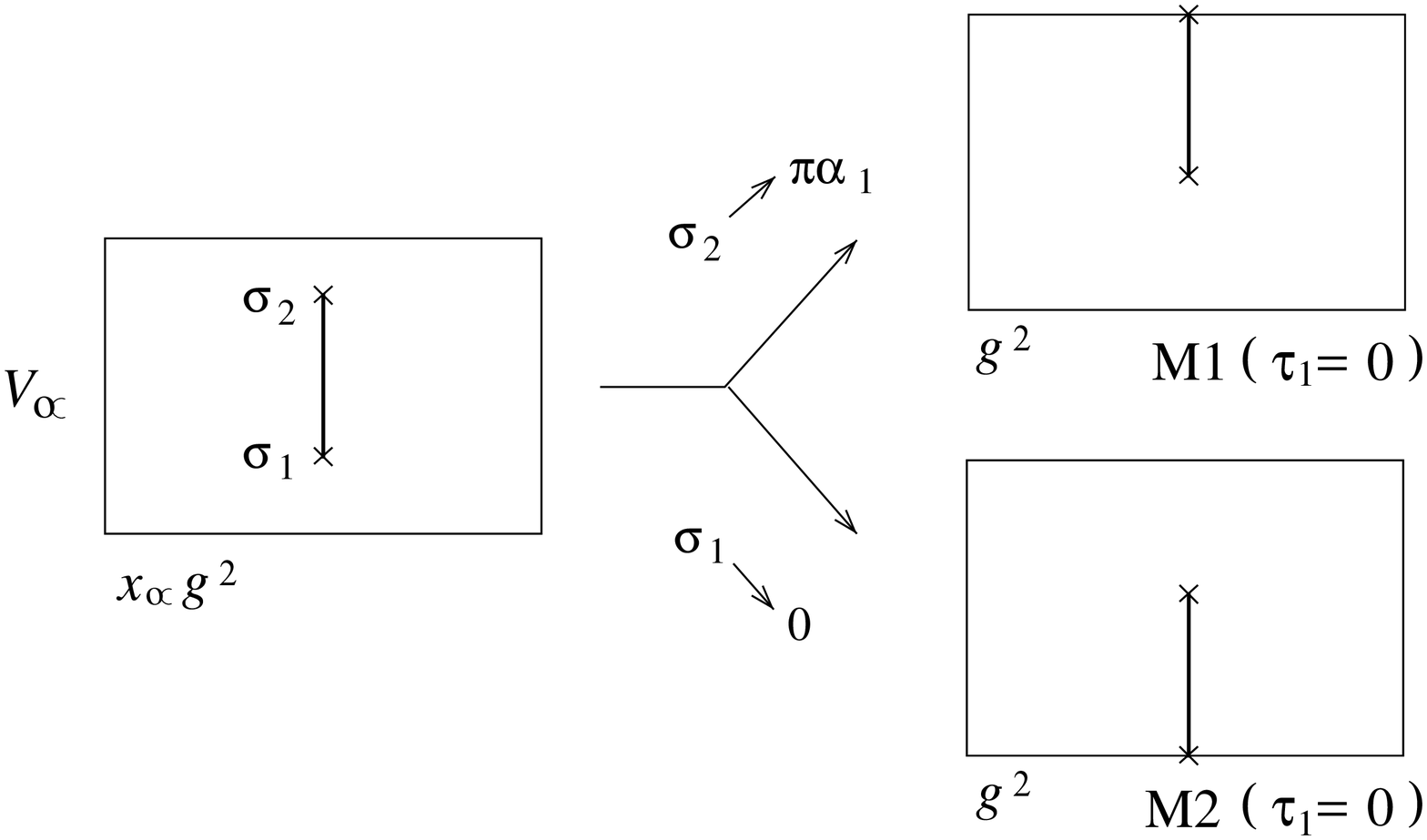}}
 \caption{Tree diagram given by $V_\propto$ vertex which 
fill in the gap between the two configurations at $\tau_1=0$ of the 
nonorientable diagrams M1 and M2 in Fig.~\protect\ref{fig:loop}.}  
 \label{fig:valpha}
\end{figure}

Next consider the two nonorientable diagrams, M1 and M2 
terms in  Fig.~\ref{fig:loop}. These two diagrams alone do not 
give the full nonorientable amplitude, again. Indeed, the two diagrams 
do not connect with each other at the moduli $\tau_1=0$, so that another diagram 
should exist which interconnects these two configurations at $\tau_1=0$. 
Such a diagram is just given by the `tree' diagram drawn in 
Fig.~\ref{fig:valpha} which is obtained by using the $V_\propto$ vertex.
%\begin{wrapfigure}[6]{r}{6.6cm}
%\end{wrapfigure}
Clearly the configuration of this diagram coincides at $\sigma_2=\pi\alpha_1$ 
(and $\sigma_1=\pi\alpha_F$) with that of the M1 diagram at $\tau_1=0$, and at $\sigma_1=0$ 
(and $\sigma_2=\pi\alpha_F$) with that of the M2 diagram at $\tau_1=0$. 
The amplitude ${\cal A}_{V\!{\propto}}$ of 
this diagram is proportional to $x_\propto$, and so the smooth connection 
condition for these amplitudes will determine $x_\propto$ as we shall see 
later.
The amplitude ${\cal A}_{V\!{\propto}}$ is given by
\begin{eqnarray}
 {\cal A}_{V\!{\propto}}\kdelta
 &=& 2\cdot \xa {g^2\over2}\,\Valp{1,2}
 \sket{\varphi^0(k_2)}_2 \sket{\varphi^0(k_1)}_1 \nn
 &=& \xa g^2 \int d\sigma_1 d\sigma_2 \,\valp{1,2;\sigma_1,\sigma_2}\,b_{\sigma_1}b_{\sigma_2}
 \sket{\varphi^0(k_2)}_2 \sket{\varphi^0(k_1)}_1 \nn
 &=& \xa g^2 \int d\sigma_1 d\sigma_2 \,
 \left\langle b_{\sigma_1}b_{\sigma_2}\,c(Z_2)e^{ik_2\cdot X(Z_2)}
   c(Z_1)e^{ik_1\cdot X(Z_1)}\right>_{V_{\propto}}.
\hspace{3em}
\label{eq:Valphaamp}
\end{eqnarray}

\begin{figure}[tb]
   \vspace{2ex}
   \epsfxsize= 11cm   %or \epsfysize= HEIGHT cm
   \centerline{\epsfbox{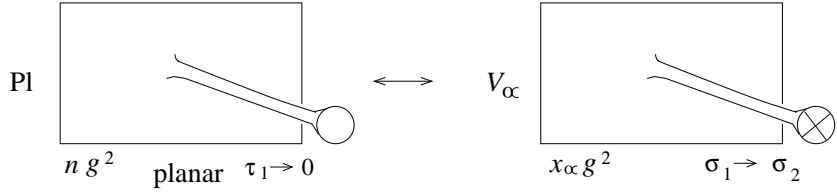}}
 \caption{Singular configurations for the planar diagram and 
$V_\propto$ diagram.}  
 \label{fig:planar}
\end{figure}

Finally, we note that the planar diagram in Fig.~\ref{fig:loop} as $\tau_1\rightarrow 
0$ and the above diagram of $V_\propto$ vertex as $\sigma_1-\sigma_2\rightarrow0$, both have 
singularities owing to the closed tachyon and dilaton vanishing into 
vacuum. As is shown in Fig.~\ref{fig:planar},
%\begin{wrapfigure}[6]{r}{6.6cm}
%\end{wrapfigure}
this is almost the same situation as what we have encountered in the 
disk and $RP^2$ amplitudes for closed tachyon 2-point function in the 
previous paper I. The former planar amplitude is proportional to $ng^2$
and the latter $V_\propto$ amplitude to $x_\propto$. The condition for the 
dilaton contributions to cancel between the two amplitudes will 
determine $n$ of $SO(n)$, as we shall explicitly see later.

\section{Conformal mapping of $\rho$ plane to torus}

In order to calculate these amplitudes in Eqs.~(\ref{eq:Xamp}), 
(\ref{eq:UUamp}) and (\ref{eq:Valphaamp}), we need conformal mapping of 
the usual unit disk $\abs{w_r}\leq1$ of participating string $r$ to the 
torus for each case. The string world sheets of the `light-cone type' 
diagrams like Figs.~\ref{fig:loop}, \ref{fig:nonplanar} etc, are called 
$\rho$ plane, on which the complex coordinate $\rho=\tau+i\sigma$ is identified 
with $\rho=\alpha_r\rho_r+i\beta_r$ ($\beta_r$: a real constant) in each string $r$ 
strip (Re$\rho_r\leq0$), where $\rho_r=\tau_r+i\sigma_r$ is the image of the unit 
disk $\abs{w_r}\leq1$ of string $r$ by a simple (conformal) mapping 
$\rho_r = \ln w_r$. Therefore we have only to know the conformal mapping 
of the $\rho$ plane to the torus for each case.

The conformal mapping of the $\rho$ plane to the torus $u$ plane 
with periods 1 and $\tau$ is generically given by the following 
(generalized) Mandelstam mapping:\cite{rf:Mandelstam,rf:GSW,rf:FGST}
\begin{equation}
\label{eq:rho(u)}
 \rho(u) = \sum_{r=1,2}\alpha_r \ln \vartheta_1(u-Z_r|\tau)+Au
\end{equation}
where $\vartheta_i(\nu|\tau)$ $(i=1,\cdots,4)$ are Jacobi $\vartheta$ functions 
with periods $1$ and $\tau$. This $\rho$ satisfies a quasi-periodicity  
\begin{equation}
\label{eq:periodicity} 
 \rho(u+m\tau+n)-\rho(u) = 2\pi im\sum_{r=1,2}\alpha_r Z_r+A(m\tau+n).
\end{equation}
The derivative $d\rho/du$ is truly a doubly periodic function, or
{\it elliptic function},\cite{rf:WhitWatson} 
which is analytic except for the poles at $u=Z_r$ ($r=1,2$):
\begin{equation}
\label{eq:drho/du}
 \frac{d\rho}{du} = \sum_{r=1,2}\alpha_r g_1(u-Z_r|\tau)+A,
\end{equation}
where $g_i$ is the logarithmic derivative of the Jacobi $\vartheta$ function:
\begin{equation}
g_i(\nu|\tau)\equiv{\vartheta'_i(\nu|\tau)\over\vartheta_i(\nu|\tau)},  \qquad 
\vartheta'_i(\nu|\tau)\equiv{\partial\vartheta_i(\nu|\tau)\over\partial\nu}.
\label{eq:gi}
\end{equation} 
$Z_r$ corresponds to the image of the external string $r$ at 
$\tau_r=-\infty$ ($w_r=0$). Two interaction points $\zz(i)$ $(i=1,\,2)$ are 
given by the zeros of this function:
\begin{equation}
\label{eq:intpt}
 \frac{d\rho}{du}(\zz(i))=0\ \ \ 
  \Rightarrow\ \ \ 
    \sum_{r=1,2}\alpha_r g_1(\zz(i)-Z_r|\tau)+A=0.
\end{equation}
By a general property of elliptic functions,\cite{rf:WhitWatson} 
a sum rule holds:
\begin{equation}
\label{eq:sumrule} 
 Z_1+Z_2=\zz(1)+\zz(2) \ (\hbox{mod periods } 1\ \hbox{and}\ \tau)
\end{equation}

We now examine the conformal mappings for the cases of 
planar, nonorientable, nonplanar, $V_{\propto}$ and $UU$ diagrams, separately, 
and will find relations which determine the torus moduli $\tau$, the parameter 
$A$ and interaction points $\zz(i)$ in terms of the string length and 
the moduli parameters of the $\rho$ plane. 

\subsection{Planar diagram (P)}

\begin{figure}[tb]
   \epsfxsize= 10cm   %or \epsfysize= HEIGHT cm
   \centerline{\epsfbox{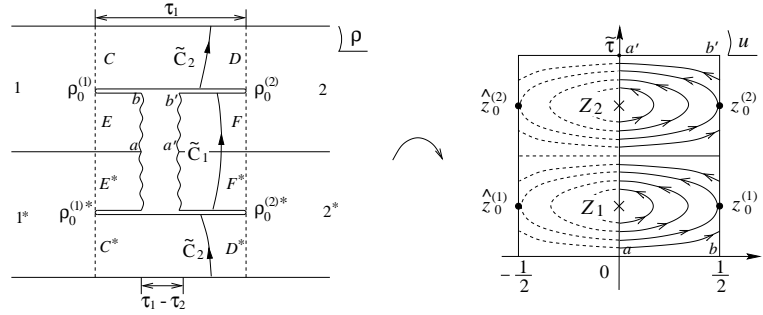}}
 \caption{Conformal mapping of planar diagram $\rho$ plane to a torus 
$u$ plane. In the $\rho$ plane, the edge curve $a$-$b$ of string $E$ is identified with the curve $a'$-$b'$ of string $F$, which are mapped to the bottom 
and top lines $a$-$b$ and $a'$-$b'$ on the $u$ plane, respectively.}
 \label{fig:Pmap}
\end{figure}

The mapping for the planar diagram case is drawn in Fig.~\ref{fig:Pmap}. 
%\begin{wrapfigure}[6]{r}{6.6cm}
%\end{wrapfigure}
In this case the period $\tau$ is purely imaginary and denoted by 
$\ttau$,  and the mapping of this Fig.~\ref{fig:Pmap}\ is given by the above 
Mandelstam mapping (\ref{eq:rho(u)}) with $\tau$ replaced by $\ttau$. 
The interaction points $\rho_0^{(i)}$ in the $\rho$ plane are mapped to 
$\zz(i)$ in the $u$ plane. By using the shift invariance on the torus plane 
and the sum rule (\ref{eq:sumrule}), 
we can parametrize the positions of strings (punctures) and interaction 
points by two real parameters $x$ and $y$ as
\begin{equation}
\label{eq:pl-paramet}
 \cases{
         Z_1=\ttau\left(\half-x\right) \cr
         Z_2=\ttau\left(\half+x\right) \cr}, \qquad 
 \cases{
         \zz(1)=\half+\ttau\left(\half-y\right) \cr
         \zz(2)=\half+\ttau\left(\half+y\right) \cr}.
\end{equation}

By the help of the periodicity (\ref{eq:periodicity}), we can determine 
the parameters $A$ and $x$ as follows. First, taking $u+1=\zz(i)$, for 
instance, we have $u=\hat\zz(i)$ and then $\rho(u+1)=\rho_0^{(i)}$ and $\rho 
(u)=\rho_0^{(i)*}$, so that
\begin{equation}
 \rho(u+1)-\rho(u) = A = \rho_0^{(i)}-\rho_0^{(i)*}= 2\pi i \alpha_F \quad \Rightarrow\quad 
 A = 2\pi i \alpha_F\,.
\label{eq:pl-peri1}
\end{equation}
Next, note that the bottom line ${\rm Im}u=0$ and the top line ${\rm 
Im}u=\abs{\ttau}$ with $0\leq{\rm Re}u\leq1/2$, are mapped to the 
wavy curves $a$-$b$ and $a'$-$b'$ of strings $E$ and $F$ on the $\rho$ plane 
in Fig.~\ref{fig:Pmap}. Therefore we have 
\begin{eqnarray}
 &&\tau_1-\tau_2 = \rho(u+\ttau)-\rho(u)=2\pi i\alpha_1
 (Z_1-Z_2)+A\ttau \nn
&&\qquad \Rightarrow\quad  \frac{\tau_1}{\alpha_1}-\frac{\tau_2}{\alpha_1}
 = 2\pi i\ttau\,\left(\frac{\alpha_F}{\alpha_1}-2x\right)\,.
\label{eq:pl-peritau}
\end{eqnarray}

The time length $\tau_1$ of the intermediate C-D propagator can be 
connected to the torus parameter $\ttau$ as follows:
\begin{eqnarray}
 \tau_1 &=& \rhoz(2)-\rhoz(1) = \rho(\zz(2))-\rho(\zz(1)) %\nn
% &=& 2\alpha_1\ln\frac{\vartheta_1(\zz(1)-Z_2|\ttau)}
 = 2\alpha_1\ln\frac{\vartheta_1(\zz(1)-Z_2|\ttau)}
 {\vartheta_1(\zz(1)-Z_1|\ttau)}+A(\zz(2)-\zz(1)) \nn
 &&\qquad \Rightarrow\quad  \frac{\tau_1}{\alpha_1} = 
 2\ln\frac{\vartheta_2(\ttau(x+y)|\ttau)}{\vartheta_2(\ttau(x-y)|\ttau)}
 +4\pi i\frac{\alpha_F}{\alpha_1}\ttau y\,.
\label{eq:pl-tau1}
\end{eqnarray}
The Eq.~(\ref{eq:intpt}) determining the interaction points, or $y$, 
now reads
\begin{equation}
\label{eq:pl-y}
 g_2(\ttau(x+y)|\ttau)+g_2(\ttau(x-y)|\ttau) 
 = -2\pi i\frac{\alpha_F}{\alpha_1}\,,
\end{equation}
which is identical with the extremum condition $\partial\tau_1/\partial y=0$ 
of $\tau_1$ in Eq.~(\ref{eq:pl-tau1}).

\subsection{Nonorientable diagrams (M1 and M2)}

\begin{figure}[tb]
   \epsfxsize= 13cm   %or \epsfysize= HEIGHT cm
   \centerline{\epsfbox{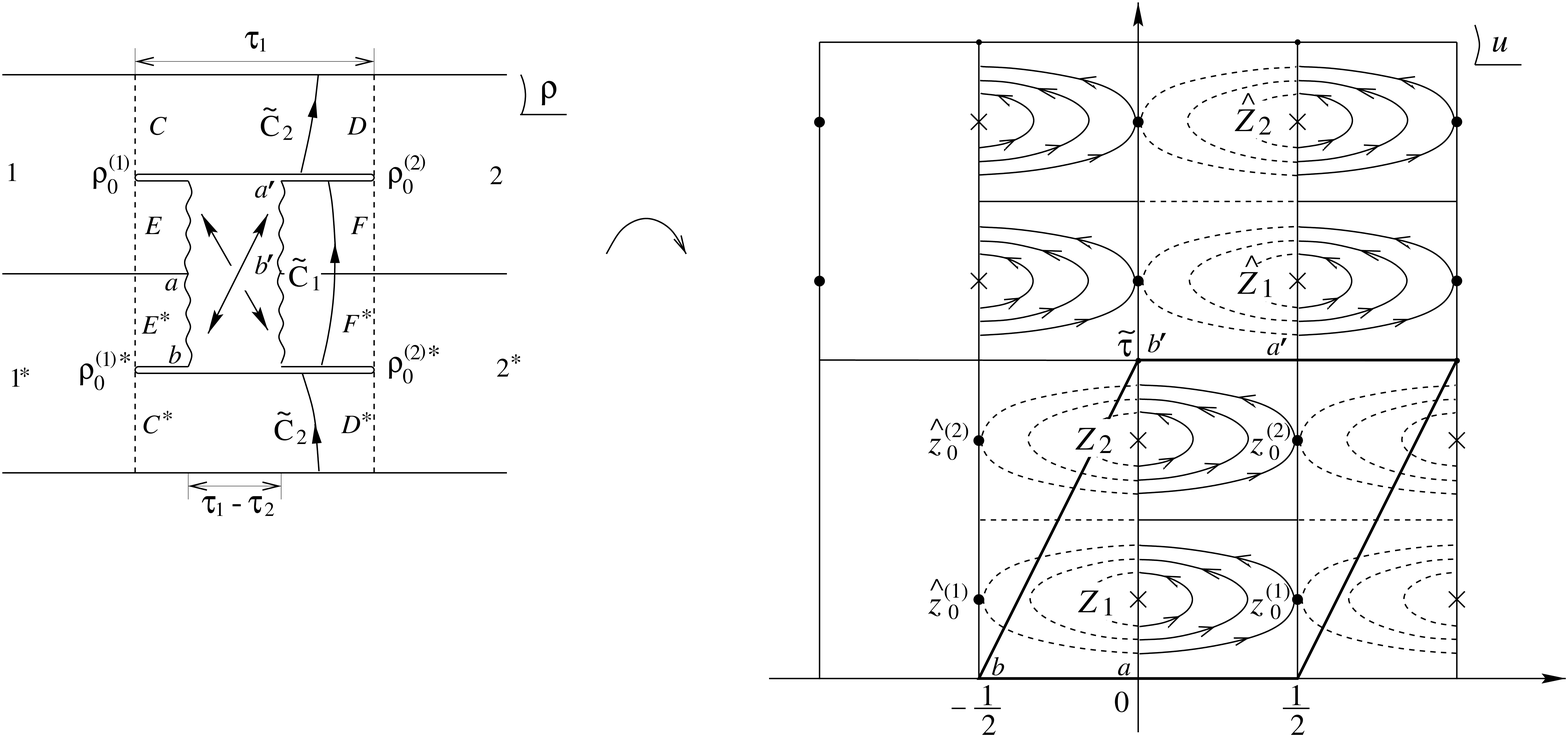}}
 \caption{Conformal mapping of the $\rho$ plane of nonorientable diagram M1 
to a torus. In the $\rho$ plane, the curve $a$-$b$ is identified with the curve $a'$-$b'$, whose images in the $u$ plane are also shown.} 
\label{fig:Mmap}
\end{figure}

The mapping of the nonorientable diagram M1 to torus is drawn 
in Fig.~\ref{fig:Mmap}. 
%\begin{wrapfigure}[6]{r}{6.6cm}
%\end{wrapfigure}
In this case the fundamental region of the torus is given as indicated 
in the same figure Fig.~\ref{fig:Mmap}, and so the period $\tau$ is now 
given by $\ttau+1/2$. Accordingly, the mapping of this 
Fig.~\ref{fig:Mmap} is given by the above Mandelstam mapping 
(\ref{eq:rho(u)}) with $\tau=\ttau+1/2$. We can parametrize the positions 
of strings (punctures) and interaction points by the same equations as 
Eq.~(\ref{eq:pl-paramet}) in this case also.

Similarly to the previous planar case, 
the periodicity (\ref{eq:periodicity}) determines the parameters $A$ and
$x$; from the period 1 we have the same relation as before,
\begin{equation}
\label{eq:no-peri1}
 A = 2\pi i \alpha_F.
\end{equation}
Noting that two points separated by $\ttau+1/2$, e.g., the points 
$a$ and $a'$, on the $u$ plane 
correspond to those separated by $\tau_1-\tau_2+i\pi\alpha_F$ on the $\rho$ plane as 
seen in Fig.~\ref{fig:Mmap},
and using the period $\ttau+\half$ of $\rho(u)$, we obtain
\begin{eqnarray}
 \tau_1-\tau_2+i\pi\alpha_F &=& 
 = \rho(u+(\ttau+\shalf))-\rho(u)
 = 2\pi i\alpha_1(Z_1-Z_2)+A(\ttau+\shalf) \nn
&&\hspace{-2em}\Rightarrow\quad  \frac{\tau_1}{\alpha_1}-\frac{\tau_2}{\alpha_1}
 = 2\pi i\ttau\,\left(\frac{\alpha_F}{\alpha_1}-2x\right)\,.
\label{eq:no-peritau}
\end{eqnarray}
Equations for $\tau_1$ and the interaction point $y$ take the same form as 
those for the previous planar case aside from the period:
\begin{eqnarray}
\label{eq:no-tau1}
&& \frac{\tau_1}{\alpha_1} = 
 2\ln\frac{\vartheta_2(\ttau(x+y)|\ttau+\half)}{\vartheta_2(\ttau(x-y)|\ttau+\half)}
 +4\pi i\frac{\alpha_F}{\alpha_1}\ttau y\,, \\
\label{eq:no-y}
&& g_2(\ttau(x+y)|\ttau+\shalf)+g_2(\ttau(x-y)|\ttau+\shalf) 
 = -2\pi i\frac{\alpha_F}{\alpha_1}\,.
\end{eqnarray}

\subsection{Nonplanar diagram (NP)}

\begin{figure}[tb]
   \epsfxsize= 11cm   %or \epsfysize= HEIGHT cm
   \centerline{\epsfbox{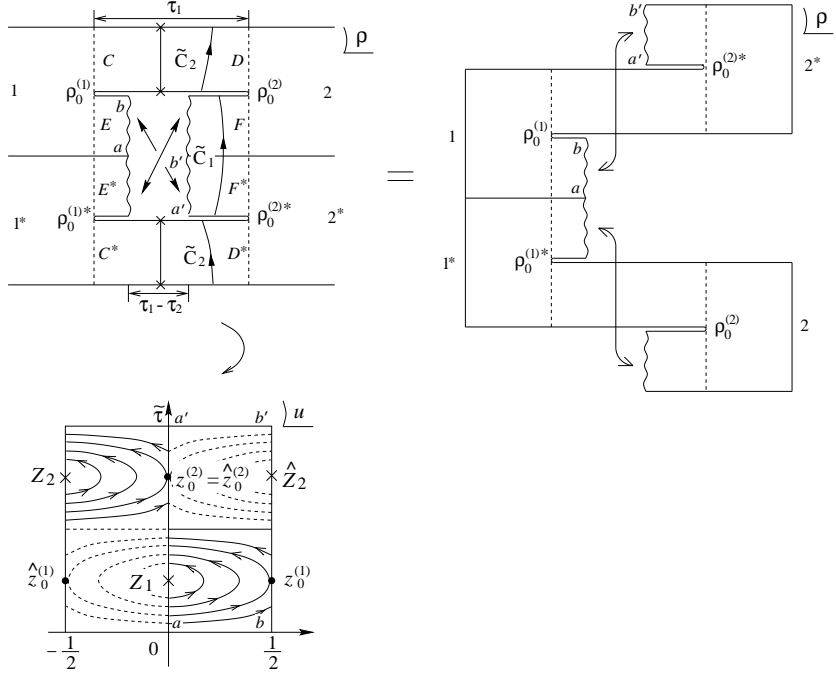}}
 \caption{Conformal mapping of the $\rho$ plane of nonplanar diagram NP 
to a torus. In the $\rho$ plane, the line connecting two crosses denotes 
twisting, and the curve $a$-$b$ is identified with the curve $a'$-$b'$, 
whose images in the $u$ plane are also shown.} 
\label{fig:NPmap}
\end{figure}

The mapping of the nonplanar diagram NP to torus is drawn 
in Fig.~\ref{fig:NPmap}. 
%\begin{wrapfigure}[6]{r}{6.6cm}
%\end{wrapfigure}
The period $\tau$ in this case is $\ttau$, the same as in planar case, and 
so is the Mandelstam mapping (\ref{eq:rho(u)}) with $\tau=\ttau$. 
However, the strings (punctures) and interaction points are placed 
differently from the planar case as shown in  Fig.~\ref{fig:NPmap}. 
So we parametrize their positions by real parameters $x$ and $y$ as
\begin{equation}
\label{eq:np-paramet}
 \cases{
         Z_1=\phantom{-\half+{}}\ttau\left(\half-x\right) \cr
         Z_2=-\half+\ttau\left(\half+x\right) \cr}, \qquad 
 \cases{
         \zz(1)=\half+\ttau\left(\half-y\right)      \cr
         \zz(2)=\phantom{\half+{}}\ttau\left(\half+y\right) \cr}.       
\end{equation}
From the period 1 and $\rhoz(1)-{\rhoz(1)}^*=2\pi i\alpha_F$, we again obtain 
the same relation as before,
\begin{equation}
\label{eq:np-peri1}
 A = 2\pi i \alpha_F.
\end{equation}
Noting that two points separated by $\ttau$, e.g., the points $a$ and 
$a'$, on the $u$ plane correspond to those separated by 
$\tau_1-\tau_2+\pi i\alpha_1$ on the $\rho$ plane as seen in Fig.~\ref{fig:NPmap}, 
and using the period $\ttau$ of $\rho(u)$, we find
\begin{eqnarray}
 \tau_1-\tau_2+\pi i\alpha_1 &=& \rho(u+\ttau)-\rho(u)=2\pi i\alpha_1
 (Z_1-Z_2)+A\ttau \nn
% &=& 2\pi i\alpha_1(\shalf-2\ttau x)+2\pi i\alpha_F \ttau \nn
&&\hspace{-3em}\Rightarrow\quad 
 \frac{\tau_1}{\alpha_1}-\frac{\tau_2}{\alpha_1}
 = 2\pi i\ttau\,\left(\frac{\alpha_F}{\alpha_1}-2x\right).
\label{eq:np-peritau}
\end{eqnarray}
Equations for $\tau_1$ and the interaction point $y$ become in this case
\begin{eqnarray}
 \tau_1+ \pi i(\alpha_1-\alpha_F) &=& {\rhoz(2)}^{*}-\rhoz(1) 
 = \rho(\hzz(2))-\rho(\zz(1)) \nn
% &=& \alpha_1\ln\frac{\vartheta_1(\zz(2)-Z_1|\ttau)\vartheta_1(\zz(1)-Z_2|\ttau)}
% {\vartheta_1(\zz(2)-Z_2|\ttau)\vartheta_1(\zz(1)-Z_1|\ttau)}+A(\zz(2)-\zz(1)) \nn
 &=& 2\alpha_1\ln\frac{\vartheta_1(\zz(1)-Z_2|\ttau)}
 {\vartheta_1(\zz(1)-Z_1|\ttau)}+A(\zz(2)-\zz(1)) \nn
% &=& 2\alpha_1\ln\frac{\vartheta_1(\ttau(x+y)|\ttau)}
%            {\vartheta_2(\ttau(x-y)|\ttau)}
% +4\pi i\alpha_F\ttau y+\pi i(\alpha_1-\alpha_F) \nn
&&\hspace{-4em}\Rightarrow\quad 
\label{eq:np-tau1}
 \frac{\tau_1}{\alpha_1} = 
 2\ln\frac{\vartheta_1(\ttau(x+y)|\ttau)}{\vartheta_2(\ttau(x-y)|\ttau)}
 +4\pi i\frac{\alpha_F}{\alpha_1}\ttau y -\pi i \,,\\
&&\hspace{-4em} g_1(\ttau(x+y)|\ttau)+g_2(\ttau(x-y)|\ttau) 
 = -2\pi i\frac{\alpha_F}{\alpha_1}\,.
\hspace{5em}
\label{eq:np-y}
\end{eqnarray}

%%%%%%%%%%%%%%%%%%%%%%%%%%%%%%%%%%%%%%%%%%%%%%%%%%%%%%%%%%%%%%%%%%%%%%%%
\subsection{$V_{\propto}$ diagram}

\begin{figure}[tb]
   \epsfxsize= 11cm   %or \epsfysize= HEIGHT cm
   \centerline{\epsfbox{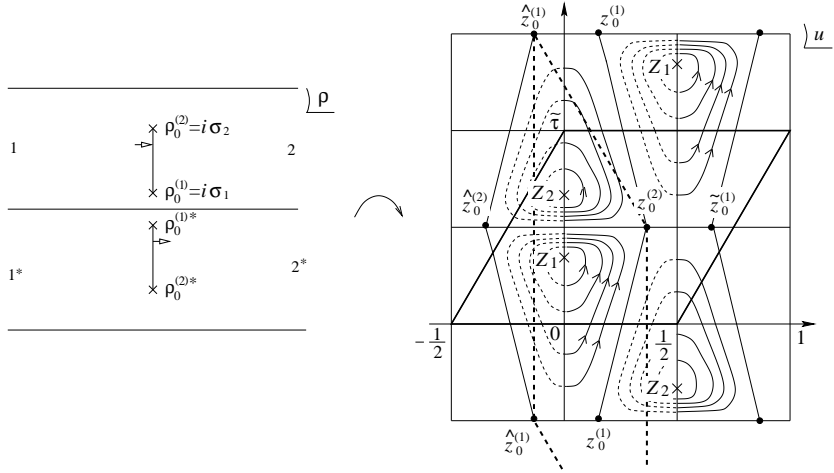}}
 \caption{Conformal mapping of the $\rho$ plane of $V_{\propto}$ diagram 
to a torus. In the $\rho$ plane, the lines connecting two crosses denote 
crosscaps across which the left edge in the upper plane is identified 
with the right edge in the lower plane. In the $u$ plane, the 
parallelogram enclosed by bold solid line denotes a fundamental region 
of the torus corresponding to the moduli $\ttau+1/2$, while that 
enclosed by bold dotted line denotes another choice of fundamental 
region corresponding to the moduli $\tau/4+1/2$ used later.} 
\label{fig:Valphamap}
\end{figure}

The diagram obtained by using $V_{\propto}$ vertex once is a tree diagram 
from the SFT viewpoint, but is actually a one-loop diagram from the CFT point of view. 
The mapping of the $V_{\propto}$ diagram to torus is drawn 
in Fig.~\ref{fig:Valphamap}. 
%\begin{wrapfigure}[6]{r}{6.6cm}
%\end{wrapfigure}
The period $\tau$ in this case is $\ttau+1/2$, as is seen from the 
fundamental region of the torus in Fig.~\ref{fig:Valphamap}. The two 
interaction points in the fundamental region are 
\begin{equation}
\label{eq:zeropt}
\cases{
         \tzz(1)=\zz(1)+(\ttau+\half)=\hzz(2)+1 \cr
         \zz(2)=\hzz(1)+(\ttau+\half)           \cr},
\end{equation}
and we use the parametrization 
\begin{equation}
\label{eq:va-paramet}
 \cases{
         Z_1=\ttau\left(\half-x\right) \cr
         Z_2=\ttau\left(\half+x\right) \cr}, \qquad 
 \cases{
         \tzz(1)=\frac{\ttau}{2}+\left(\half+y\right) \cr
         \zz(2)=\frac{\ttau}{2}+\left(\half-y\right) \cr}.
\end{equation}
Since $\zz(1)+2\ttau$ and $\zz(1)$ on the $u$ plane correspond to a 
single point $\rhoz(1)$ on the $\rho$ plane, we find, using the period 
$\ttau+1/2$ of $\rho(u)$,
\begin{eqnarray}
 0 &=& \rho(\zz(1)+2\ttau)-\rho(\zz(1))
 =\rho(\zz(1)+2(\ttau+\shalf)-1)-\rho(\zz(1)) \nn
 &=& 4\pi i\alpha_1(Z_1-Z_2)+A(2(\ttau+\shalf)-1)
 = -8\pi i\alpha_1\ttau x+2\ttau A \nn
&& \hspace{-2em}\Rightarrow\quad  A = 4\pi i \alpha_1x.
\label{eq:va-periodtau}
\end{eqnarray}
Then, $2\pi i\alpha_1 (Z_1-Z_2)=-\ttau A$, and hence the periodicity relation 
(\ref{eq:periodicity}) can be rewritten as
\begin{equation}
\rho(u+m(\ttau+\shalf)+n)-\rho(u) 
 = A\left(\frac{m}{2}+n\right).
\label{eq:va-periodicity}
\end{equation}
Using this periodicity and Eq.~(\ref{eq:zeropt}), we obtain
\begin{eqnarray}
 A &=&
 \rho(\tzz(1))-\rho(\hzz(2))=\rho(\zz(1)+(\ttau+\shalf))-\rho(\hzz(2)) \nn
 &=& \rhoz(1)+\frac{A}{2}-{\rhoz(2)}^{*}=i(\sigma_1+\sigma_2)+\frac{A}{2} \nn
&& \hspace{-2em}\Rightarrow\quad 
\label{eq:va-period1}
 \sigma_1+\sigma_2 = 2\pi\alpha_1 x
\end{eqnarray}
and also 
\begin{eqnarray}
 i(\sigma_2-\sigma_1) &=& \rho(\zz(2))-\rho(\zz(1))
 =\rho(\zz(2))-\rho(\tzz(1)-(\ttau+\shalf)) \nn
 &=& \rho(\zz(2))-\rho(\tzz(1))+\frac{A}{2} \nn
 &=& 2\alpha_1\ln\frac{\vartheta_1(\tzz(1)-Z_2|\ttau+\half)}
 {\vartheta_1(\tzz(1)-Z_1|\ttau+\half)}+A(\zz(2)-\tzz(1))+\frac{A}{2} \nn
&& \hspace{-5.5em}\Rightarrow\quad 
\label{eq:va-sig-sig}
 i{\sigma_2-\sigma_1\over\alpha_1} =  
 -2\ln\frac{\vartheta_2(\ttau x+y|\ttau+\half)}
{\vartheta_2(\ttau x-y|\ttau+\half)}
 +2\pi i x(1-4y)\,.
\end{eqnarray}
The equation determining the interaction point $y$ is again the same as 
the stationarity condition $\partial(\sigma_1-\sigma_2)/\partial y=0$:
\begin{equation}
\label{eq:va-y}
 g_2(\ttau x+y|\ttau+\shalf)+g_2(\ttau x-y|\ttau+\shalf) 
 = -4\pi ix\,.
\end{equation}

%%%%%%%%%%%%%%%%%%%%%%%%%%%%%%%%%%%%%%%%%%%%%%%%%%%%%%%%%%%%%%%%%%%%%%%%%
\subsection{$UU$ diagram}

\begin{figure}[tb]
   \epsfxsize= 11cm   %or \epsfysize= HEIGHT cm
   \centerline{\epsfbox{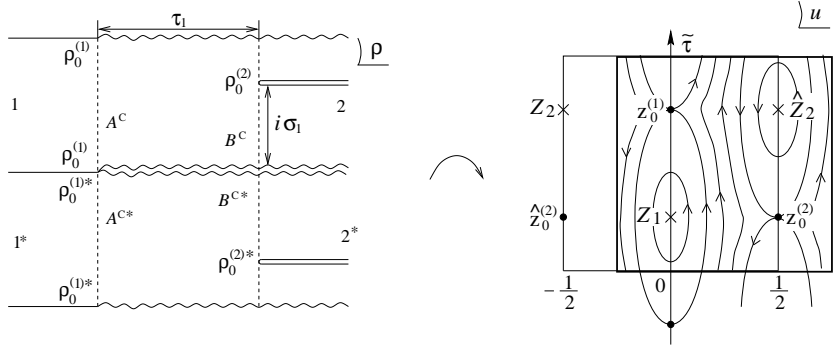}}
 \caption{Conformal mapping of the $\rho$ plane of $UU$ diagram 
to a torus.}
 \label{fig:UUmap}
\end{figure}

The diagram obtained by using $U$ vertex twice is again a tree diagram 
from the SFT viewpoint but is a one-loop diagram from the CFT viewpoint.
The mapping of the $UU$ diagram to torus is drawn 
in Fig.~\ref{fig:UUmap}. 
%\begin{wrapfigure}[6]{r}{6.6cm}
%\end{wrapfigure}
The period $\tau$ in this case is $\ttau$, the same as in planar case, and 
we parametrize the positions of punctures and interaction points as
\begin{equation}
\label{eq:uu-paramet}
 \cases{
         Z_1=\phantom{\half+{}}\ttau\left(\half-x\right) \cr
         Z_2=-\half+\ttau\left(\half+x\right) \cr}, \qquad 
 \cases{
         \zz(1)=\phantom{\half+{}}\ttau\left(\half+y\right) \cr
         \zz(2)=\half+\ttau\left(\half-y\right)    \cr}.    
\end{equation}
From periods $\ttau$ and 1, we obtain
\begin{eqnarray}
 \pi i\alpha_1 &=& \rho(\zz(1))-\rho(\zz(1)-\ttau) =
 2\pi i\alpha_1(Z_1-Z_2)+A\ttau \nn
&\Rightarrow& \quad
\label{eq:uu-periodtau}
 A = 4\pi i\alpha_1 x \,,\\
 2i\sigma_1 &=& \rho(\zz(2))-\rho(\hzz(2)) =
 \rho(\hzz(2)+1)-\rho(\hzz(2)) = A \nn
&\Rightarrow& \quad
\label{eq:uu-period1}
 \sigma_1 = 2\pi\alpha_1 x\,.
\end{eqnarray}
As for the relation between $\tau_1$ and the torus moduli $\ttau$, we find
\begin{eqnarray}
 \tau_1+i\sigma_1 &=& \rho(\zz(2))-\rho(\zz(1)-\ttau)
 = \rho(\zz(2))-\rho(\zz(1))+\pi i\alpha_1 \nn
 &=& 2\alpha_1\ln\frac{\vartheta_1(\zz(1)-Z_2|\ttau)}
 {\vartheta_1(\zz(1)-Z_1|\ttau)}+A(\zz(2)-\zz(1))+\pi i\alpha_1 \nn
% &=& -2\alpha_1\ln\frac{\vartheta_1(\ttau(x+y)|\ttau)}
% {\vartheta_2(\ttau(x-y)|\ttau)}+A\left(\half-2\ttau y\right)
% +\pi i\alpha_1 \nn
&& \hspace{-3.5em}\Rightarrow\quad 
\label{eq:uu-tau1}
 \frac{\tau_1}{\alpha_1} = 
 -2\ln\frac{\vartheta_1(\ttau(x+y)|\ttau)}{\vartheta_2(\ttau(x-y)|\ttau)}
 -8\pi ix\ttau y+\pi i\,.
\end{eqnarray}
Stationarity $\partial\tau_1/\partial y=0$ determines the interaction point $y$:
\begin{equation}
\label{eq:uu-y}
 g_1(\ttau(x+y)|\ttau)+g_2(\ttau(x-y)|\ttau) 
 = -4\pi ix\,.
\end{equation}

\section{Jacobian}

In the explicit computation of the amplitudes below, we will need to make 
change of the variables from the moduli parameters on the $\rho$ plane to the 
torus moduli $\tau$ and $x$. Here we compute the Jacobian for this change 
of variables, for each case of the diagrams.

\subsection{general}

Since the Jacobi $\vartheta$ functions satisfy
\begin{equation}
{\partial\vartheta_i(\nu|\tau)\over\partial\tau} = -{i\over4\pi}{\partial^2\vartheta_i(\nu|\tau)\over\partial\nu^2}\,,  
\end{equation} 
we have the following relation using $g_i(\nu|\tau)$ defined in 
Eq.~(\ref{eq:gi}):
\begin{equation}
{\partial\over\partial\tau}\ln \vartheta_i(\nu|\tau) 
= -{i\over4\pi}{\vartheta''_i\over\vartheta_i}(\nu|\tau)
= -{i\over4\pi}(g'_i+g^2_i)(\nu|\tau)\,. 
\end{equation}
Using this equation, we can $\tau$-differentiate Eq.(\ref{eq:rho(u)}) with 
$u$ generally dependent of $\tau$:
\begin{equation}
\label{eq:drdt(u)}
 \frac{\partial\rho(u)}{\partial\tau} = 
 \frac{\partial u}{\partial\tau}\frac{d\rho}{du}
 -\sum_{r=1,2}\alpha_r\frac{\partial Z_r}{\partial\tau}g_1(u-Z_r)
 -\sum_{r=1,2}\frac{i\alpha_r}{4\pi}\left(g_1'(u-Z_r)+{g_1}^2(u-Z_r)\right),
\end{equation} 
where the period $\tau$ has been omitted for brevity. 
When $u=\zz(i)$ (in the fundamental region), $d\rho/du(\zz(i))=0$ because of 
Eq.~(\ref{eq:intpt}) and we have 
\begin{equation}
\label{eq:drhodtau(zzero)}
 \frac{\partial\rho(\zz(i))}{\partial\tau} = 
 -\sum_{r=1,2}\alpha_r\frac{\partial Z_r}{\partial\tau}g_1(\zz(i)-Z_r)
 -\sum_{r=1,2}\frac{i\alpha_r}{4\pi}\left(g_1'(\zz(i)-Z_r)
 +{g_1}^2(\zz(i)-Z_r)\right).
\end{equation} 
Using further the identities
\begin{eqnarray}
&& \vartheta_1(\zz(2)-Z_1)
         = \vartheta_1(\zz(1)-Z_2) 
           \qquad \hbox{when}\quad Z_1+Z_2+1=\zz(1)+\zz(2) \nn
% =\cases{
%         \vartheta_1(\zz(1)-Z_2) 
%           &for \quad $Z_1+Z_2+1=\zz(1)+\zz(2)$ \cr
%         -\vartheta_1(\zz(1)-Z_2) 
%           &for \quad $Z_1+Z_2=\zz(1)+\zz(2)$ \cr} \nn
 &&\ \ \Rightarrow\ \ g_1(\zz(2)-Z_1)= - g_1(\zz(1)-Z_2),\quad 
 g_1'(\zz(2)-Z_1)= g_1'(\zz(1)-Z_2),
\hspace{2em}
\end{eqnarray}
where $\zz(1)$ is understood to be $\tzz(1)$ for the $V_\propto$ case, 
we obtain
\begin{eqnarray}
\label{eq:rhoz-rhoz}
 &&\frac{\partial}{\partial\tau}\left(\rho(\zz(2))-\rho(\zz(1))\right) \nn
 &&= \alpha_1\frac{\partial(Z_1-Z_2)}{\partial\tau}\left(g_1(\zz(1)-Z_1)
 +g_1(\zz(1)-Z_2)\right) \nn
 &&\qquad {} + \frac{i\alpha_1}{2\pi}\left(g_1'(\zz(1)-Z_1)
 -g_1'(\zz(1)-Z_2)+{g_1}^2(\zz(1)-Z_1)-{g_1}^2(\zz(1)-Z_2)\right) \nn
 &&= \frac{i\alpha_1}{2\pi}\big(g_1(\zz(1){-}Z_1)+g_1(\zz(1){-}Z_2)\big)
 \left\{g_1(\zz(1){-}Z_1)-g_1(\zz(1){-}Z_2)-2\pi i
 \frac{\partial(Z_1{-}Z_2)}{\partial\tau}\right\} \nn
 &&\qquad {} + \frac{i\alpha_1}{2\pi}\left[\,g_1'(\zz(1)-Z_1)
 -g_1'(\zz(1)-Z_2)\,\right].
\end{eqnarray}
In the present case of tachyon amplitude, we shall see below that the 
amplitudes for P, NO (M1 and M2) and NP diagrams contain delta function 
factor $\delta(\tau_1-\tau_2)$ (which is always the case when the external 
string states have no excitations of $\alpha_n^-$ modes (See \S7)).  Then, 
as seen in Eqs.~(\ref{eq:pl-peritau}), (\ref{eq:no-peritau}) and 
(\ref{eq:np-peritau}) in those cases of P, NO (M1 and M2) and NP diagrams, 
$\tau_1=\tau_2$ means
\begin{equation}
 A = 2\pi i\alpha_F = 4\pi i\alpha_1 x\,,
\end{equation}
the same expression as Eqs.~(\ref{eq:va-periodtau}) and 
(\ref{eq:uu-periodtau}) for the $V_{\propto}$ and $UU$ amplitude cases. Thus,
for all amplitude cases, we have 
\begin{equation}
 -2\pi i\frac{\partial(Z_1-Z_2)}{\partial\tau} = 4\pi ix = \frac{A}{\alpha_1}\,,
\end{equation}
which together with Eq.(\ref{eq:intpt}) implies that the part 
enclosed by a curly bracket in Eq.~(\ref{eq:rhoz-rhoz}) vanishes. 
Consequently we obtain
\begin{equation}
\label{eq:rhoz-rhoz-simple}
 \frac{\partial}{\partial\tau}\left(\rho(\zz(2))-\rho(\zz(1))\right) 
 = \frac{i\alpha_1}{2\pi}\left[\,g_1'(\zz(1)-Z_1|\tau)
 -g_1'(\zz(1)-Z_2|\tau)\,\right]. 
\end{equation} 

\subsection{P, NO, NP}

In these cases under the condition $\tau_1=\tau_2$, remaining two 
free parameters are
\begin{eqnarray}
\label{eq:v3-para}
 \alpha_F &=& 2\alpha_1 x\,, \nn
 \tau_1 &=& \cases{
    \rho(\zz(2))-\rho(\zz(1)) &for P, NO \cr
    \rho(\zz(2))-\rho(\zz(1))-\pi i\alpha_1+2\pi i\alpha_1x &for NP \cr}.
\end{eqnarray}  
Taking account of Eq.(\ref{eq:rhoz-rhoz-simple}), 
the Jacobian for the change of variables $(\tau_1, \alpha_F)\ \rightarrow\ (\tau, x)$ 
has the form
\begin{eqnarray}
\label{eq:v3-jacobian}
 d\tau_1 d\alpha_F &=& \frac{\partial(\tau_1, \alpha_F)}{\partial(\tau, x)}
 d\tau dx = \frac{\partial\tau_1}{\partial\tau}\frac{\partial\alpha_F}
 {\partial x}d\tau dx \nn
 &=& 2\alpha_1\cdot\frac{i\alpha_1}{2\pi}
 \left[\,g_1'(\zz(1)-Z_1|\tau)-g_1'(\zz(1)-Z_2|\tau)\,\right]d\tau dx .
\end{eqnarray} 
Here $\tau=\ttau$ for P, NP and $\tau=\ttau+\half$ for NO, but $d\tau=d\ttau$ 
in any case. More explicitly, using Eqs.~(\ref{eq:pl-paramet}), 
(\ref{eq:np-paramet}), $g_1(u+1/2)=g_2(u)$ and $g'_i(-u)=g'_i(u)$, 
\begin{equation}
\label{eq:pnonp-jacobian}
 d\tau_1 d\alpha_F = -{{i\alpha_1}^2\over\pi}{\times}
 \cases{
    \displaystyle \left[\,
    g_2'\left(\ttau(x+y)|\ttau\right)-g_2'\left(\ttau(x-y)|\ttau\right)
    \,\right]d\ttau dx &for P \cr
    \displaystyle \left[\,
    g_2'\left(\ttau(x+y)|\ttau+\shalf\right)
    -g_2'\left(\ttau(x-y)|\ttau+\shalf\right)\,\right]
    d\ttau dx &for NO \cr
    \displaystyle \left[\,
    g_1'\left(\ttau(x+y)|\ttau\right)-g_2'\left(\ttau(x-y)|\ttau\right)
    \,\right]d\ttau dx &for NP \cr}.
\end{equation} 

\subsection{$V_{\propto}$}

Two free parameters in this $V_{\propto}$ case are
\begin{eqnarray}
\label{eq:va-para}
 i\sigma_{+} &\equiv& i(\sigma_1+\sigma_2) = 2\pi i\alpha_1 x \nn
 i\sigma_{-} &\equiv& i(\sigma_2-\sigma_1) = 
    \rho(\zz(2))-\rho(\tzz(1))+2\pi i\alpha_1x\,.
\end{eqnarray}  
The Jacobian for the change of variables $(\sigma_1, \sigma_2)\ \rightarrow\ (\tau, x)$ 
is given by
\begin{eqnarray}
\label{eq:va-jacobian}
 d\sigma_1 d\sigma_2 &=& -\half d\sigma_{-}d\sigma_{+} = 
 \half\frac{\partial(i\sigma_{-},i\sigma_{+})}{\partial(\tau, x)}
 d\tau dx = \half\frac{\partial i\sigma_{-}}{\partial\tau}
 \frac{\partial i\sigma_{+}}{\partial x}d\tau dx \nn
 &=& i\pi\alpha_1\cdot\frac{i\alpha_1}{2\pi}\left[\,
 g_1'(\tzz(1)-Z_1|\tau)-g_1'(\tzz(1)-Z_2|\tau)\,\right]d\tau dx .
\end{eqnarray} 
More explicitly, using $\tau=\ttau+1/2$ and Eq.~(\ref{eq:va-paramet}),
\begin{equation}
\label{eq:va-jacobian2}
 d\sigma_1 d\sigma_2 = 
 -\frac{{\alpha_1}^2}{2}\left[\,
 g_2'\left(\ttau x+y|\ttau+\shalf\right)
 -g_2'\left(\ttau x-y|\ttau+\shalf\right)\,\right]d\ttau dx .
\end{equation} 

\subsection{$UU$}

Free parameters in the $UU$ case are
\begin{eqnarray}
\label{eq:uu-para}
 \sigma_1 &=&2\pi\alpha_1 x \nn
 \tau_1 &=& \rho(\zz(2))-\rho(\zz(1))+\pi i\alpha_1-2\pi i\alpha_1x,
\end{eqnarray}  
so that the Jacobian for the change of variables $(\sigma_1, \tau_1)\ 
\rightarrow\ (\tau, x)$ reads
\begin{eqnarray}
\label{eq:uu-jacobian}
 d\tau_1 d\sigma_1 &=& 
 \frac{\partial(\tau_1,\sigma_1)}{\partial(\tau,x)}d\tau dx 
 = \frac{\partial\tau_1}{\partial\tau}
 \frac{\partial\sigma_1}{\partial x}d\tau dx \nn
 &=& 2 \pi\alpha_1\cdot\frac{i\alpha_1}{2\pi}\left[\,
 g_1'(\zz(1)-Z_1|\tau)-g_1'(\zz(1)-Z_2|\tau)\,\right]d\tau dx.
\end{eqnarray} 
More explicitly, using $\tau=\ttau$ and Eq.~(\ref{eq:uu-paramet}),
\begin{equation}
\label{eq:uu-jacobian2}
 d\tau_1 d\sigma_1 = 
 i{\alpha_1}^2\left[\,g_1'(\ttau(x+y)|\ttau)
 -g_2'(\ttau(x-y)|\ttau)\,\right]d\ttau dx .
\end{equation} 

\section{CFT correlation functions on the torus}

\subsection{Correlation functions and GGRT}
To compute the CFT correlation functions on the torus, we further 
map the $u$ plane to the $\trho=-2\pi iu$ plane such that 
the $\trho= \tau+i\sigma$ plane becomes of the usual open string strip with 
${\rm Im}\trho\,({=}\sigma)=0$ and $\pi$ being the open string boundaries and, 
therefore, the usual Fourier expansion $\phi(\trho)=\sum_n \phi_ne^{-n\trho}$ 
can be used there for the string coordinate and ghost fields 
$\phi=X, b$ and $c$.
Then, the CFT correlation functions can be calculated by the following 
formulas: for the cases of P and NP diagrams with period $\ttau$,
\begin{equation}
\label{eq:vertical-def1}
       \bigl<\,\prod_{r}\calO_r(u_r)\,\bigr>_{\ttau} 
       =(-2\pi i)^{\sum_r\!d_r}\tr\bigl[(-)^{N_{\rm FP}+1}\,w^{(L_{0}
       -{c/24})}\,\prod_{r}\calO_r(-2\pi iu_r)\,\bigr],
%       \left<\,\calO_1(u_1)\,\cdots\,\calO_N(u_N)\,\right>_{\ttau} 
%       \equiv(-2\pi i)^{\sum_r d_r}\tr\left[(-)^{N_{\rm FP}+1}\,w^{(L_{0}
%       -{c/24})}\,\calO_1(-2\pi iu_1)\,\cdots\,\calO_N(-2\pi iu_N)\,\right],
\end{equation} 
and for the NO (M1 and M2) cases with period $\ttau+\half$,
\begin{equation}
\label{eq:vertical-def2}
       \bigl<\,\prod_{r}\calO_r(u_r)\,\bigr>_{\ttau+\half}\! 
       =(-2\pi i)^{\sum_r\!d_r}\tr\bigl[\Omega(-)^{N_{\rm FP}+1}w^{(L_{0}
       -{c/24})}\prod_{r}\calO_r(-2\pi iu_r)\bigr],
\end{equation} 
where $w\equiv e^{2\pi i\ttau}$, $N_{\rm FP}$ is the ghost number operator 
(whose explicit expression is given shortly), 
$c$ is the central charge of the system, 
and $\Omega$ is the (open 
string) twist operator, $\Omega:u\rightarrow u+\half$ 
$(\trho \rightarrow\trho-\pi i)$. Here we 
have assumed that the operators $\calO_r$ are {\it primary fields} with 
conformal weights $d_r$, as being always the case in our computations 
below.

As noted before, these formulas are {\it not} mere definitions of the
torus CFT correlation functions but the result of the 
GGRT\cite{rf:AKT1} 
for the cases of P, NP and NO diagrams. Actually, once the period is 
specified, the functional form of the torus correlation function is 
unique but the overall normalization factor is {\it not}. GGRT 
determines those normalization factors also. 

For the remaining $UU$ and $V_\propto$ cases, these formula does not give the
correct normalization factor. First consider the $UU$ diagram case, 
in which the 
$\bra{v_{UU}}$ vertex is given in Eq.~(\ref{eq:UUvertex}) by contracting
two tree level $U$ vertices by closed reflector $\ketRc(AB)$. In view of
the $u$ plane in Fig.~\ref{fig:UUmap} in which the vertical lines 
represent the intermediate closed string, we note that the direction of 
the time evolution on the $u$ plane is not vertical but horizontal
in the $UU$ diagram case. So the 
trace operation should be taken in the horizontal direction in this $UU$ 
diagram case, and the correct mapping from the $u$ plane to $\trho$ 
plane should be $\trho=2\pi u(i/\ttau)+$const.\ such that the image of the
vertical line ${\rm Re}u={}$const., $0\leq{\rm Im}u\leq 
\abs{\ttau}(=\ttau/i)$ has the correct width $2\pi$ in $\sigma={\rm
Im}\trho$. Using this mapping and denoting $-1/\ttau\equiv\tau$, the correct
formula in this $UU$ case reads
\begin{equation}
%\label{eq:horizontal-def}
       \bigl<\,\prod_{r}\calO_r(u_r)\,\bigr>_{UU,\tau} 
       =(-2\pi i\tau)^{\sum_r\!d_r}{\tr}'\bigl[(-)^{N_{\rm FP}+1}\,q^{2(L_{0}
       -{c/24})}\prod_{r}\calO_r(-2\pi i\tau u_r)\bigr],
\label{eq:UUGGRT}
\end{equation} 
where $q\equiv e^{\pi i\tau}$, and ${\tr}'$ means that no trace is taken over the
momentum; the `trace' $\tr_0[\cdots]$ in the zero mode sector of $X$ is 
simply $\bra{-k_1/2}\cdots\ket{-k_1/2}$ where $k_1$ is 
the momentum of the external string 1.
We emphasize again that the difference between the CFT correlation
functions (\ref{eq:vertical-def1}) and (\ref{eq:UUGGRT}) computed
vertically and horizontally, respectively, in fact appears only in
their numerical coefficients and their function forms are exactly the
same.  

For completeness, we present here the proof for this formula 
(\ref{eq:UUGGRT}). It can be proven by using the GGRT for the case of 
open-string loop diagram.\cite{rf:AKT1} 
The closed string $A^\c$ can be treated 
essentially as a product of two open strings $A$ and $\bar A$, 
corresponding to the holomorphic and anti-holomorphic pieces. This 
identification is, however, slightly violated in the zero-mode sector of
$X$, since there exists only a single zero-mode $x$ (or $p$) common to 
the holomorphic and anti-holomorphic parts. To make the identification 
exact, we can extend the zero mode sector such that both the holomorphic
and anti-holomorphic sectors have their own zero modes $p$ and $\bar p$,
and identify the original state $\ket{p_1}$ to be $\ket{p,\bar p}$ with 
$p+\bar p =p_1$ and $p-\bar p=0$. Then, taking account of $dp\,d\bar p = 
d(p+\bar p)\,d((p-\bar p)/2)$, we can identify the closed reflector 
$\ketRc(AB)$ as 
\begin{equation}
\ketRc(AB) = (2\pi)^d\delta^d\bigl({\hat p_A-{\hat p}_{\bar A}\over2}\bigr)
\ketRo(AB)\ketRo({\bar A}{\bar B})
\label{eq:closedRfl}
\end{equation}
and the vertex $\v2(1A)$, for instance, becomes to have momentum
conservation factor $\delta^d(p_A+\bar p_A + k_1)$ instead of 
$\delta^d(p_A+k_1)$. With this device, we now apply the 
GGRT\cite{rf:LPP2,rf:AKT1} for the open string case to the present $UU$ 
vertex contracted by the closed string reflector:
with $\rho_1\equiv\tau_1+i\sigma_1$,
\begin{eqnarray}
&& \bra{v_{UU}(\tau_1,\sigma_1)}\nn %\ket{\varphi_2}\ket{\varphi_1}  \nn
 &&\quad = \v2(1A)\v2(2B)  e^{-L_0\rho_1}e^{-\bar L_0\bar\rho_1}
  \ketRc(AB)\nn %\,\ket{\varphi_2}\ket{\varphi_1} \nn
 &&\quad = \sbra{u(\bar{A},1,A)} e^{-L_0\rho_1}\sbra{u(B,2,\bar{B})}
e^{-\bar L_0\bar\rho_1}(2\pi)^d\delta^d\bigl({\hat p_A-{\hat p}_{\bar A}\over 
 2}\bigr)
  \ketRo(AB)\ketRo({\bar A}{\bar B}) \nn
 &&\quad = \sbra{u(\bar{A},1,A)} e^{-L_0\rho_1}\sbra{u(B,2,\bar{B})}\ketRo(AB)
e^{-\bar L_0\bar\rho_1}(2\pi)^d\delta^d\bigl(\hat p_{\bar A}+{k_1\over2}\bigr)
  \ketRo({\bar A}{\bar B})\nn
 &&\quad = \sbra{v(\bar{A},1,2,\bar{B})}e^{-\bar L_0\bar\rho_1}
(2\pi)^d\delta^d\bigl(\hat p_{\bar A}+{p_1\over2}\bigr)
  \ketRo({\bar A}{\bar B})\,,
\end{eqnarray}
where we have used the momentum conservation 
$\sbra{u(\bar{A},1,A)}(\hat p_A+\hat p_{\bar A}+k_1)=0$ 
in going to the third expression, and the tree level GGRT  
$\sbra{v(\bar{A},1,2,\bar{B})}
=\sbra{u(\bar{A},1,A)} e^{-L_0\rho_1}\sbra{u(B,2,\bar{B})}\ket{R(A,B)}$
in going to the last line. Since every quantities are now of
open-string, we can apply  the loop level GGRT proven in
Ref.~\citen{rf:AKT1} to the last expression and obtain  
\begin{eqnarray}
&& \bra{v_{UU}(\tau_1,\sigma_1)}\ket{\Phi_2}\ket{\Phi_1}  \nn
 &&\qquad = \int{d^d{p}\over(2\pi)^d}\bra{{p}} 
(2\pi)^d\delta^d\bigl(\hat p_{\bar A}+{k_1\over2}\bigr)\tr_{n\not=0}
\bigl[(-)^{N_{\rm FP}+1}\,q^{2L_0}
  \prod_i h_i[\Phi_i]\bigr] 
\ket{{p}} \nn
 &&\qquad = \bra{-{k_1\over2}} 
\tr_{n\not=0}\bigl[(-)^{N_{\rm FP}+1}\,q^{2L_0}
  \prod_i h_i[\Phi_i]\bigr] \ket{-{k_1\over2}}.
\end{eqnarray}
Note that the bra and ket zero-mode states here correspond to 
the `open strings' of anti-holomorphic parts $\bar B$ and $\bar A$,
respectively. Therefore this expression corresponds to the cutting of
the torus on the $u$ plane as indicated by bold line in 
Fig.~\ref{fig:UUmap}, and also shows that the time evolution direction 
is horizontal as claimed above. 

Finally, consider the $V_\propto$ case. Actually we have not given a precise 
definition of the $\bra{v_\propto}$ vertex in the preceding papers I and II, 
since it refers to CFT on the torus. Nevertheless, we have used in I 
the GGRT that the two glued vertices 
\begin{eqnarray}
\bra{v_1(1,4^\c;\sigma_0,\theta,\tau)} 
&\equiv&\v2(12)\v\infty(34{\sigma_0})e^{-\tau(L+\bar L)}e^{i\theta(L-\bar L)}\ketRc(23)\,, \nn
\bra{v_2(1,4^\c;\sigma_1,\sigma_2,\tau)}&\equiv&
-\v\propto(12{\sigma_1}{\sigma_2})\v2(34)e^{-\tau L}\ketRo(23)\,,
\end{eqnarray}
become identical at $\tau=0$. Since $\bra{v_\infty}$ and $\bra{u}$ are already 
defined, this identity fixes the normalization of the $\bra{v_\propto}$ vertex. 
The minus sign in the expression of $\bra{v_2}$ here is because the 
contraction was taken by $\ketRo(32)=-\ketRo(23)$ in I. However, 
the contraction by $\ketRo(23)$ is more natural in the sign from the 
GGRT view point, so we here require that $\bra{v_1(1,4^\c;\sigma_0,\theta,\tau)}= 
-\bra{v_2(1,4^\c;\sigma_1,\sigma_2,\tau)}$ holds at $\tau=0$ by changing the sign 
convention of the $\bra{v_\propto}$ vertex and hence of the coupling constant 
$x_u$ from I and II. In order for this identity to hold, the $\bra{v_\propto}
$ vertex should be defined by referring to the `horizontal' computation 
as in the $UU$ case. Indeed the first glued vertex $\bra{v_1}$ is given 
by contraction using the closed string reflector $\ketRc(23)$ which 
contains the delta function constraining the intermediate state momentum
as in Eq.~(\ref{eq:closedRfl}). To perform horizontal calculation, we 
note that the `vertical' period $\ttau+1/2$ is equivalent to the 
`horizontal' period 
\begin{equation}
{1-(\ttau+1/2)\over1-2(\ttau+1/2)}={\tau\over4}+\half,
\label{eq:modulartrf}
\end{equation} 
which corresponds to taking the torus fundamental region to be the region 
enclosed by the dotted bold line in Fig.~\ref{fig:Valphamap}, and the 
fundamental region on the $u$ plane is mapped to $\trho$-plane by 
$\trho=2\pi iu/2\ttau+{\rm const.}=-\pi i\tau u+{\rm const.}$ such that the 
image of the vertical line ${\rm Re}u=-y$, $-\abs{\ttau}/2\leq{\rm 
Im}u\leq3\abs{\ttau}/2$ has the correct width $2\pi$ in $\sigma={\rm Im}\trho$.
Using this mapping, the correct formula for defining the $V_\propto$ vertex 
is:
\begin{eqnarray}
&&\bigl<\,\prod_{r}\calO_r(u_r)\,\bigr>_{V_\propto,{\tau\over4}+\half}  \nn
&&\quad  =(-\pi i\tau)^{\sum_r\!d_r}\bra{0}\tr_{n\not=0}
\bigl[\Omega(-)^{N_{\rm FP}+1}\,q^{\half(L_{0}
       -{c/24})}\prod_{r}\calO_r(-\pi i\tau u_r)\bigr] \ket{0},
\label{eq:vadef}
\end{eqnarray} 
where $q\equiv e^{\pi i\tau}$, and $\bra{0}\cdots\ket{0}$ means that 
the expectation value is taken with momentum 0 state 
in the zero mode sector of $X$. If we have adopted the `vertical' 
definition (\ref{eq:vertical-def2}) for this case also, the weight 
would be different by an intriguing factor $2^{d/2}/(2\pi)^di$:
\begin{equation}
\bigl<\,\prod_{r}\calO_r(u_r)\,\bigr>_{\ttau+\half}
={2^{d/2}\over(2\pi)^di}
\bigl<\,\prod_{r}\calO_r(u_r)\,\bigr>_{V_\propto,{\tau\over4}+\half}
\label{eq:varel}
\end{equation}
for the operator $\prod_{r}\calO_r(u_r)
=b_{\sigma_1}b_{\sigma_2}\,c(Z_2)e^{ik_2\cdot X(Z_2)}
c(Z_1)e^{ik_1\cdot X(Z_1)}$ relevant here, as we shall see below.

Let us now evaluate the ghost and $X$ parts, separately.

\subsection{ghost part}

As is seen in Eqs.~(\ref{eq:CFTfnX}), (\ref{eq:CFTfnUU}) and 
(\ref{eq:Valphaamp}), the correlation functions we need in this paper 
have the following form as their ghost parts:
\begin{equation}
 \left<\,b(u_1)b(u_2)c(Z_2)c(Z_1)\,\right>_J
\quad \hbox{where}\ \  
 J = \cases{
      \ttau       & for P and NP \cr
      \ttau+\half & for NO (M1, M2) \cr
      \tau        & for $UU$ \cr
      {\tau\over4}+\half & for $V_\propto$ \cr}
\end{equation}

\subsubsection{P and NP}

First consider the planar and nonplanar diagram cases with period $\ttau$, 
to which the formula (\ref{eq:vertical-def1}) applies. Substituting into
it the expansion of the ghost fields on the $\trho=-2\pi iu$ plane
\begin{equation}
c(\trho)=\sum_n c_n e^{-n\trho},\qquad 
b(\trho)=\sum_n b_n e^{-n\trho},
\end{equation}
we evaluate the ghost correlation function as follows:\footnote{Note that 
the time ordering is always implied in any CFT correlation functions. 
The operators are rearranged in the order of time in the first 
equation here.}
\begin{eqnarray}
 &&\left<\,b(u_1)\,b(u_2)\,c(Z_2)\,c(Z_1)\,\right>_{\ttau} 
 =\left<\,b(u_2)\,c(Z_2)\,b(u_1)\,c(Z_1)\,\right>_{\ttau} 
\nn
 &&=(-2\pi i)^{2\cdot 2-2\cdot 1}\tr\!\left[(-)^{N_{\rm FP}+1}\,
 w^{(L_0^{\rm FP}+\frac{1}{12})}\,b(-2\pi iu_2)\,
 c(-2\pi iZ_2)\,b(-2\pi iu_1)\,c(-2\pi iZ_1)\right] \nn
 &&= -(2\pi i)^2 w^{\frac{1}{12}}\tr\Bigl[(-)^{N_{\rm FP}}
 w^{L_0^{\rm FP}}\bigl\{b_0c_0b_0c_0+\sum_{n\not=0}
\bigl(b_0c_0b_nc_{-n}e^{2\pi in(u_1-Z_1)}
 \nn
 &&\ \ + b_0c_0c_{-n}b_ne^{2\pi in(u_1-Z_2)}\!
 +c_0b_0b_nc_{-n}e^{2\pi in(u_2-Z_1)}\!+b_0c_0b_nc_{-n}e^{2\pi in(u_2-Z_2)}
 \bigr)\!\bigr\}\!\Bigr]\!,
\hspace{2em}
\label{eq:4.6}
\end{eqnarray}
where use has been made of the ghost central charge $c_{\rm FP}=-26$ and
\begin{eqnarray}
 N_{\rm FP} &=& c_0b_0+\sum_{n\geq1}\left(c_{-n}b_n-b_{-n}c_n\right),
 \nn
 L_0^{\rm FP}&=&\sum_{n\geq1}n\left(c_{-n}b_n+b_{-n}c_n\right) 
 = L_{0\,{\rm CFT}}^{\rm FP} +1 .
\end{eqnarray}
The trace $\tr$ can be calculated mode by mode, 
$\tr=\tr_0\times\prod_{n=1}^\infty\tr_n$. Noting, in particular, the zero-mode 
part trace formula 
$\tr_0[(-)^{N_{\rm FP}}b_0c_0] =-\tr_0[(-)^{N_{\rm FP}}c_0b_0] = 1$ and 
$\tr_0[(-)^{N_{\rm FP}}1] = \tr_0[(-)^{N_{\rm FP}}c_0] 
= \tr_0[(-)^{N_{\rm FP}}b_0] = 0$, as explained in 
Ref.~\citen{rf:AKT1}, we obtain
\begin{eqnarray}
 && \tr\bigl[(-)^{N_{\rm FP}}w^{L_0^{\rm FP}}b_0c_0\bigr]= 
    \left[\,f(w)\,\right]^2 
   \qquad \qquad \bigl(f(w)\equiv\prod_{n=1}^\infty(1-w^n)\bigr) \nn
 && \tr\bigl[(-)^{N_{\rm FP}}w^{L_0^{\rm FP}}b_0c_0\sum_{l\not=0}
   \pmatrix{b_lc_{-l}\cr c_{-l}b_l\cr}
       e^{2\pi ilu}\bigr]=\left[\,f(w)\,\right]^2 \Bigl(
    \pm\frac{i}{2\pi}g_1(u|\ttau)-\half\Bigr).
\hspace{2em}\label{eq:4.8}
\end{eqnarray}
Thus, together with the help of Eq.~(\ref{eq:drho/du}), 
the RHS of Eq.~(\ref{eq:4.6}) reduces to
\begin{eqnarray}
\label{eq:vev-ttau}
% &&\left<\,b(u_1)\,b(u_2)\,c(Z_2)\,c(Z_1)\,\right>_{\ttau} \nn
  &=& 2\pi i\left[\,w^{\frac{1}{24}}f(w)\,\right]^2 
 \bigl\{g_1(u_1{-}Z_1|\ttau)-g_1(u_1{-}Z_2|\ttau)
       -g_1(u_2{-}Z_1|\ttau)+g_1(u_2{-}Z_2|\ttau)\bigr\} \nn
 &=& \frac{2\pi i}{\alpha_1}\left[\,w^{\frac{1}{24}}f(w)\,\right]^2
     \left\{\,\frac{d\rho}{du}(u_1)-\frac{d\rho}{du}(u_2)\right\}
 \equiv\calG
     \left\{\,\frac{d\rho}{du}(u_1)-\frac{d\rho}{du}(u_2)\right\}.
\end{eqnarray}

\subsubsection{NO (M1 and M2)}

Next we consider the nonorientable diagram case with 
period $\ttau+1/2$, to which the formula (\ref{eq:vertical-def2}) 
applies. In this case, the twist operator $\Omega$ is additionally inserted, 
and its effect can simply be taken into account by making replacements 
$w\rightarrow-w$ and $\ttau\rightarrow\ttau+1/2$. So we find\footnote{Note that the Fock 
vacuum $\ket{1}\equiv c_1\ket{0}$ is {\it even} under the twist $\Omega$, $\Omega 
\ket{1}=+\ket{1}$, so $SL(2;C)$ vacuum $\ket{0}$ is {\it odd}.}
\begin{eqnarray}
 && \tr\bigl[\Omega(-)^{N_{\rm FP}}w^{L_0^{\rm FP}}b_0c_0\bigr]= 
    \left[\,f(-w)\,\right]^2, \nn
 && \tr\bigl[\Omega(-)^{N_{\rm FP}}w^{L_0^{\rm FP}}\!b_0c_0\sum_{l\not=0}
    b_lc_{-l}e^{2\pi ilu}\bigr]=\left[\,f(-w)\,\right]^2 \left(
    \frac{i}{2\pi}g_1(u|\ttau+\shalf)-\half\right), 
\hspace{3em}
\end{eqnarray}
etc., and hence 
\begin{eqnarray}
\label{eq:vev-ttau+half}
&&\hspace{-1em}
\left<\,b(u_1)\,b(u_2)\,c(Z_2)\,c(Z_1)\,\right>_{\ttau+\half} \nn
 &&= 2\pi i\left[w^{\frac{1}{24}}f(-w)\right]^2 
 \bigl\{g_1(u_1-Z_1|\ttau+\shalf)-g_1(u_1-Z_2|\ttau+\shalf) \nn
&&\hspace{9.5em}
   -g_1(u_2-Z_1|\ttau+\shalf)+g_1(u_2-Z_2|\ttau+\shalf)\bigr\} \nn
 &&= \frac{2\pi i}{\alpha_1}\left[w^{\frac{1}{24}}f(-w)\right]^2
     \left\{\,\frac{d\rho}{du}(u_1)-\frac{d\rho}{du}(u_2)\right\}
 \equiv\calG^N
     \left\{\,\frac{d\rho}{du}(u_1)-\frac{d\rho}{du}(u_2)\right\}.
\hspace{2em}
\end{eqnarray}

%%%%%%%%%%%%%%%%%%%%%%%%%%%%%%%%%%%%%
\subsubsection{$UU$}

Thirdly we consider the $UU$ diagram case, to which the formula 
(\ref{eq:UUGGRT}) applies. Comparing this formula with 
the previous one (\ref{eq:vertical-def1}) for P and NP cases, 
we immediately see that we obtain the desired result in this case 
by making replacements $w\rightarrow q^2$, $u\rightarrow\tau u$ and $(-2\pi i)^2\rightarrow(-2\pi i\tau)^2$ 
(for the conformal factor) in the above first result (\ref{eq:vev-ttau}):
\begin{eqnarray}
\label{eq:vev-tau}
&&\hspace*{-2em}\left<b(u_1)b(u_2)c(Z_2)c(Z_1)\right>_{\tau} \nn
% &&=(-2\pi i\tau)^{2\cdot 2-2\cdot 1}\tr\left[\,(-)^{N_{\rm FP}+1}\,
% q^{2(L_0^{\rm FP}+\frac{1}{12})}\,b(-2\pi i\tau u_1)\,b(-2\pi i\tau u_2)\,
% c(-2\pi i\tau Z_2)\,c(-2\pi i Z_1)\,\right] \nn
 &&= 2\pi i\tau^2\left[\,q^{\frac{1}{12}}f(q^2)\,\right]^2 
 \bigl\{g_1(\tau(u_1-Z_1)|\tau)-g_1(\tau(u_1-Z_2)|\tau) \nn
&&\hspace{9.5em} -g_1(\tau(u_2-Z_1)|\tau)+g_1(\tau(u_2-Z_2)|\tau)\bigr\}. 
\end{eqnarray}
But, the functions $f$ and $g_1$ have simple transformation properties 
under the Jacobi imaginary transformation $\tau\rightarrow-1/\tau=\ttau$:
\begin{equation}
 \left[\,w^{\frac{1}{24}}f(w)\,\right]^2 
%   = \left(-\frac
%    {\ln q}{\pi}\right)\left[\,q^{\frac{1}{12}}f(q^2)\,\right]^2 
    = -i\tau\left[\,q^{\frac{1}{12}}f(q^2)\,\right]^2, \qquad 
 g_1(\tau u|\tau) = \frac{1}{\tau}g_1(u|\ttau)-2\pi iu.
\label{eq:Jacobi}
\end{equation}
Owing to these, the present correlation function (\ref{eq:vev-tau}) 
in fact turns out to equal the previous one (\ref{eq:vev-ttau}) 
up to an overall factor $i$:
\begin{equation}
\label{eq:vev-relation}
 \left<\,b(u_1)\,b(u_2)\,c(Z_2)\,c(Z_1)\,\right>_{\tau} 
 = i\,\left<\,b(u_1)\,b(u_2)\,c(Z_2)\,c(Z_1)\,\right>_{\ttau}.
\end{equation}
This reflects the modular invariance of the theory, but note that 
the factor $i$ difference remains here contrary to the vacuum energy. 

\subsubsection{$V_\propto$}

Final is the $V_\propto$ diagram case, to which the formula 
(\ref{eq:vadef}) applies. Comparison of this formula with 
Eq.~(\ref{eq:vertical-def2}) for NO case shows that
the result in this case can be obtained 
by making replacements $w\rightarrow q^{1/2}$, $u\rightarrow(\tau/2)u$ and 
$(-2\pi i)^2\rightarrow(-\pi i\tau)^2$ 
in the above result (\ref{eq:vev-ttau+half}) for NO case:
\begin{eqnarray}
\label{eq:vev-tau/4+half}
&&\hspace*{-1em}\left<b(u_1)b(u_2)c(Z_2)c(Z_1)\right>_{{\tau\over4}+\half} \nn
 &&= 2\pi i\left({\tau\over2}\right)^2\left[\,q^{\frac{1}{48}}f(-\sqrt q)\,\right]^2 
 \bigl\{g_1({\textstyle{\tau\over2}}(u_1-Z_1)|{\textstyle{\tau\over4}+\half})
-g_1({\textstyle{\tau\over2}}(u_1-Z_2)|{\textstyle{\tau\over4}+\half}) \nn
&&\hspace{9em} -g_1({\textstyle{\tau\over2}}(u_2-Z_1)|{\textstyle{\tau\over4}+\half})+g_1({\textstyle{\tau\over2}}(u_2-Z_2)|{\textstyle{\tau\over4}+\half})\bigr\}. 
\hspace{2em}
\end{eqnarray}
Since the functions $f$ and $g_1$ have the following transformation laws 
under the modular transformation (\ref{eq:modulartrf})
\begin{equation}
 \left[\,w^{\frac{1}{24}}f(-w)\,\right]^2 
    = -i{\tau\over2}\left[\,q^{\frac{1}{48}}f(-\sqrt q)\,\right]^2, \qquad 
 g_1({\textstyle{\tau\over2}}u|{\textstyle{\tau\over4}+\half}) 
= \frac{2}{\tau}g_1(u|\ttau+\shalf)-4\pi iu,
\label{eq:Jacobihalf}
\end{equation}
the present correlation function (\ref{eq:vev-tau/4+half}) again 
turns out to equal the previous one (\ref{eq:vev-ttau+half}) 
up to a factor $i$:
\begin{equation}
\label{eq:vev-relationhalf}
 \left<\,b(u_1)\,b(u_2)\,c(Z_2)\,c(Z_1)\,\right>_{{\tau\over4}+\half} 
 = i\,\left<\,b(u_1)\,b(u_2)\,c(Z_2)\,c(Z_1)\,\right>_{\ttau+\half}.
\end{equation}

\subsection{X part}

\subsubsection{covariant case}

In our SFT, manifest Lorentz covariance is lost by the choice $\alpha=(2)p^+$. 
However, the violation occurs only in the zero mode $p^{\pm}$ sector 
and all the other parts still retains the manifest covariance. 
So we first calculate the $X$ correlation function in the manifest 
covariant case, and later will clarify where and how the covariant 
result is modified. 

What we need calculate is the 2-point $X$ correlation function:
\begin{equation}
 \left<e^{ik_2\cdot X(Z_2)}e^{ik_1\cdot X(Z_1)}\right>_\ttau 
 = (-2\pi i)^{\frac{k^2_1}{2}+\frac{k^2_2}{2}}\tr\left[
 w^{(L_0^X-{d\over24})}:e^{ik_2\cdot X(\trho_2)}:\,
 :e^{ik_1\cdot X(\trho_1)}:\right],
\label{eq:2ptX}
\end{equation}
where use has been made of the formula (\ref{eq:vertical-def1}) 
and $\trho_r=-2\pi iZ_r$ ($r=1,2$). 
On the $\trho$ plane, the coordinate fields $X^\mu$ are expanded as
\begin{eqnarray}
 X^\mu(\trho)=x^\mu-i\alpha_0^\mu\trho+i\sum_{n\not=0}\frac{1}{n} 
 \alpha_n^\mu e^{-n\trho},
\end{eqnarray}
and the normal ordered operator $:e^{ik\cdot X(\trho)}:$ is given by
\begin{eqnarray}
 && :e^{ik\cdot X(\trho)}:\ = e^{\half\trho k^2}e^{ik\cdot \hat{x}}
    e^{\trho k\cdot \hat{p}}\prod_{n=1}^\infty e^{\frac{1}{n}k\cdot 
    \alpha_{-n}e^{n\trho}}e^{-\frac{1}{n}k\cdot\alpha_{n}e^{-n\trho}}.
\end{eqnarray}
Note that $e^{ik\cdot(x^\mu-i\alpha_0^\mu\trho)} = 
e^{\half\trho k^2}e^{ik\cdot\hat{x}}e^{\trho k\cdot\hat{p}}$ has been 
used. Inserting this into Eq.~(\ref{eq:2ptX}), we can evaluate the trace
$\tr$ part mode by mode, $\tr=\tr_0\times\prod_{n=1}^\infty\tr_n$. Using 
\begin{equation}
\cases{
  \displaystyle\tr_0[\cdots]=\int\frac{d^d p}{(2\pi)^d}\bra{p}\cdots\ket{p} \cr
  \displaystyle\tr_n[\cdots]=\int\frac{d^d z_n d^d\bar{z_n}}{\pi^d}
  e^{-{\abs{z_n}}^2}\bra{z_n}\cdots\ket{z_n}                         \cr}
\label{eq:modebymode}
\end{equation}
(where $\hat p\ket{p}=p\ket{p}$ normalized as 
$\VEV{p|q}=(2\pi)^d\delta^d(p-q)$, and 
$\ket{z_n}\equiv e^{z_n\cdot\alpha_{-n}/\sqrt n}\ket0$), we find
\begin{eqnarray}
&&\tr_0[\cdots] = \delta^d (k_1+k_2)w^{-\frac{d}{24}}
  \bigl({-2\pi\over\ln w}\bigr)^{\frac{d}{2}}\exp{\bigl[
  k_1\sdot k_2\bigl(\frac{(\trho_1-\trho_2)^2}{2\ln w}
  -\frac{\trho_1-\trho_2}{2}\bigr)\bigr]} ,
\label{eq:covAns} \\
&&\prod_{n=1}^{\infty}\tr_n[\cdots] 
= \prod_{n=1}^{\infty}\frac{1}{(1-w^n)^d}\exp
  \bigl[-k_1\sdot k_2\bigl(\frac{e^{n(\trho_1-\trho_2)}
  +w^n e^{-(\trho_1-\trho_2)}-2w^n}{n(1-w^n)}\bigr)\bigr].
\hspace{3em}
\label{eq:covAnsnz}
\end{eqnarray}
Substituting these into Eq.~(\ref{eq:2ptX}), we obtain
\begin{eqnarray}
&&\left<\,e^{ik_2\cdot X(Z_2)}\,e^{ik_1\cdot X(Z_1)}
 \,\right>_\ttau \nn[-1.5ex]
 &&\quad\  = (-2\pi i)^{-k_1\cdot k_2}\delta^d (k_1+k_2)
 \left[\,w^{1\over24}f(w)\,\right]^{-d}\
 \left({-2\pi}\over{\ln w}\right)^{\frac{d}{2}}
 \left[\,\psi(e^{\trho_1-\trho_2},w)\,\right]^{k_1\cdot k_2}\!\!,
\hspace{3em}
\label{eq:covresult}
\end{eqnarray}
where $\psi(\rho,w)$ is the function defined in Eq.~(8.A.10) in 
GSW:\cite{rf:GSW}
\begin{equation}
\psi(\rho,w)= {1-\rho\over\sqrt\rho}\exp\left({\ln^2\rho\over2\ln w}\right)
\prod_{n=1}^\infty{(1-w^n\rho)(1-w^n/\rho)\over(1-w^n)^2}\,.
\end{equation}

\subsubsection{P, NP and NO cases}

Let us see where and what modifications are necessary in our case of 
P, NP and NO diagrams. 

First we should note that, in these cases of P, NP and NO diagrams, 
the loop momentum $p$ in the zero-mode trace calculation $\tr_0[\cdots]$,
Eq.~(\ref{eq:modebymode}), equals the minus of the momentum 
of the intermediate open string $F$, i.e., $p=-p_F$,\footnote{The minus 
sign is understandable if we note that the contours $C$ and $C'$ giving 
the zero modes $p_F=i\oint_C(dw/2\pi i)\partial X(w)$ of string $F$ and 
$p=i\oint_{C'}(d\trho/2\pi i)\partial X(\trho)$ on the $\trho$ plane, have 
opposite directions by the mapping $w\rightarrow\rho\rightarrow u\rightarrow\trho$.}
and so $-2p^+$ is just the string 
length $\alpha_F$ of the string $F$. Since the conformal mappings to the 
torus for those cases {\it depend} on $\alpha_F$, the $\alpha_F$ dependence 
appears everywhere not only in the zero-momentum trace part, so that the
integration over $\alpha_F$ cannot be performed in the zero-momentum trace 
part alone.

The zero-mode part of the trace 
$\tr[w^{(L_0^X-{d\over24})}:e^{ik_2\cdot X(\trho_2)}:
 :e^{ik_1\cdot X(\trho_1)}:]$ reads
\begin{eqnarray}
% && \left<\,e^{ik_2\cdot X(Z_2)}\,e^{ik_1\cdot X(Z_1)}\,\right>_\ttau \nn
%&&\tr\left[\,
% w^{(L_0^X-{d\over24})}\,:e^{ik_2\cdot X(\trho_2)}:\,
% :e^{ik_1\cdot X(\trho_1)}:\right]_{\hbox{\scriptsize zero-mode part}} \nn
&&\tr_0\left[w^{(\frac{p^2}{2}
 -\frac{d}{24})}\,e^{ik_2\cdot \hat{x}}\,e^{\trho_2 k_2\cdot \hat{p}}
 \,e^{ik_1\cdot \hat{x}}\,e^{\trho_1 k_1\cdot \hat{p}}
 \,e^{\half(\trho_1 k_1^2+\trho_2 k_2^2)}\right] \nn
&&\ =\int{d^dp\over(2\pi)^d}\bra{p}w^{(\frac{p^2}{2}
 -\frac{d}{24})}\,e^{ik_2\cdot \hat{x}}\,e^{\trho_2 k_2\cdot \hat{p}}
 \,e^{ik_1\cdot \hat{x}}\,e^{\trho_1 k_1\cdot \hat{p}}
 \,e^{\half(\trho_1 k_1^2+\trho_2 k_2^2)}\ket{p} \nn
&&\ =\int{d^dp\over(2\pi)^d}(2\pi)^d\delta^d(k_1+k_2)w^{(\frac{p^2}{2}
 -\frac{d}{24})}
e^{(\trho_1 k_1+\trho_2 k_2)\cdot p
   -\half(\trho_1-\trho_2)k_1\cdot k_2},
% && \ \ \times\prod_{n=1}^{\infty}\tr_n\left[w^{(\alpha_{-n}\cdot \alpha_n)}
% \exp(\frac{1}{n}k_2\cdot\alpha_{-n}e^{n\trho_2})
% \exp(-\frac{1}{n}k_2\cdot\alpha_{n}e^{-n\trho_2}) 
% \exp(\frac{1}{n}k_1\cdot\alpha_{-n}e^{n\trho_1})
% \exp(-\frac{1}{n}k_1\cdot\alpha_{n}e^{-n\trho_1})\right]
%\hspace{3em}
\label{eq:A.2}
\end{eqnarray}
where in going to the last line we have used 
$\hat p\ket{p}=p\ket{p}$ and $\exp(ik\hat x)\ket{p}=\ket{p+k}$, 
%\begin{eqnarray}
%&=&\int{d^dp\over(2\pi)^d}\VEV{p|p+k_1+k_2}w^{(\frac{p}{2}
% -\frac{d}{24})}\,e^{\trho_2 k_2\cdot (p+k_1)}
% \,e^{\trho_1 k_1\cdot p}
% \,e^{\half(\trho_1 k_1^2+\trho_2 k_2^2)} \nn
%\end{eqnarray}
and further $\VEV{p|p+k_1+k_2}=(2\pi)^d\delta^d(k_1+k_2)$ and 
$k_1^2=k_2^2=-k_1\cdot k_2$. 
%, we can rewrite it into
%\begin{equation}
%&&\ =\int{d^dp\over(2\pi)^d}(2\pi)^d\delta^d(k_1+k_2)w^{(\frac{p}{2}
% -\frac{d}{24})}
%%\exp\left((\trho_1 k_1+\trho_2 k_2)\cdot p
%%   -\half(\trho_1-\trho_2)k_1\cdot k_2\right).
%e^{(\trho_1 k_1+\trho_2 k_2)\cdot p
%   -\half(\trho_1-\trho_2)k_1\cdot k_2}.
%\end{equation}
Taking account of the above remark for the P, NP and NO diagram cases,
we here insert $1=\int d\alpha_F\delta(\alpha_F+2p^+)$ and keep the $\int d\alpha_F$ 
integration undone. Then 
\begin{equation}
 =\int d\alpha_F\,\delta(k_1+k_2) w^{-d/24}\int d^dp\,\delta(\alpha_F+2p^+)
e^{\half(\ln w)p^2+(\trho_1-\trho_2)k_1\cdot p
   +\half(\trho_2-\trho_1)k_1\cdot k_2}.
\end{equation}
Completing the square in the exponent and
%\begin{equation}
%{\ln w\over2}\bigl(p+{(\trho_1-\trho_2)k_1\over\ln w}\bigr)^2\cdot p
%   +\bigl({(\trho_2-\trho_1)^2\over2\ln w} +
%   +{\trho_2-\trho_1\over2}\bigr)k_1\cdot k_2 .
%\end{equation}
making a shift of the integration variable 
$p\rightarrow p-(\trho_1-\trho_2)k_1/\ln w$, we find
\begin{eqnarray}
&& =\int d\alpha_F\,\delta(k_1+k_2) w^{-d/24}\exp\bigl[
   \bigl({(\trho_2-\trho_1)^2\over2\ln w} +
   +{\trho_2-\trho_1\over2}\bigr)k_1\cdot k_2\bigr] \nn
&&\quad \times\int d^dp
%e^{\half\ln w\,p^2}
\exp\bigl({\ln w\over2}p^2\bigr)\,
\delta\big(\alpha_F+2p^+-{\trho_1-\trho_2\over\ln w}\alpha_1\bigr)\,.
\label{eq:A.3}
\end{eqnarray}
Using $d^dp=d^{d-2}{\mib p}\,dp^+dp^-$, $p^2=-2p^+p^-+{\mib p}^2$
and $\int dp^- e^{-(\ln w)p^+p^-}=({2\pi\over{-}\ln w})\delta(p^+)$, 
the momentum integration in the second line yields
\begin{equation}
\left({-2\pi\over\ln w}\right)^{d\over2} 
\delta\big(\alpha_F-{\trho_1-\trho_2\over\ln w}\alpha_1\bigr).
\end{equation}
If the $\alpha_F$ dependence appeared only in this zero-mode sector, then the 
$\alpha_F$ integration of the delta function would trivially give 1 and the 
expression (\ref{eq:A.3}) just reproduced the covariant case answer 
(\ref{eq:covAns}), as it should be. 
Therefore, the zero-mode trace in our case is given by
\begin{equation}
\tr_0[\cdots] = \int d\alpha_F\,\delta\big(\alpha_F-{\trho_1-\trho_2\over\ln w}\alpha_1\bigr)
\times\bigl[\hbox{covariant result Eq.~(\ref{eq:covAns})}\bigr].
\label{eq:tr0}
\end{equation}

Secondly, note that the open string coordinate $X(\rho_r)$ on the original 
$\rho$ plane is usually an abbreviation for the real coordinate 
$(X(\rho_r)+X(\bar \rho_r))/2$, and is, therefore, mapped to the coordinate 
$(X(\trho_r)+X(\bar{\trho}_r))/2$ on the final $\trho=-2\pi iu$ plane, which 
coincides with $X(\trho_r)$ if $\trho_r$ lies on the real axis. This 
was actually the case in the P and NO diagram cases both for $r=1$ and 2. 
In the NP diagram case, however, $\trho_2$ does not lie on the real 
axis, so that we should make a replacement
\begin{eqnarray}
 X^\mu(\trho_2) &\rightarrow& \half\left(X^\mu(\trho_2)+X^\mu(\bar{\trho_2})\right)
=\half\left(X^\mu(\trho_2)+X^\mu(\trho_2-2\pi i)\right) \nn
 &=& x^\mu-i\alpha_0^\mu(\trho_2-\pi i)+i\sum_{n\not=0}\frac{1}{n} 
 \alpha_n^\mu e^{-n\trho_2}.
\end{eqnarray}
This change amounts to making a replacement $\trho_2\rightarrow\trho_2-\pi i$ 
only in the zero mode part in the above calculation, and hence the 
replacement of $\psi(\rho,w)$ by
\begin{eqnarray}
 \psi(\rho,w)\,&\rightarrow&\,\psi'(\rho,w) 
 = {1-\rho\over\sqrt{-\rho}}\exp\left({\ln^2(-\rho)\over2\ln w}\right)
 \prod_{n=1}^\infty{(1-w^n\rho)(1-w^n/\rho)\over(1-w^n)^2} \nn
 &=& \psi^T(-\rho,w)\,,
\end{eqnarray}
where
\begin{equation}
 \psi^T(\rho,w)\equiv{1+\rho\over\sqrt\rho}\exp\left({\ln^2\rho\over2\ln w}\right)
 \prod_{n=1}^\infty{(1+w^n\rho)(1+w^n/\rho)\over(1-w^n)^2}\,.
\end{equation}
The $\delta$ function factor in Eq.~(\ref{eq:tr0}) is also modified 
by this shift $\trho_2\rightarrow\trho_2-\pi i$. But, thanks to this change, 
$\trho_1-\trho_2=-2\pi i(Z_1-Z_2)$ becomes $4\pi i\ttau x$ in coincidence 
with the P and NO cases. (Compare Eqs.~(\ref{eq:pl-paramet}) and 
(\ref{eq:np-paramet}).) \ Therefore, the $\delta$ function factor in 
Eq.~(\ref{eq:tr0}) can be rewritten uniformly in the three cases, P, NO 
and NP, into 
\begin{eqnarray}
\delta\big(\alpha_F-{\trho_1-\trho_2\over\ln w}\alpha_1\bigr)
\,\rightarrow\,\delta\left(\frac {\tau_1-\tau_2}{\ln w}\right)
 = (-\ln w)\delta(\tau_1-\tau_2),
\end{eqnarray}
where use has been made of $\ln w=2\pi i\ttau<0$ and 
Eqs.~(\ref{eq:pl-peritau}), (\ref{eq:no-peritau}) and  
(\ref{eq:np-peritau}).

Finally, for the case of NO diagram where the period is $\ttau+1/2$, 
we should include the twist operator $\Omega$. 
Since $\Omega\alpha_n\Omega^{-1}=(-)^n\alpha_n$, the whole effect is simply to make 
a replacement $w\rightarrow-w$ in the {\it non-zero mode} sector 
in Eq.~(\ref{eq:covresult}); that is, 
\begin{eqnarray}
   \psi(\rho,w)&\rightarrow&{1-\rho\over\sqrt\rho}\exp\left({\ln^2\rho\over2\ln w}\right)
   \prod_{n=1}^\infty{(1-(-w)^n\rho)(1-(-w)^n/\rho)\over(1-(-w)^n)^2}
    \equiv\psi^N(\rho,w), \nn
   f(w)&\rightarrow&f(-w). 
\end{eqnarray}

Putting all these modifications together, and using a variable 
$\rho_{12}\equiv e^{4\pi i\ttau x}$ with which
\begin{eqnarray}
 e^{\trho_1-\trho_2}=e^{-2\pi i(Z_1-Z_2)}
=\cases{ \rho_{12} &for P and NO cases \cr
         -\rho_{12} &for NP case \cr},
\end{eqnarray}
we find the 2-point $X$ correlation functions for P, NP and NO:
\def\ds{\displaystyle}
\begin{equation}
 \left<e^{ik_2\cdot X(Z_2)}e^{ik_1\cdot X(Z_1)}\right>
=\int d\alpha_F\,(-\ln w)\delta(\tau_2-\tau_1)\times\cases{ 
   {\cal F} &for P \cr
   {\cal F}^T &for NP \cr
   {\cal F}^N &for NO \cr
%   \hspace{5em} {\cal F}^N  &for $V_{\propto}$ \cr
},
\label{eq:Xcorr}
\end{equation}
where
\begin{eqnarray}
  {\cal F}&=& (-2\pi i)^{-k_1\cdot k_2}\delta^d(k_1+k_2)
   \left({-2\pi}\over{\ln w}\right)^{\frac{d}{2}}\,
   \left[\,w^{1\over24}f(w)\,\right]^{-d}
   \left[\,\psi(\rho_{12},w)\,\right]^{k_1\cdot k_2}, \nn
  {\cal F}^T &=& (-2\pi i)^{-k_1\cdot k_2}\delta^d(k_1+k_2)
   \left({-2\pi}\over{\ln w}\right)^{\frac{d}{2}}\,
   \left[\,w^{1\over24}f(w)\,\right]^{-d}
   \left[\,\psi^T(\rho_{12},w)\,\right]^{k_1\cdot k_2},  \nn
  {\cal F}^N &=& (-2\pi i)^{-k_1\cdot k_2}\delta^d(k_1+k_2)
    \left({-2\pi}\over{\ln w}\right)^{\frac{d}{2}}\,
   \left[w^{1\over24}f(-w)\right]^{-d}
   \left[\,\psi^N(\rho_{12},w)\,\right]^{k_1\cdot k_2}\!\!\!\!.
\hspace{4em}
\label{eq:Ffns}
\end{eqnarray}

\subsubsection{$UU$}

As explained in Eq.~(\ref{eq:UUGGRT}), the correlation function 
for $UU$ case is given by 
\begin{equation}
 \left<\,e^{ik_2\cdot X(Z_2)}\,e^{ik_1\cdot X(Z_1)}\,\right>_\tau 
 = (-2\pi i\tau)^{\frac{k^2_1}{2}+\frac{k^2_2}{2}}{\tr}'\left[\,
 q^{2(L_0^X-{d\over24})}\,:e^{ik_2\cdot X(\trho_2^{\rm h})}:\,
 :e^{ik_1\cdot X(\trho_1^{\rm h})}:\right] 
\end{equation}
where $\trho_r^{\rm h}=-2\pi i\tau Z_r$ and the zero-mode part trace is 
an expectation value:
\begin{eqnarray}
 \tr_0[\cdots]=\bra{-k_1/2}\cdots\ket{-k_1/2}. 
\end{eqnarray}
This zero-mode part `trace' is immediately found by the help of 
Eq.~(\ref{eq:A.2}) to give
\begin{equation}
\tr_0[\cdots]=(2\pi)^d\delta^d(k_1+k_2)
(q^2)^{-\frac{d}{24}} (q^{-{1\over4}})^{k_1\cdot k_2}.
\end{equation}
The trace over the non-zero modes is essentially the same as before and 
is given by Eq.~(\ref{eq:covAnsnz}) by replacements $w\rightarrow q^2$ and 
$\trho_r\rightarrow\trho_r^{\rm h}$: so we have
\begin{eqnarray}
&& \left<e^{ik_2\cdot X(Z_2)}e^{ik_1\cdot X(Z_1)}\right>_\tau\nn
 &&\quad\ \ = (-2\pi i\tau)^{-k_1\cdot k_2}(2\pi)^d \delta^d (k_1+k_2)
 \left[q^{1\over12}f(q^2)\right]^{-d}
 \left[\,\widetilde\psi(e^{\trho_1^{\rm h}-\trho_2^{\rm h}},q^2)
 \,\right]^{k_1\cdot k_2},
\hspace{3em}
\label{eq:UUX}
\end{eqnarray}
where 
\begin{equation}
\widetilde\psi(e^{\trho_1^{\rm h}-\trho_2^{\rm h}},w) =
 q^{-{1\over4}}e^{\half(\trho_1^{\rm h}-\trho_2^{\rm h})}
 \exp\bigl(-{(\trho_1^{\rm h}-\trho_2^{\rm h})^2\over2\ln q^2}\bigr)
 \psi(e^{\trho_1^{\rm h}-\trho_2^{\rm h}},q^2). 
\end{equation}
Noting the relations $q=e^{i\pi\tau}$ and 
\begin{equation}
 e^{\trho_1^{\rm h}-\trho_2^{\rm h}} 
 = e^{-2\pi i\tau(Z_1-Z_2)}=e^{-4\pi ix+\pi i\tau} = qz_{12}^{-1}  
\qquad (z_{12}\equiv e^{4\pi ix}\equiv e^{2\pi i\nu_{12}}),
\end{equation}
in this $UU$ case, 
this function $\widetilde\psi$ can be rewritten as follows: 
\begin{eqnarray}
\widetilde\psi(e^{\trho_1^{\rm h}-\trho_2^{\rm h}},w) 
&=& q^{-{1\over4}}\left(qz_{12}^{-1}\right)^{\half}
 e^{-{1\over4\pi i\tau}(2\pi i\nu_{12}-\pi i\tau)^2}\psi(qz_{12}^{-1},q^2) \nn
 &=& e^{-\pi i\nu_{12}^2/\tau}\psi(qz_{12}^{-1},q^2)
 = e^{-\pi i\nu_{12}^2/\tau}\psi(qz_{12},q^2)
\end{eqnarray}
where in the last step we have used the properties (\ref{eq:psiproperty}),
$\psi(q^2 z,q^2)=-\psi(z,q^2)$ and $\psi(z^{-1},q^2)=-\psi(z,q^2)$
of the $\psi$ function, explained in Appendix.

The expression (\ref{eq:UUX}) with this $\widetilde\psi$ can be rewritten 
in terms of $\rho$ and $w$. Using the relation (\ref{eq:psiTrel}),
\begin{equation}
 e^{-\pi i\nu_{12}^2/\tau}\psi(qz_{12},q^2)
 = -i\tau\psi^T(\rho_{12},w),
\end{equation}
in the Appendix and the first relation in 
Eq.~(\ref{eq:Jacobi}), we find 
\begin{eqnarray}
 &&\hspace{-1em} 
\left<\,e^{ik_2\cdot X(Z_2)}\,e^{ik_1\cdot X(Z_1)}\,\right>_{UU} \nn
 &&\ = (-2\pi i\tau)^{-k_1\cdot k_2}(2\pi)^d \delta^d (k_1+k_2)
 \left[w^{1\over24}f(w)\right]^{-d}\left({-2\pi\over\ln w}\right)^{d\over2}
 \left[\,-i\tau\psi^T(\rho_{12},w)\,\right]^{k_1\cdot k_2} \nn
 &&\ = (2\pi)^d(\tau)^{-k_1\cdot k_2}(-i\tau)^{k_1\cdot k_2}{\cal F}^T 
  = (-i)^{k_1\cdot k_2}(2\pi)^d {\cal F}^T,
\label{eq:XcorrUU}
\end{eqnarray}
with ${\cal F}^T$ defined before in Eq.~(\ref{eq:Ffns}).

\subsubsection{$V_\propto$}

The correlation function for $V_\propto$ case is also calculated similarly to 
the $UU$ case. By the formula (\ref{eq:vadef}) for this case, the zero-mode 
part trace $\tr_0[\cdots]$ is the expectation value $\bra{0}\cdots\ket{0}$ with 
zero-momentum state, which can be read from 
Eq.~(\ref{eq:A.2}) to give
\begin{equation}
\tr_0[\cdots]=(2\pi)^d\delta^d(k_1+k_2)
(q^{1/2})^{-\frac{d}{24}} 
e^{-{1\over4}(\trho_1^{\rm h}-\trho_2^{\rm h})k_1\cdot k_2},
\end{equation}
with $-\pi i\tau Z_r=\trho_r^{\rm h}/2$ used. 
As in the ghost part calculation, comparison of the formulas 
(\ref{eq:vadef}) and (\ref{eq:vertical-def2}) shows that 
the trace over the non-zero modes in this case 
can be obtained by making replacements $w\rightarrow q^{1/2}$, $u\rightarrow(\tau 
/2)u$ and $(-2\pi i)\rightarrow(-\pi i\tau)$ in the above result for NO case: so we have
\begin{eqnarray}
&&\left<e^{ik_2\cdot X(Z_2)}e^{ik_1\cdot X(Z_1)}\right>_{UU,{\tau\over4}+\half}
= (-\pi i\tau)^{-k_1\cdot k_2}(2\pi)^d \delta^d (k_1+k_2) \nn
&&\qquad  \times\left[q^{1\over48}f(-\sqrt q)\right]^{-d}
 \left[\,
\exp\bigl(-{(\trho_1^{\rm h}-\trho_2^{\rm h})^2\over8\ln q^{1/2}}\bigr)
%e^{-{(\trho_1^{\rm h}-\trho_2^{\rm h})^2\over8\ln q^{1/2}}}
  \widetilde\psi^N(e^{\half(\trho_1^{\rm h}-\trho_2^{\rm h})},q^{1/2})
 \,\right]^{k_1\cdot k_2}. 
\hspace{3em}
\label{eq:VaX}
\end{eqnarray}
%where 
%\begin{equation}
%  \widetilde\psi^N(e^{\half(\trho_1^{\rm h}-\trho_2^{\rm h})},q^{1/2})
%=
%\exp\left(-{(\trho_1^{\rm h}-\trho_2^{\rm h})^2\over8\ln q^{1/2}}\right)
%%e^{-{(\trho_1^{\rm h}-\trho_2^{\rm h})^2\over8\ln q^{1/2}}}
%\psi^N(e^{\half(\trho_1^{\rm h}-\trho_2^{\rm h})},q^{1/2}).
%\end{equation}
Note that $Z_1-Z_2=-2\ttau x=\nu_{12}/\tau$ in this $V_\propto$ case, so that 
\begin{eqnarray}
 e^{\half(\trho_1^{\rm h}-\trho_2^{\rm h})} 
 = e^{-\pi i\tau(Z_1-Z_2)} = z_{12}^{-1/2}, \quad 
 \exp\bigl(-{(\trho_1^{\rm h}-\trho_2^{\rm h})^2\over8\ln q^{1/2}}\bigr)
   = e^{-\pi i\nu_{12}^2/\tau}.
\hspace{2em}
\end{eqnarray}
We see that the quantity inside of 
$[\cdots]^{k_1\cdot k_2}$ in Eq.~(\ref{eq:VaX}) just equals 
$(\tau/2)\psi^N(\rho_{12},w)$ by the identity (\ref{eq:psiNrel}). Then using also 
the first relation in Eq.~(\ref{eq:Jacobihalf}), we find 
\begin{eqnarray}
&&\left<e^{ik_2\cdot X(Z_2)}e^{ik_1\cdot X(Z_1)}\right>_{UU,{\tau\over4}+\half}
= (-\pi i\tau)^{-k_1\cdot k_2}(2\pi)^d \delta^d (k_1+k_2) \nn
&&\qquad \times\left[w^{1\over24}f(-w)\right]^{-d}\left({-\pi\over\ln w}\right)^{d\over2}
 \left[\,{\tau\over2}\psi^N(e^{\half(\trho_1^{\rm h}-\trho_2^{\rm h})},q^{1/2})
 \,\right]^{k_1\cdot k_2} 
 = {(2\pi)^d\over2^{d/2}}{\cal F}^N.
\hspace{4em}
% &&\ = (-2\pi i\tau)^{-k_1\cdot k_2}(2\pi)^d \delta^d (k_1+k_2)
% \left[w^{1\over24}f(w)\right]^{-d}\left({-2\pi\over\ln w}\right)^{d\over2}
% \left[\,-i\tau\psi^T(\rho_{12},w)\,\right]^{k_1\cdot k_2} \nn
% &&\ = (2\pi)^d(\tau)^{-k_1\cdot k_2}(-i\tau)^{k_1\cdot k_2}{\cal F}^T 
%  = (-i)^{k_1\cdot k_2}(2\pi)^d {\cal F}^T,
\label{eq:XcorrVa}
\end{eqnarray}
Incidentally, this relation together with Eq.~(\ref{eq:vev-relationhalf}) 
confirms the relation (\ref{eq:varel}) announced before.

\section{Explicit evaluation of tachyon amplitude}

\subsection{Amplitudes for P, NP, NO}

\begin{figure}[tb]
   \epsfxsize= 13cm   %or \epsfysize= HEIGHT cm
   \centerline{\epsfbox{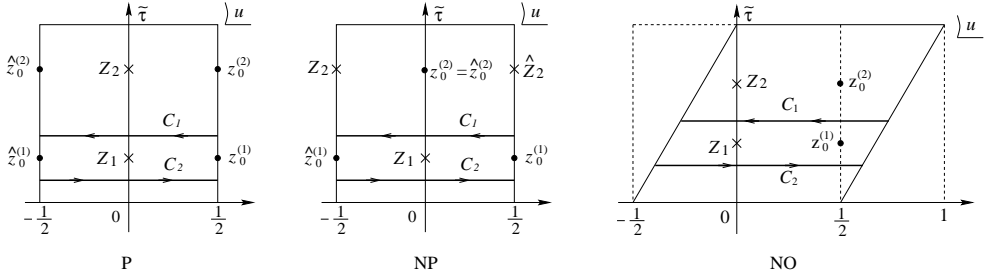}}
 \caption{The contours $C_1$ and $C_2$ giving antighost factors for P, 
NP and NO cases.} 
\label{fig:cont}
\end{figure}

Let us now evaluate the amplitudes (\ref{eq:Xamp}) for P, NP and NO 
explicitly. First the ghost part of the CFT correlation function in 
Eq.~(\ref{eq:CFTfnX}) is evaluated by mapping the integration contours
on the $\rho$ plane for the antighost factors $b_{\tau_r}$ to those on the 
torus $u$ plane:
\begin{eqnarray}
 && \left\langle\,b_{\tau_1}b_{\tau_2}\,c(Z_2)\,c(Z_1)\,\right>_J \nn
 &&\qquad = \int_{C_1}{du_1\over2\pi i}\left({du_1\over d\rho}\right)
   \int_{C_2}{du_2\over2\pi i}\left({du_2\over d\rho}\right)
   \left\langle\,b(u_1)\,b(u_2)\,c(Z_2)\,c(Z_1)\,\right>_J \nn
 &&\qquad = {\cal G}^J \int_{C_1}{du_1\over2\pi i}\int_{C_2}{du_2\over2\pi i}
   \left({du_1\over d\rho}\right)\left({du_2\over d\rho}\right)
   \left\{\,{d\rho\over du_1} - {d\rho\over du_2}\,\right\} \nn
  &&\qquad = {\cal G}^J \left(\int_{C_1}{du_1\over2\pi i}\,
   \int_{C_2}{du_2\over2\pi i}{du_2\over d\rho}
   -\int_{C_2}{du_2\over2\pi i}\,
   \int_{C_1}{du_1\over2\pi i}{du_1\over d\rho}\right)
\label{eq:bt1bt2}
\end{eqnarray}
where we have used the results (\ref{eq:vev-ttau}) and 
(\ref{eq:vev-ttau+half}) for the ghost correlation functions and 
${\cal G}^J$ denotes
\begin{eqnarray}
{\cal G}^J = \cases{
  {\cal G}\equiv\frac{2\pi i}{\alpha_1}\left[\,w^{\frac{1}{24}}f(w)\,\right]^2
  & for P, NP \ \ $(J=\ttau)$ \cr
  {\cal G}^N 
  \equiv\frac{2\pi i}{\alpha_1}\left[\,w^{\frac{1}{24}}f(-w)\,\right]^2
  & for NO \ \ $(J = \ttau+\half)$ \cr
}.
\end{eqnarray}       
The contours $C_1$ and $C_2$ on the $u$ plane are depicted in 
Fig.~\ref{fig:cont}.
%\begin{wrapfigure}[6]{r}{6.6cm}
%\end{wrapfigure}
Noting that $du_2/d\rho$ is independent of $u_1$ and vice versa, 
and $\int_{C_r}du_r\,1 = \mp1$ for $r=1$ and $2$, respectively,
we can evaluate the contour integrations in Eq.~(\ref{eq:bt1bt2}) 
as follows:
\begin{equation}
 {-1\over2\pi i}\int_{C_2}{du\over2\pi i}{du\over d\rho}
   -{1\over2\pi i}\int_{C_1}{du\over2\pi i}{du\over d\rho}
 = {-1\over2\pi i}\int_{C_1 +C_2}{du\over2\pi i}{du\over d\rho}
  = {-1\over2\pi i}\oint_{\zz(1)}{du\over2\pi i}{du\over d\rho}. 
\label{eq:evalcont}
\end{equation}   
Noting that the integrand $du/d\rho$ has a pole at $u=\zz(1)$ and the residue 
there
\begin{eqnarray}
 && \mbox{Res}_{\zz(1)}\left({du\over d\rho}\right)
  ={1\over\alpha_1}\left[\,g_1'(\zz(1)-Z_1)-g_1'(\zz(1)-Z_2)\,\right]^{-1}
 \equiv R,
\label{eq:defR}
\end{eqnarray}
we find 
\begin{equation}
 \left\langle\,b_{\tau_1}b_{\tau_2}\,c(Z_2)\,c(Z_1)\,\right>_J 
  = -{R\over2\pi i}\,{\cal G}^J .
\label{eq:v3-ghost}
\end{equation}

Next we rewrite the remaining $X$ part and integrals $d\tau_1d\tau_2$ in 
Eq.~(\ref{eq:Xamp}) into
\begin{eqnarray}
 \int d\tau_1 d\tau_2\,\left<\,e^{ik_2\cdot X(Z_2)}\,
  e^{ik_1\cdot X(Z_1)}\,\right>_J 
  &=& \int d\tau_1 d\tau_2 d\alpha_F (-\ln w)\delta(\tau_2-\tau_1){\cal F}^J \nn
 &=& \int d\ttau dx {i\over\pi}\alpha_1 R^{-1}(-\ln w){\cal F}^J ,
\label{eq:v3-Xt1t2}
\end{eqnarray}
where we have used the results (\ref{eq:Xcorr}) for the 
2-point $X$ correlation functions, the Jacobian (\ref{eq:v3-jacobian}) 
for $(\tau_1,\alpha_F)\rightarrow(\tau,x)$ and the definition (\ref{eq:defR}) of $R$.

Putting Eqs.~(\ref{eq:v3-ghost}) and (\ref{eq:v3-Xt1t2}) together, 
the amplitudes (\ref{eq:Xamp}) for P, NO and NP are 
finally obtained as
\begin{eqnarray}
 \pmatrix{
   {\cal A}_{\rm P}  \cr
   {\cal A}_{\rm NP} \cr
   {\cal A}_{\rm NO} \cr} %={\cal A}'_{\rm NO} \\
% &=& -{g^2\over2}\int d\ttau dx {-C\over2\pi i\alpha_1}{i\over\pi}
%  (\alpha_1)^2 C^{-1}(-\ln w)\left(
% \begin{array}{c}
%   n\,{\cal G}\,{\cal F} \\
%   {\cal G}^N {\cal F}^N \\
%   {\cal G}\,{\cal F}^T
% \end{array}\right) \nn
 &=& -i{\alpha_1\over(2\pi)^2}g^2\,\int d\ttau dx\,(-\ln w)
\pmatrix{
   n\,{\cal G}\,{\cal F} \cr
   {\cal G}\,{\cal F}^T  \cr
   {\cal G}^N {\cal F}^N \cr
}/[(2\pi)^d \delta^d(k_1+k_2)] \nn
&=& -ig^2\int d({\ttau\over i}) dx
 \pmatrix{
   n\left[\,w^{1\over24}f(w)(-2\pi\ln w)^{\half}\,\right]^{-(d-2)}
   \left[\,\psi(\rho_{12},w)\,\right]^{-2} \cr
   \left[\,w^{1\over24}f(w)(-2\pi\ln w)^{\half}\,\right]^{-(d-2)}
   \left[\,\psi^T(\rho_{12},w)\,\right]^{-2} \cr
   \left[\,w^{1\over24}f(-w)(-2\pi\ln w)^{\half}\,\right]^{-(d-2)}
   \left[\,\psi^N(\rho_{12},w)\,\right]^{-2} \cr
}. \nn[-.5ex]
\label{eq:PNONPamp}
\end{eqnarray}
In going to the last expressions, the explicit forms (\ref{eq:Ffns}) 
for ${\cal F}$'s and Eqs.~(\ref{eq:vev-tau}) and (\ref{eq:vev-ttau+half}) 
for ${\cal G}$'s, are used together with 
the on-shell condition $k_1\cdot k_2=-2$ and relations 
$w = e^{2\pi i\ttau}$, $\rho_{12} = e^{4\pi i\ttau x}$.

\subsection{Full nonplanar amplitude and $UU$ diagram}

The nonplanar amplitude ${\cal A}_{\rm NP}$ in Eq.~(\ref{eq:PNONPamp}) 
gives a correct nonplanar on-shell tachyon amplitude\cite{rf:GSW} 
if it covers the full integration region $0\leq\ttau/i<\infty$ and $0\leq2x(=\nu)
\leq1$. However, the diagram in Fig.~\ref{fig:NPmap} does not cover the 
full region. The integration over $x$ is all right since 
Eq.~(\ref{eq:np-peritau}) with $\tau_1-\tau_2=0$ implies $2x=\alpha_F/\alpha_1$ 
which indeed runs over $0\leq2x\leq1$ as $\alpha_F$ runs from 0 to $\alpha_1$. But 
Eq.~(\ref{eq:np-tau1}) implies that, as $\tau_1$ runs from 0 to $\infty$, 
$\ttau/i$ runs from a certain value $\ttau_0(x)/i(>0)$ to $\infty$, where 
$\ttau_0(x)$ is the root for $\ttau$ of the Eq.~(\ref{eq:np-tau1}) with 
$\tau_1=0$. So it is necessary to cover the missing region $0\leq\ttau/i\leq 
\ttau_0(x)/i$. 

As was announced already in \S 2, this missing region is covered by the 
$UU$ diagram contribution (\ref{eq:UUamp}), which we now evaluate 
explicitly. 

The ghost part of the CFT correlation function (\ref{eq:CFTfnUU}) is 
calculated in quite the same way as in Eq.~(\ref{eq:bt1bt2})
by using Eq.~(\ref{eq:vev-relation}) 
together with (\ref{eq:vev-ttau}):
\begin{eqnarray}
&& \left\langle\,b_0\bar{b}_0\,c(Z_2)\,c(Z_1)\,\right>_\tau\nn
 &&\qquad = \int_{C_1}{du_1\over2\pi i}\left({du_1\over d\rho}\right)
   \int_{C_2}{du_2\over2\pi i}\left({du_2\over d\rho}\right)
   \left\langle\,b(u_1)\,b(u_2)\,c(Z_2)\,c(Z_1)\,\right>_\tau\nn
 && \qquad = i{\cal G}\left(\int_{C_1}{du_1\over2\pi i}
   \int_{C_2}{du_2\over2\pi i}{du_2\over d\rho}
   -\int_{C_2}{du_2\over2\pi i}
   \int_{C_1}{du_1\over2\pi i}{du_1\over d\rho}\right). %\nn
% = {\ttau\over2\pi\alpha_1}C\,{\cal G}
\label{eq:UU-ghost}
\end{eqnarray}
\begin{wrapfigure}{r}{\halftext}
%\begin{figure}[htb]
   \epsfxsize= 3.5cm   %or \epsfysize= HEIGHT cm
   \centerline{\epsfbox{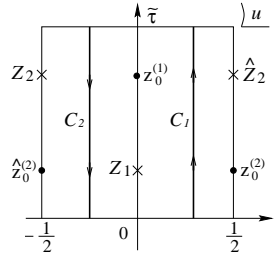}}
 \caption{The contours $C_1$ and $C_2$ for UU case.}
 \label{fig:UUcont}
%\end{figure}
\end{wrapfigure}
The contours $C_1$ and $C_2$ in this case are drawn in Fig.~\ref{fig:UUcont}.
Note that $\int_{C_r}du_r\,1= \pm\ttau$ for $r=1,2$ instead of $\mp1$ in 
the previous case. Evaluating the remaining integration by the pole 
residue as done in Eq.~(\ref{eq:evalcont}), we obtain 
\begin{equation}
\left\langle\,b_0\bar{b}_0\,c(Z_2)\,c(Z_1)\,\right>_\tau 
 = {\ttau\over2\pi}R\,{\cal G}.
\end{equation}

\begin{figure}[tb]
   \epsfxsize= 10cm
   \centerline{\epsfbox{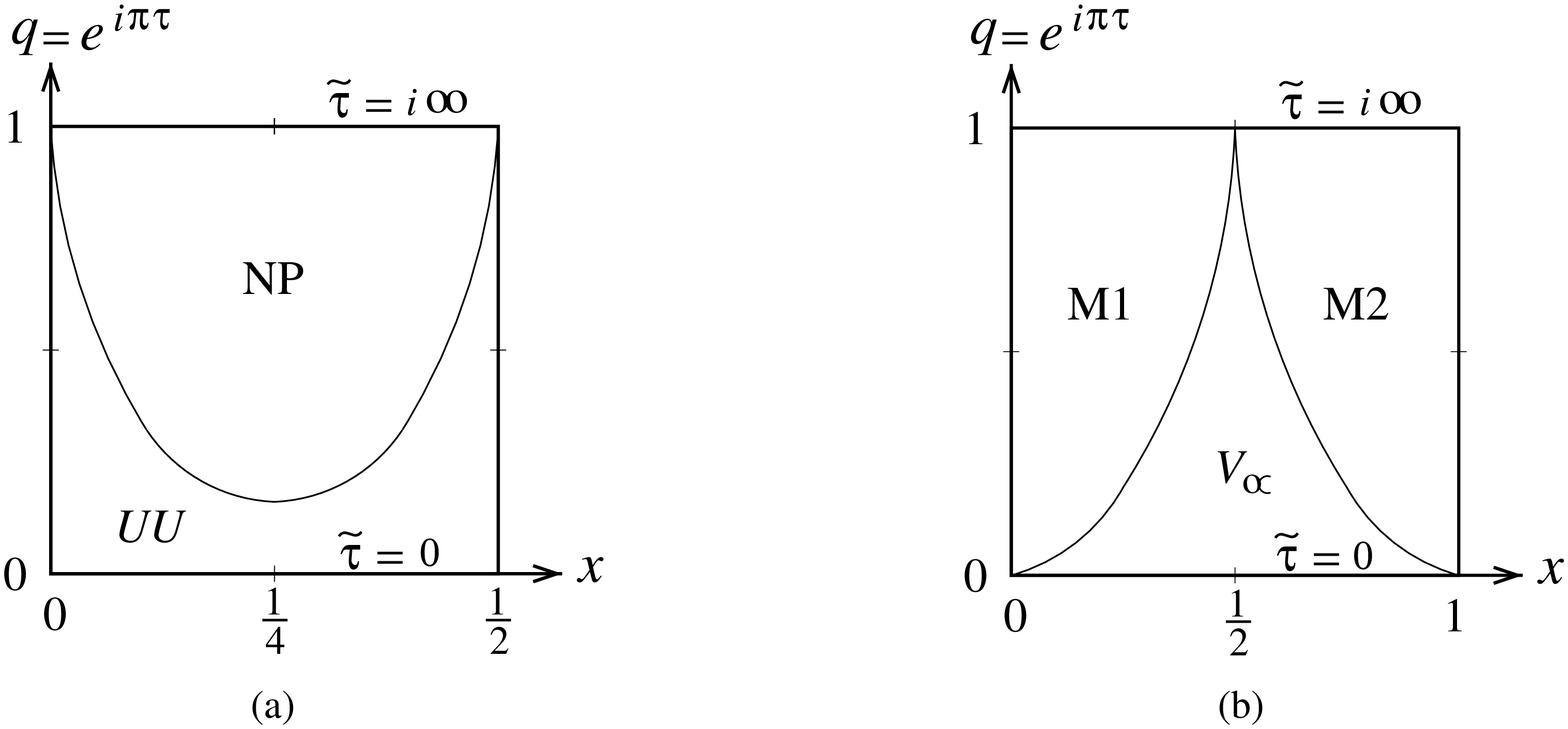}}
 \caption{A schematic view for the moduli regions covered by individual 
diagrams for the full (a) nonplanar and (b) nonorientable amplitudes.}
 \label{fig:moduli}
\end{figure}

The $X$-part of the correlation function was obtained in the 
Eq.~(\ref{eq:XcorrUU}) above, and the Jacobian for the change of variables 
$(\tau_1,\sigma_1)\rightarrow(\ttau,\,x)$ was given in Eq.~(\ref{eq:uu-jacobian}), 
which reads $d\tau_1d\sigma_1=i\alpha_1R^{-1}d\ttau dx$ by using $R$ defined in 
Eq.~(\ref{eq:defR}). Putting these altogether, the $UU$ amplitude 
(\ref{eq:UUamp}) turns out to be
\begin{eqnarray}
 {\cal A}_{UU} 
% &&\quad = {\xu^2 g^2\over2\pi}\int d\ttau dx \,i(\alpha_1)^2 C^{-1}{\ttau\over2\pi\alpha_1}C
%  (-i)^{k_1\cdot k_2}(2\pi)^d {\cal G}\,{\cal F}^T \nn
  &=& i\alpha_1(-i)^{k_1\cdot k_2}(2\pi)^{d-2}\xu^2 g^2\int d\ttau dx
  \,\ttau\,{\cal G}\,{\cal F}^T\,/[(2\pi)^d \delta^d(k_1+k_2)] \nn
  &=& {\xu^2\over(2\pi)^{1-d}}g^2\int d({\ttau\over i}) dx
 \left[\,w^{1\over24}f(w)(-2\pi\ln w)^{\half}\,\right]^{-(d-2)}\!\!
 \left[\,\psi^T(\rho_{12},w)\,\right]^{-2}\!\!\!\!,
\hspace{3em}
\end{eqnarray}
the same form as the above ${\cal A}_{\rm NP}$ in Eq.~(\ref{eq:PNONPamp}). 
So if the coefficients are the same with {\it opposite} sign, i.e.,
\begin{equation}
-{\xu^2\over(2\pi)^{1-d}}g^2 = -ig^2 \quad \rightarrow 
\quad \xu^2=i(2\pi)^{1-d}= {i\over(2\pi)^{25}}\,,
\label{eq:xu2}
\end{equation}
then the full nonplanar amplitude is reproduced in this SFT. 
The reason of opposite sign is as follows. Originally, 
the variable $\sigma_1$ runs from 0 to $2\pi\alpha_1$ and $\tau_1$ from 0 to $\infty$, 
which are mapped into the present variables $x$ and $\ttau$ by 
the relations (\ref{eq:uu-period1}) and (\ref{eq:uu-tau1}). From those 
relation we find the integration region correspondence
\begin{equation}
\int_0^{2\pi\alpha_1}d\sigma_1\int_0^\infty d\tau_1\  \leftrightarrow \  
\int_0^{1/2}dx\int_{\ttau_0(x)/i}^0d({\ttau\over i}) =
-\int_0^{1/2}dx\int^{\ttau_0(x)/i}_0 d({\ttau\over i})\,. 
\end{equation}
Namely a minus sign appears since the increasing direction of $\tau_1$ is 
opposite to that of $\ttau/i$, contrary to the previous NP case. 
This could have been inferred from the diagrams in Fig.~\ref{fig:nonplanar}, and indeed the relations (\ref{eq:uu-tau1}) and (\ref{eq:np-tau1}) 
between $\tau_1$ and $\ttau$ for $UU$ and NP cases, respectively, have 
just opposite signs. Thus, with the coupling strength $x_u^2$ in 
Eq.~(\ref{eq:xu2}), we obtain the correct nonplanar amplitude ${\cal 
A}_{\rm NP}$ in Eq.~(\ref{eq:PNONPamp}) with the full integration region
$\int_0^\infty d(\ttau/i)\int_0^{1/2}dx$. In Fig.~\ref{fig:moduli} (a), 
we present a schematic view to show how the two diagrams NP and UU cover
the full moduli region $0\leq\ttau/i<\infty$, $0\leq x\leq1/2$.
%\begin{wrapfigure}[n]{r}{6.6cm}
%\end{wrapfigure}

\subsection{Full nonorientable amplitude and $V_{\propto}$ diagram}

In the same way, the amplitude ${\cal A}_{\rm NO}$ 
in Eq.~(\ref{eq:PNONPamp}) to which contribute the two 
nonorientable diagrams M1 and M2, does not yet cover the 
full region of the moduli, $0\leq\ttau/i<\infty,\ 0\leq x\leq1$. 
As explained in \S 2, the contributions of M1 and M2 diagrams have a 
gap in the moduli space and the $V_{\propto}$ diagram just gives the 
contribution filling the gap. 

The ghost part of the CFT correlation function in the $V_{\propto}$ 
amplitude (\ref{eq:Valphaamp}) is calculated in a similar way to the 
preceding two cases (\ref{eq:bt1bt2}) and (\ref{eq:UU-ghost}). 
Noting that the anti-ghost factors $b_{\sigma_i}$ ($i=1,2$) in this case are
defined on the $\rho$ plane as
\begin{equation}
 b_{\sigma_i}\equiv{d\rho_0^{(i)}\over d\sigma_i}\int_{C_i}{d\rho\over2\pi i}b(\rho)
 = i\int_{C_i}{d\rho\over2\pi i}b(\rho),
\end{equation}
and mapping them into the torus $u$ plane, we evaluate the ghost part as
follows:
\begin{eqnarray}
&& \left\langle\,b_{\sigma_1}b_{\sigma_2}\,c(Z_2)\,c(Z_1)\,\right>_{{\tau\over4}+\half} \nn
 &&\qquad = i^2\int_{C_1}{du_1\over2\pi i}\left({du_1\over d\rho}\right)
   \int_{C_2}{du_2\over2\pi i}\left({du_2\over d\rho}\right)
   \left\langle\,b(u_1)\,b(u_2)\,c(Z_2)\,c(Z_1)\,\right>_{{\tau\over4}+\half} \nn
% &&\quad = -{\cal G}^N \int_{C_1}{du_1\over2\pi i}\int_{C_2}{du_2\over2\pi i}
%   \left({du_1\over d\rho}\right)\left({du_2\over d\rho}\right)
%   \left\{\,{d\rho\over du_1} - {d\rho\over du_2}\,\right\} \nn
 &&\qquad = -i{\cal G}^N \left(\int_{C_1}{du_1\over2\pi i}
   \int_{C_2}{du_2\over2\pi i}{du_2\over d\rho}
   -\int_{C_2}{du_2\over2\pi i}
   \int_{C_1}{du_1\over2\pi i}{du_1\over d\rho}\right) \nn
% &&\qquad = -{\cal G}^N \left(
%   {-2\ttau\over2\pi i}\int_{C_2}{du\over2\pi i}{du\over d\rho}
%   -{-2\ttau\over2\pi i}\int_{C_1}{du\over2\pi i}{du\over d\rho}\right) \nn
 &&\qquad = -i{\cal G}^N {2\ttau\over2\pi i}\int_{C_1 -C_2}{du\over2\pi i}{du\over d\rho}
  = -i{\cal G}^N {2\ttau\over2\pi i}\bigl(\oint_{\zz(1)}-\oint_{\zz(2)}\bigr)
   {du\over2\pi i}{du\over d\rho} \nn
 &&\qquad = -{4\ttau\over2\pi}R\,{\cal G}^N,
\end{eqnarray}
\begin{wrapfigure}[15]{r}{\halftext}
%\begin{figure}[htb]
   \vspace{-2ex}
   \epsfxsize= 5cm   %or \epsfysize= HEIGHT cm
   \centerline{\epsfbox{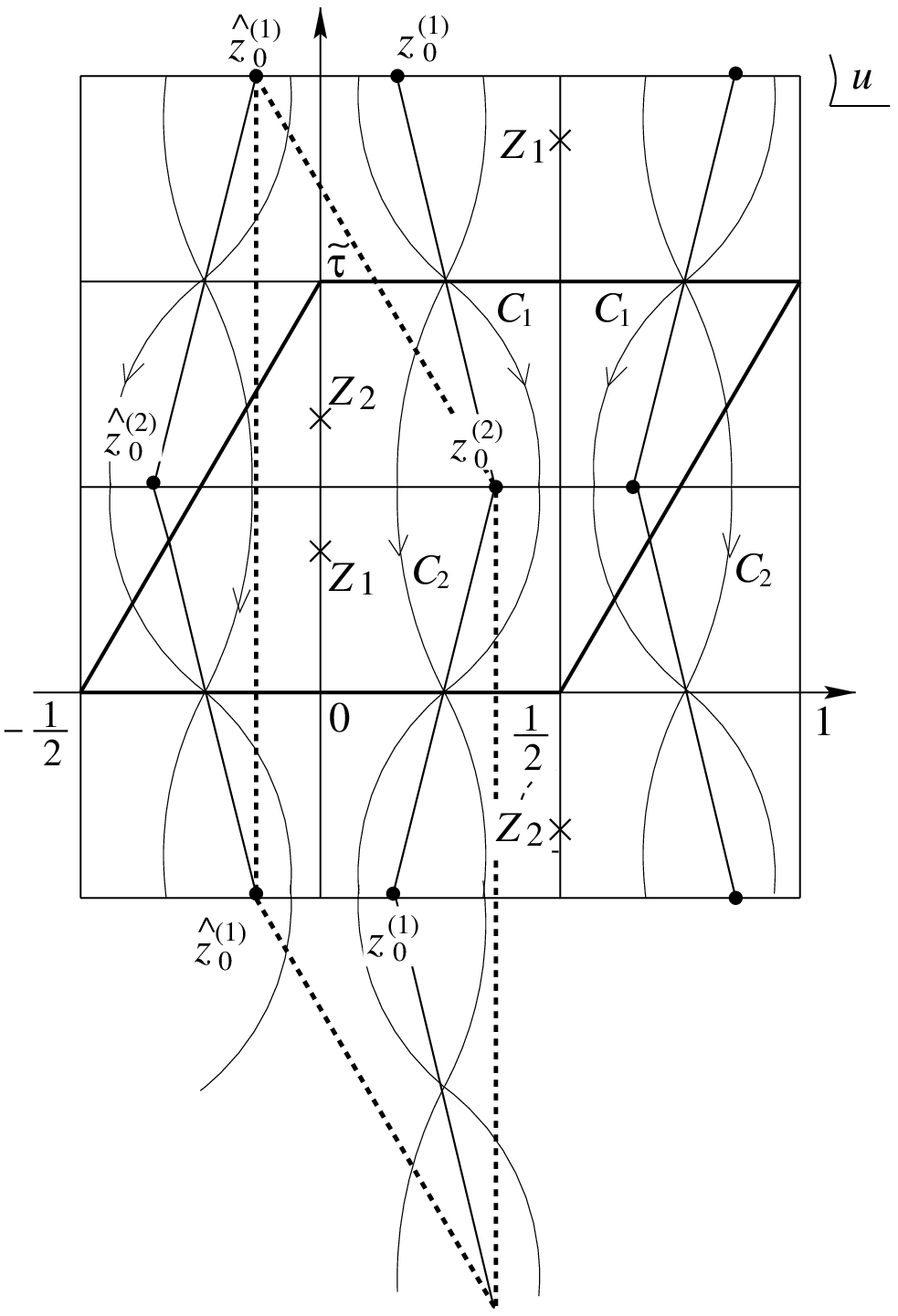}}
 \caption{The contours $C_1$ and $C_2$ for $V_\propto$ case.}
 \label{fig:Valphacont}
%\end{figure}
\end{wrapfigure}
where the contours $C_1$ and $C_2$ on the $u$ plane are those shown in 
Fig.~\ref{fig:Valphacont}, and we have used 
Eqs.~(\ref{eq:vev-relationhalf}) and (\ref{eq:vev-ttau+half}), 
$\int_{C_r}du_r\,1 = -2\ttau$ for both $r=1$ and 2 and the fact that the 
residues of $du/d\rho$ at the poles $\zz(1)$ and $\zz(2)$ are $\pm R$, 
respectively. 

The $X$-part of the correlation function was given in 
Eq.~(\ref{eq:XcorrVa}), and the Jacobian for the change of variables 
$(\sigma_1,\sigma_2)\rightarrow(\ttau,\,x)$ was calculated in Eq.\ 
(\ref{eq:va-jacobian2}), which is written as $d\sigma_1d\sigma_2=-(\alpha 
_1/2)R^{-1}d\ttau dx$ by using the residue $R$. Putting these 
altogether, the $V_\propto$ amplitude (\ref{eq:Valphaamp}) is found to be
\begin{eqnarray}
 &&{\cal A}_{V_\propto}
% &=& \xa g^2 \int d\ttau dx\,{-1\over2}(\alpha_1)^2 C^{-1}{-4\ttau\over2\pi i\alpha_1}C
% \,{\cal G}^N {\cal F}^N
 = {(2\pi)^d\over2^{d/2}}{\alpha_1\over\pi}\xa g^2 \nn
&&\ \times\int d\ttau dx\,\ttau\,{\cal G}^N {\cal F}^N 
 \,/[(2\pi)^d \delta^d(k_1+k_2)] \nn
 &&\ ={(2\pi)^d\over2^{d/2}}2i\xa g^2\int\!d({\ttau\over i}) dx
 \left[\,w^{1\over24}f(-w)(-2\pi\ln w)^{\half}\,\right]^{-(d-2)}\!\!
 \left[\,\psi^N(\rho_{12},w)\,\right]^{-2}\!\!\!.
\hspace{3em}
\label{eq:Va-amp}
\end{eqnarray}
This again correctly gives the same form as the nonorientable amplitude 
${\cal A}_{\rm NO}$ in Eq.~(\ref{eq:PNONPamp}). In this case the moduli 
integration directions are the same as is seen shortly, 
therefore, the full nonorientable amplitude is reproduced in our SFT if
their coefficients are the same with the same sign:
\begin{equation}
{(2\pi)^d\over2^{d/2}}2i\xa g^2=-ig^2 \quad \rightarrow\quad \xa=
-{2^{d/2}\over2(2\pi)^d}\,.
\label{eq:vacoupling}
\end{equation}
Originally the $V_\propto$ vertex has a moduli integration 
$\int_{0\leq\sigma_1\leq\sigma_2\leq\pi\alpha_1}d\sigma_1d\sigma_2$ which corresponds to 
$\int_0^{\pi\alpha_1}d\sigma_+\int_0^{\sigma_+}d\sigma_-$ (for M1 diagram) plus 
$\int_{\pi\alpha_1}^{2\pi\alpha_1}d\sigma_+\int_0^{2\alpha_1\pi-\sigma_+}d\sigma_-$ (for M2 diagram), 
in terms of $\sigma_\pm\equiv\sigma_2\pm\sigma_1$. 
Inspection of Eqs.~(\ref{eq:va-period1}) and (\ref{eq:va-sig-sig}) 
shows that this integration region corresponds just to 
$\int_0^1dx\int_0^{\ttau_0(x)/i}d(\ttau/i)$, where the boundary value 
$\ttau_0(x)$ is given by the root for $\ttau$ 
of Eq.~(\ref{eq:va-sig-sig}) with 
$\sigma_-=2\pi\alpha_1x$ for $x\leq1/2$ and $\sigma_-=2\pi\alpha_1(1-x)$ for $x\geq1/2$. 
More explicitly, from Eq.~(\ref{eq:va-sig-sig}),
we see that $y=0$ corresponds to $\sigma_-=2\pi\alpha_1x$ and $y=1/2$ to 
$\sigma_-=2\pi\alpha_1(1-x)$, so that the boundary value $\ttau_0(x)$ is determined 
by Eq.~(\ref{eq:va-y}) with $y=0$ and $1/2$:
\begin{equation}
-2\pi ix = \cases{
 g_2(\ttau x|\ttau+1/2)|_{\ttau=\ttau_0(x)}
 &for $x\leq1/2$ \cr
 g_2(\ttau x+1/2|\ttau+1/2)|_{\ttau=\ttau_0(x)}
 &for $x\geq1/2$ \cr}\,.
\end{equation}
The former equation for $x\leq1/2$ just coincides with Eq.~(\ref{eq:no-y}) 
with $y=0$ for nonorientable M1 case, where $y=0$ for M1 case 
corresponds to the $\tau_1=0$ boundary as is seen in Eq.~(\ref{eq:no-tau1}).
Thus M1 diagram covers the moduli region 
$\int_0^{1/2}dx\int_{\ttau_0(x)/i}^\infty d(\ttau/i)$, 
and, similarly, the M2 diagram covers the region 
$\int_{1/2}^1dx\int_{\ttau_0(x)/i}^\infty d(\ttau/i)$ (although we have not written 
the mapping explicitly for the latter case).
In Fig.~\ref{fig:moduli} (b), it is shown how the full moduli region $0\leq 
\ttau/i<\infty$, $0\leq x\leq1$ is covered by these three diagrams.

\subsection{Singularities of planar and $V_\propto$ amplitudes}

The planar amplitude ${\cal A}_{\rm P}$ in Eq.~(\ref{eq:PNONPamp}) and 
the $V_\propto$ amplitude (\ref{eq:Va-amp}) both have a singularity coming 
from the configuration drawn in Fig.~\ref{fig:planar}. These 
singularities should cancel each other between these two amplitudes in 
order for the theory to be consistent. 

To analyze the singularity, we first rewrite the planar amplitude 
${\cal A}_{\rm P}$ in Eq.~(\ref{eq:PNONPamp}) in terms of the variables 
$(q,\nu)$ in place of $(w,x)$:
\begin{equation}
\cases{q=e^{i\pi\tau}=e^{-i\pi/\ttau} \cr
\nu_{12}=2x \quad \cr}
\quad \leftrightarrow \quad 
\cases{w=e^{2\pi i\ttau} \cr
\rho_{12} = e^{4\pi i\ttau x}=e^{2\pi i\nu_{12}}\cr}.
\end{equation}
Noting that the integrand is rewritten as 
\begin{eqnarray}
 && \psi(\rho_{12},w) = \left({-2\pi\over\ln q}\right)\sin \pi\nu_{12}
 \prod {1-2q^{2n}\cos 2\pi\nu_{12} +q^{4n}\over(1-q^{2n})^2} 
 \equiv\left({-2\pi\over\ln q}\right)\,F(\nu_{12},q^2)\,, \nn
 && \left[\,w^{1\over24}f(w)(-2\pi\ln w)^{\half}\,\right] 
 = (2\pi)\left[\,q^{1\over12}f(q^2)\,\right],
\label{eq:funF}
\end{eqnarray}
and also using 
\begin{equation}
  dx = \half d\nu_{12}\,, \qquad 
  d(\ttau/i) = {d\tau\over i\tau^2} = {\pi\over(\ln q)^2}{dq\over q}\,,
\end{equation}
we find 
\begin{eqnarray}
 {\cal A}_{\rm P}
 &=& -ing^2\int{dq\over q}d\nu_{12}\,\half{\pi\over(\ln q)^2}(2\pi)^{-(d-2)}
   \!\Bigl({-2\pi\over\ln q}\Bigr)^{-2}\!
   \left[\,q^{1\over12}f(q^2)\,\right]^{-(d-2)}\!
   \left[\,F(\nu_{12},q^2)\,\right]^{-2} \nn
 &=& -{ing^2\over(2\pi)^d}{\pi\over2}\int{dq\over q}d\nu_{12}\,q^{-{d-2\over12}}
   \left[\,f(q^2)\,\right]^{-(d-2)}
   \left[\,F(\nu_{12},q^2)\,\right]^{-2} \nn
 &=&
   -{ing^2\over(2\pi)^{26}}{\pi\over2}\int_0^1{dq\over q^3}\int_0^1d\nu_{12}\,
   \left[\,f(q^2)\,\right]^{-24}
   \left[\,F(\nu_{12},q^2)\,\right]^{-2}\quad (\hbox{at}\ d=26).
\label{eq:Pfinal}
\end{eqnarray}
In the final equation we have made the integration region explicit.

Next we rewrite the $V_\propto$ amplitude (\ref{eq:Va-amp}), 
or more precisely the full nonorientable amplitude 
${\cal A}_{V_\propto}+{\cal A}_{\rm NO}={\cal A}_{V_\propto+{\rm NO}}$, 
into a similar form using the same variables $(q,\nu)$. The integrand is 
rewritten as 
\begin{eqnarray}
 \psi^N(\rho_{12},w) 
 &=& \left({-4\pi\over\ln q}\right)\sin {\pi\nu_{12}\over2}
 \prod {1-2(-\sqrt q)^n\cos \pi\nu_{12} +q^n\over(1-(-\sqrt q)^n)^2} \nn
 &=& \left({-4\pi\over\ln q}\right)F\left({\nu_{12}\over2},-\sqrt q\right), \nn
 &&\hspace{-6em} \left[\,w^{1\over24}f(-w)(-2\pi\ln w)^{\half}\,\right] 
 = 2^{-\half}(2\pi)\left[\,q^{1\over48}f(-\sqrt q)\,\right],
\end{eqnarray}
using the same function $F$ as defined in Eq.~(\ref{eq:funF}). 
Also using the coupling relation (\ref{eq:vacoupling}), 
we obtain
\begin{eqnarray}
 {\cal A}_{V_\propto+{\rm NO}}
% &=& g^2\int{dq\over q}d\nu_{12}\,\half{\pi\over(\ln q)^2}
%   2^{d-2\over2}(2\pi)^{-(d-2)}
%   \left({-4\pi\over\ln q}\right)^{-2}
%   \left[\,q^{1\over48}f(-\sqrt q)\,\right]^{-(d-2)}
%   \left[\,F\left({\nu_{12}\over2},-\sqrt q\right)\,\right]^{-2} \nn
 &=& -{ig^2\over(2\pi)^d}2^{{d\over2}-4}\pi 
   \int{dq\over q}d\nu_{12}\,q^{-{d-2\over48}}
   \left[\,f(-\sqrt q)\,\right]^{-(d-2)}
   \left[\,F\left({\nu_{12}\over2},-\sqrt q\right)\,\right]^{-2} \nn
 &=&
   -{ig^2\over(2\pi)^{26}}2^9\pi\int{dq\over q^{3\over2}}d\nu_{12}\,
   \left[\,f(-\sqrt q)\,\right]^{-24}
   \left[\,F\left({\nu_{12}\over2},-\sqrt q\right)\,\right]^{-2}
%\quad (\hbox{at}\ d=26).
\end{eqnarray}
at $d=26$. If we perform a further change of the variables,
\begin{equation}
 {\nu_{12}\over2}=\nu'_{12},\quad \sqrt q=q'^2 \qquad 
 \rightarrow\qquad d\nu_{12}=2d\nu'_{12},\quad dq=4q'^3dq',
\end{equation}
the amplitude can be cast into almost the same form as ${\cal A}_{\rm P}$:
\begin{equation}
 {\cal A}_{V_\propto+{\rm NO}}
 = -{ig^2\over(2\pi)^{26}}2^{12}\pi 
   \int_0^1{dq'\over q'^3}\int_0^1d\nu'_{12}\,
   \left[\,f(-q'^2)\,\right]^{-24}
   \left[\,F\left(\nu'_{12},-q'^2\right)\,\right]^{-2},
\label{eq:Vafinal}
\end{equation}
where again the integration region has been made explicit. 
Note that $\nu_{12}=2x$ runs over the range [0,2] in this full 
nonorientable amplitude so that $\nu'_{12}$ runs over the same range 
[0,1] as the previous $\nu_{12}$ in the P case. 

Now we can compare the two amplitudes (\ref{eq:Pfinal}) and 
(\ref{eq:Vafinal}). The singularities occur at $q^2=0$, around which 
the integrand function is expanded as
\begin{equation}
[f(q^2)]^{-24}[F(\nu,q^2)]^{-2}
= \sin\pi\nu\left[ 1+4(\cos2\pi\nu+5)q^2+O(q^4) \right].
\end{equation}
Since this is integrated with the measure $dq/q^3$, the first and the 
second terms of this expansion yield singularities corresponding to 
the (closed) tachyon and dilaton, respectively. 
Noting that the argument $q^2$ is replaced by $-q^2$ for the 
${\cal A}_{V_\propto+{\rm NO}}$ 
case, the latter dilaton singularity can be cancelled between ${\cal 
A}_{{\rm P}}$ and ${\cal A}_{V_\propto+{\rm NO}}$ if
\begin{equation}
   {-ing^2\over(2\pi)^{26}}{\pi\over2} + (-1)
  {-ig^2\over(2\pi)^{26}}2^{12}\pi=0 \qquad \Rightarrow\qquad 
 n=2^{13}.
\label{eq:n2to13}
\end{equation}
Namely, our gauge group $SO(n)$ must be 
$SO(2^{13})$.\cite{rf:DougGrin,rf:Weinberg,rf:ItoyamaMoxhay,rf:Ohta}

At this point we recall the relation (\ref{eq:rel-2}), 
$x_\infty=nx_u^2=-4\pi ix_\propto$, derived in I. This demands, in particular, 
the equality $nx_u^2=-4\pi ix_\propto$. But, here in the above, we have determined 
all of $n$, $x_u^2$ and $x_\propto$ by computing the loop amplitudes. 
Substituting the above results (\ref{eq:xu2}), (\ref{eq:vacoupling}) and
(\ref{eq:n2to13}), we see that this equality is actually satisfied. 
This gives a rather nontrivial consistency check for the present theory.

There remains, however, a stronger singularity due to tachyon 
contribution which under the condition (\ref{eq:n2to13}) adds up 
between ${\cal A}_{{\rm P}}$ and ${\cal A}_{V_\propto+{\rm NO}}$:
\begin{eqnarray}
&&2\times{\pi\over2}{-ing^2\over(2\pi)^{26}}
   \int_0^1{dq\over q^3}\int_0^1d\nu\,\sin\pi\nu 
=-{nx_u^2g^2\over2}
   \lim_{\varepsilon\rightarrow0}\int_0^1d\nu\,\sin\pi\nu\int_{\varepsilon(8\alpha_1\sin\pi\nu)^{-1}}^1{dq\over q^3} \nn
&&\quad =-{nx_u^2g^2\over2}
   \lim_{\varepsilon\rightarrow0}{32\alpha_1^2\over\varepsilon^2}\int_0^1d\nu\,\sin^3\pi\nu 
=-{nx_u^2g^2\over2}
   \lim_{\varepsilon\rightarrow0}{32\alpha_1^2\over\varepsilon^2}{4\over3\pi}\,,
\label{eq:TachyonDiv}
\end{eqnarray}
where $x_u^2=i(2\pi)^{-25}$ is used and the singular endpoint of the $q$ 
integration has been cut off by the time length $\tau_1\geq\varepsilon$ on the $\rho 
$-plane, which corresponds to the cutoff $q\geq\varepsilon(8\alpha_1\sin\pi\nu)^{-1}$ by 
the mapping relations (\ref{eq:pl-tau1}) and (\ref{eq:pl-y}). [For $q\ll 
1$, we have $y=1/4 +O(q^2)$ from Eq.~(\ref{eq:pl-y}), and $\tau_1/2\alpha_1
=4q\sin\pi\nu$ from Eq.~(\ref{eq:pl-tau1}).]\ On the other hand, as in 
the case of closed tachyon amplitude considered in I, we still have 
another contribution to this amplitude, coming from the 
counterterm which was introduced as a `renormalization' of the zero 
intercept. The counterterm is contained in $\tQB^\o=\QB^\o+\lambda_\o g^2\alpha 
^2c_0$, and so contributes to the open tachyon amplitude as
\begin{eqnarray}
&&(2\pi)^d\delta^d(k_1+k_2){\cal A}_{\rm count}
= - \bra{R^\o(1,2)}
\ket{\varphi^0(k_2)}_2
\lambda_\o g^2\alpha_1^2 {c_0}^{(1)}
\ket{\varphi^0(k_1)}_1\nn
&&\qquad  \Rightarrow\qquad 
 {\cal A}_{\rm count}= -\lambda_\o \alpha_1^2 g^2 
= +\alpha_1^2 g^2\lim_{\epsilon\rightarrow0}
      \frac{32n\aa^2}{2\epsilon^2}\,,
\label{eq:counterterm}
\end{eqnarray}
where the value (\ref{eq:rel-1}) for $\lambda_\o=\lambda_\c/2$ has been substituted.
Unfortunately, this contribution seems {\it not} to cancel the divergence 
in Eq.~(\ref{eq:TachyonDiv}) since the coefficients differ by a factor 
$4/3\pi$. However, we should note that the problem is very subtle since they
are divergent quantities; indeed, the value for $\lambda_\o=\lambda_\c/2$ used in 
Eq.~(\ref{eq:counterterm}) was determined in I considering the cancellation 
of similar divergences in the case of closed string tachyon amplitude. 
So the identification of the cutoff parameters $\varepsilon$ on the $\rho$ plane 
in both terms may not be suitable since the present $\rho$ plane of the open 
tachyon amplitude has boundaries while that of the previous closed tachyon 
amplitude does not. It is also unclear whether there is no other suitable 
cutoff procedure with which exact cancellation can be realized. At the moment, unfortunately, we cannot show that the divergence due to the presence of 
tachyon is cancelled by the renormalization of the `zero intercept' term 
$\lambda_\o g^2\alpha^2c_0$, although there is still a chance 
of cancellation since Eqs.~(\ref{eq:TachyonDiv}) and (\ref{eq:counterterm})
have {\it opposite} signs in any case.

\section{Conclusion and discussions}

We have shown that the (open) tachyon one-loop 2-point amplitudes are 
correctly reproduced in our SFT by choosing the coupling constants suitably.
All the coupling constants of the seven interaction vertices have now 
been determined. Some relations among the coupling constants 
which are found in the previous two papers and this paper, turn out to 
be mutually consistent. 

Nevertheless the presence of tachyon is a problem in our SFT, as is 
always the case in bosonic string theories. The (closed) tachyon 
vanishing into vacuum causes divergences in various amplitudes. As we 
have seen at the end of the last section, it is not clear that even the 
on-shell amplitudes can be made finite at loop levels by the 
`renormalization' of the `zero-intercept term' proportional to $\alpha^2$ in
$\tQB$. Moreover, off the mass-shell, the amplitudes even at the tree 
level {\it cannot} be made finite by such simple counterterms. 
Indeed, in the closed tachyon 2-point amplitude considered in I, the 
cancellation condition of the divergences between the disk (D) and real 
projective plane (RP) amplitudes becomes different off the mass-shell, 
where the conformal factors $(\prod_r dZ_r/dw_r)^{k^2/8-1}$ become 
contributing and the imbalance of them between the D and RP amplitude 
cases yield the terms proportional to $k^2-8$ and $(k^2-8)^2$ at $O(q)$ 
and $O(q^2)$. Since the tachyon singularity is as singular as $\int 
dq/q^3$, even such `small' differences of $O(q)$ and $O(q^2)$ lead to 
divergences, proportional to operators $L+\bar L$ and $(L+\bar L)^2$.

In our computations of one-loop tachyon 2-point amplitudes in this paper, 
we had the delta function factor $\delta(\tau_1-\tau_2)$, with which factor 
the string diagrams such as in Fig.~\ref{fig:loop} reduced to the same 
diagrams as those appearing in the light-cone gauge SFT.\cite{rf:KK1,rf:KK} 
So the discussions became almost the same as in the light-cone gauge SFT. 
For instance, the nonplanar diagram (a) in Fig.~\ref{fig:nonplanar} 
did not cover the whole moduli region of the nonplanar one-loop amplitude 
and required the existence of the $UU$ diagram (b) in 
Fig.~\ref{fig:nonplanar}, hence explaining the reason of existence of 
the open-closed transition vertex $U$.\cite{rf:KK} 
Actually, in the light-cone
gauge SFT, the former nonplanar diagram (a) is anomalous from the 
viewpoint of Lorentz-invariance and the anomaly is cancelled by the 
diagram (b) at $\tau=0$. This was shown by Saito and Tanii\cite{rf:ST1,rf:ST2}
and Kikkawa and Sawada.\cite{rf:KikkawaSawada}

In the light-cone gauge SFT, there is a universal time (light-cone time) 
in which the interactions are local and the time lengths are the 
same along whichever the paths on the diagram they are measured. 
Namely, the diagrams in the light-cone gauge SFT are always stretched 
{\it tight}, and there appear no diagrams with propagators which 
are {\it slack} or propagating {\it backward} in the time.
The presence of such universal time was essential in the proof of 
physical equivalence of the light-cone gauge SFT to the covariant Polyakov 
formulation (hence of the modular and Lorentz invariance of 
the light-cone gauge SFT).\cite{rf:GiddingsD'Hoker}

The ultimate reason why the diagrams become tight in the light-cone 
gauge string (or particle) field theory resides in the fact that the 
vertices there have no dependence on $P^-$, the $-$ components of momenta
other than in momentum conservation $\delta$ function: 
suppose, for instance, that two vertices are connected by two 
propagators 1 and 2 simultaneously, thus making a loop. The two 
propagators can be represented as 
\begin{equation}
{1\over p_r^2+M^2} 
= \int_0^\infty dT_r\, e^{-(p_r^2+M^2)T_r}
= \int_0^\infty dT_r\, e^{[2p_r^+p_r^--({\mib p}_r^2+M^2)]T_r},
\end{equation} 
using proper times $T_r$ for each string (or particle) $r=1,2$ 
with $M^2$ being squared mass (operator). The momenta $p_r$ can be 
represented as $p_1=l$ and $p_2=k-l$ by using the loop momentum $l$ and 
a certain external momentum $k$. Then, if the two vertices have no 
dependence on the $-$ components of momenta, i.e., are {\it independent} 
of $l^-$ in this case, the $l^-$ dependence appears only in the 
propagators and we can perform the $l^-$ integration of the loop 
integral $\int d^dl/(2\pi)^d$ as follows, writing $2p^+_r=\alpha_r$:
\begin{equation}
\int{dl^-\over2\pi}\, e^{\alpha_1l^-T_1}e^{\alpha_2(k^--l^-)T_2}
=e^{\alpha_2k^-T_2}\delta(\alpha_1T_1-\alpha_2T_2)
=e^{k^-\tau_2}\delta(\tau_1-\tau_2).
\end{equation}
Namely, the equality of the light-cone times $\tau_r=\alpha_rT_r$ resulted 
for $r=1$ and $2$. The same thing happens for any more complicated diagrams 
and the diagrams become stretched tight. 

In our $\alpha=p^+$ HIKKO type SFT,\cite{rf:kugozwie} on the other hand, 
the vertices {\it has} the dependence on $P^-$ and, therefore, there 
generally appear `slack' diagrams. As was explained in some detail in 
the Appendix B of Ref.~\citen{rf:kugozwie}, if all the external string 
states contain no excitations of $\alpha_n^-$ modes, the $P^-$ dependent 
part of the vertices plays no role and can be discarded, and then only 
`tight' diagrams as in the light-cone gauge case can contribute to the 
process. This was actually the case in our computation of the tachyon 
amplitudes in this paper since the tachyon has no excitation of 
$\alpha_n^\mu$ modes at all.

If we consider more general external states, however, all the slack 
diagrams, and even those with propagators propagating backward in the 
light-cone time, become contributing. This is the case, in particular, 
when we try to prove the BRS invariance of the effective action 
$\Gamma[\Psi,\Phi]$ possessing general external string fields $\Psi,\ \Phi$. 
Therefore the BRS invariance proof at loop level would look rather 
different from the computations in the present paper. Consider, for 
instance, the one-loop effective action $\Gamma_\1loop[\Psi]$ quadratic in 
open-string field $\Psi$ in Eq.~(\ref{eq:effaction}) which contains the 
planar diagram contribution 
\begin{equation}
 \Gamma_\1loop^{\rm planar}[\Psi] 
 = i{g^2\over4} \int_{0}^{\infty}d\tau_1 d\tau_2
   \, n\bra{v_{\rm P}(\tau_1,\!\tau_2)}
   b_{\tau_1}b_{\tau_2}\of_2 \of_1 ,
\end{equation}
This term generally corresponds to slack planar diagram in 
Fig.~\ref{fig:loop} with $\tau_1\not=\tau_2$. Now act the BRS operator 
on the two external fields $\Psi_r$. Then, as being a general 
property,\cite{rf:KugoSuehiro}
BRS operator acts as an differential operator on the 
moduli parameters, $\tau_1$ and $\tau_2$ in this case, and obtain two 
surface terms with $\int d\tau_1\bra{v_{\rm P}(\tau_1,\!\tau_2{=}0)}b_{\tau_1}$ and
$\int d\tau_2\bra{v_{\rm P}(\tau_1{=}0,\!\tau_2)}b_{\tau_2}$. The former term, 
for instance, corresponds to the diagram in which the second propagator is 
collapsed. Which contribution does this term cancel with? 

\begin{figure}[tb]
   \epsfxsize= 13cm   %or \epsfysize= HEIGHT cm
   \centerline{\epsfbox{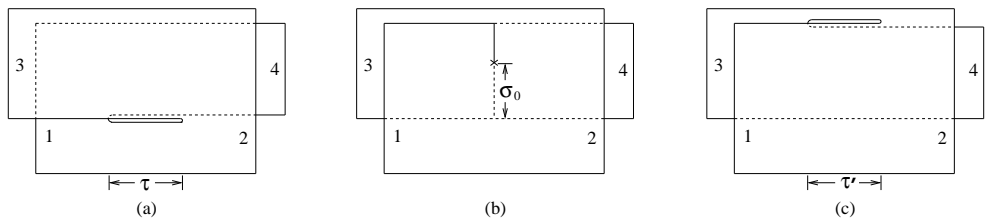}}
 \caption{A set of tree diagrams among which BRS invariance is satisfied.}
 \label{fig:tree1}
\end{figure}

Generically, the loop-level action is BRS invariant if the (tree level) 
action $S$ is. This should be the case also here. In fact, consider the 
tree diagrams drawn in Fig.~\ref{fig:tree1} 
%\begin{wrapfigure}[6]{r}{6.6cm}
%\end{wrapfigure}
and recall that the BRS 
invariance was realized among those three diagrams;\cite{rf:HIKKO1,rf:KK} 
the BRS transformation of the diagrams (a) and (c) leaves the surface 
terms of the moduli $\tau$ and $\tau'$ at $\tau=0$ and $\tau'=0$, respectively. 
But the BRS transform of the diagram (b) with quartic interaction 
$V_4^\o$, yields the surface terms of the moduli $\sigma_0$ at $\sigma_0=0$ and 
$\pi\abs{\alpha_4}$. These are the same configurations as the above two 
surface terms of the diagrams (a) at $\tau=0$ and (c) at $\tau'=0$ and 
cancel them. 

\begin{figure}[tb]
   \epsfxsize= 10cm   %or \epsfysize= HEIGHT cm
   \centerline{\epsfbox{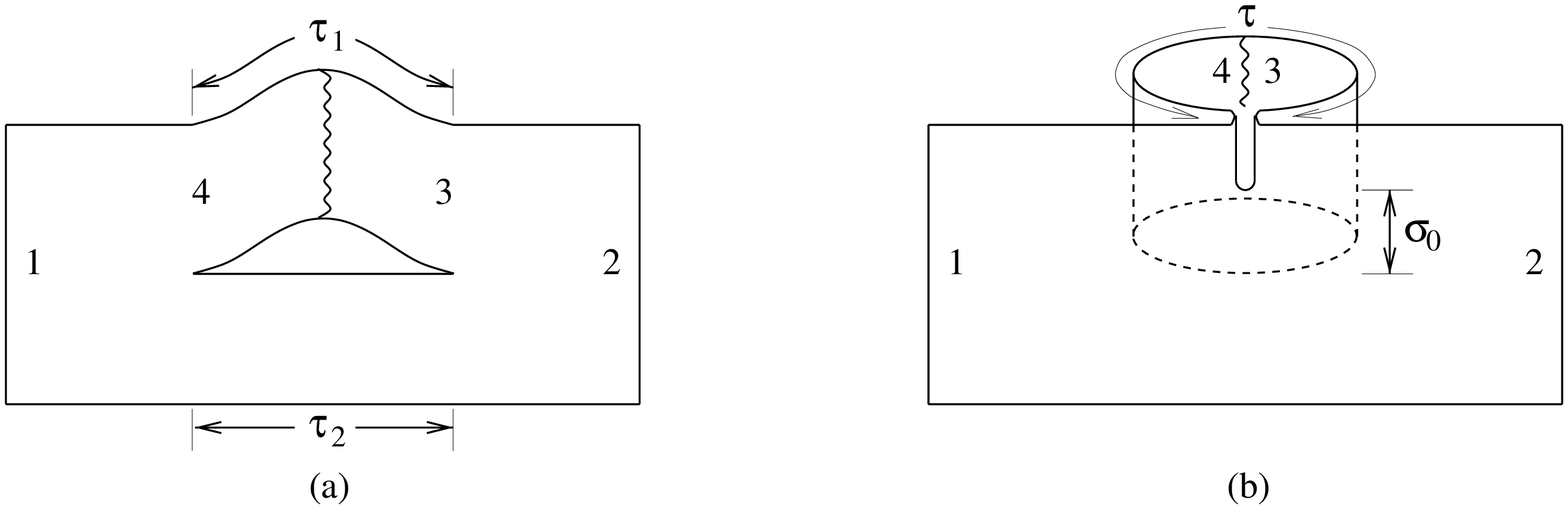}}
 \caption{Loop diagrams obtained from the tree diagrams (a) and (b) in 
 Fig.~\protect\ref{fig:tree1} by contracting strings 3 and 4.}
 \label{fig:loop1}
\end{figure}

This cancellation mechanism should also work when these tree diagrams are 
lifted to loop diagrams. Indeed, if the strings 3 and 4 have the same 
length, i.e., $\alpha_3=\abs{\alpha_4}$, we can connect the strings 3 and 4 and 
make one-loop diagrams (a) and (b) drawn in Fig.~\ref{fig:loop1}. 
%\begin{wrapfigure}[6]{r}{6.6cm}
%\end{wrapfigure}
In this case of $\alpha_3=\abs{\alpha_4}$, the diagram (c) disappears. Note that the 
diagram (a) in Fig.~\ref{fig:loop1} is nothing but the `slack' planar 
diagram, and that the diagram at $\tau_2=0$ is just the surface 
term which we have been discussing in the above. As is now clear from the 
above tree level arguments, it is cancelled by the diagram (b) in 
Fig.~\ref{fig:loop1} at $\sigma_0=0$ which is left as a surface term when 
the diagram (b) is BRS transformed. (A surface term at another endpoint 
$\sigma_0=\pi\abs{\alpha_4}$ would probably not contribute since it gives a 
disconnected diagram.) The diagram (b) gives another surface term 
at $\tau=0$, but it is as yet not clear whether it vanishes by itself or 
not. 

\begin{figure}[tb]
   \epsfxsize= 9cm   %or \epsfysize= HEIGHT cm
   \centerline{\epsfbox{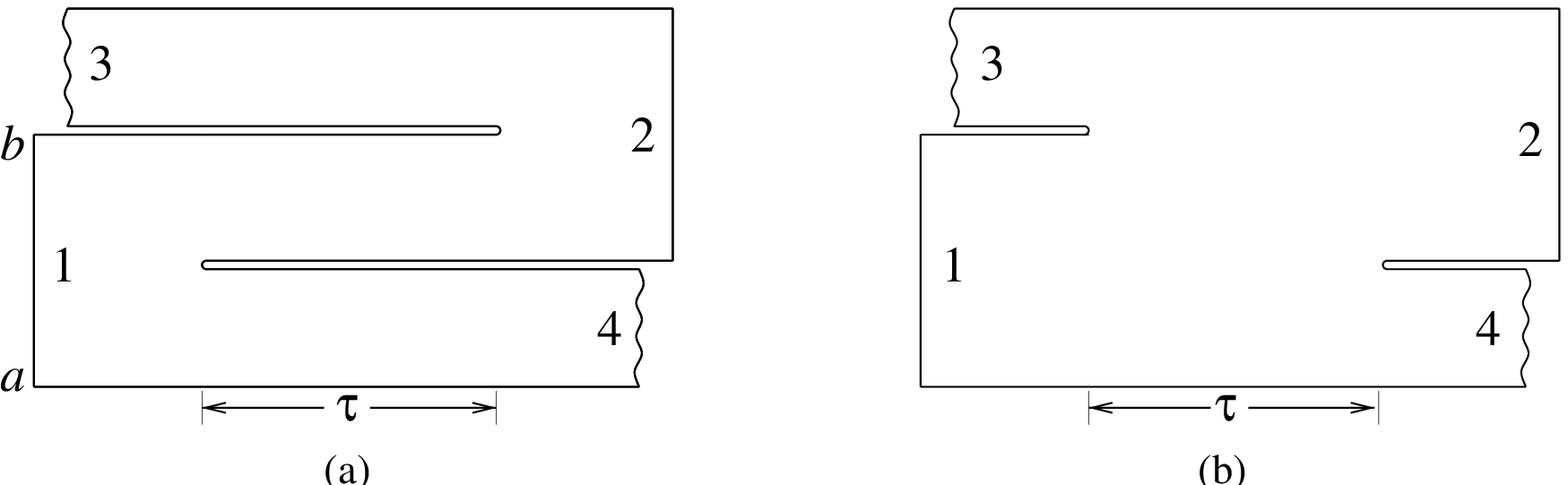}}
 \caption{Another set of tree diagrams satisfying BRS invariance.}
 \label{fig:tree2}
 \vspace{4ex}
  \epsfxsize= 10.5cm   %or \epsfysize= HEIGHT cm
   \centerline{\epsfbox{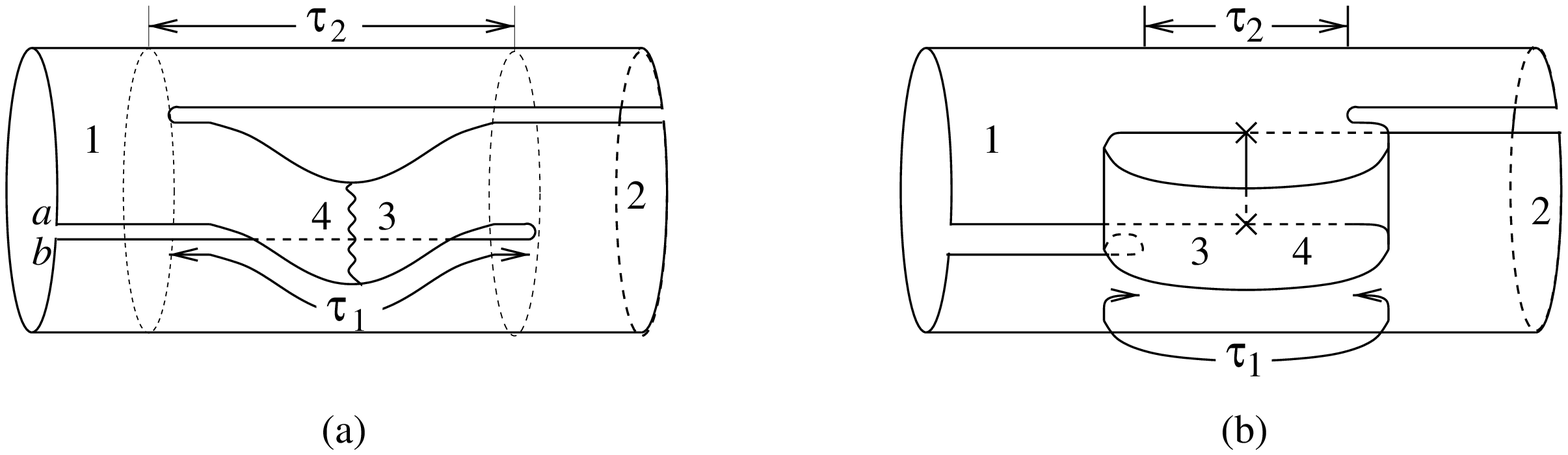}}
 \caption{Loop diagrams obtained from the tree diagrams (a) and (b) in 
 Fig.~\protect\ref{fig:tree2} by contracting strings 3 and 4.}
 \label{fig:loop2}
\end{figure}

Finally, we consider two tree diagrams in Fig.~\ref{fig:tree2}, 
the sum of 
which is clearly BRS invariant since the surface terms at $\tau=0$ from 
those two are the same and cancel. If strings 3 and 4 have the same length,
$\alpha_3=\abs{\alpha_4}$, we can again connect the strings 3 and 4 and 
obtain one-loop diagrams (a) and (b) drawn in Fig.~\ref{fig:loop2}.
%\begin{wrapfigure}[6]{r}{6.6cm}
%\end{wrapfigure}
The resultant diagram (a) is nothing but the nonplanar diagram, 
generally slack $\tau_1\not=\tau_2$. For generic external states, 
both diagrams (a) and (b) contribute and their surface terms 
at $\tau_2=0$ are the same and cancel with each other. So generically 
the BRS invariance holds with these two diagrams alone. However, if the 
external states $\Psi_r$ contain no $\alpha_n^-$ modes, then the delta function 
factor $\delta(\tau_1-\tau_2)$ appears and only the tight diagrams can contribute. 
This implies that the diagram (b) which contains backward propagation 
(i.e., $\tau_1=\alpha_4T_1<0$) does not contribute from the start, and thus 
the counterterm which can cancel the surface term at $\tau_1=\tau_2=0$ 
of diagram (a) becomes missing. This is an anomaly of the BRS invariance
in our SFT. As demonstrated in the present paper, the desired 
counterterm is supplied by the $UU$ diagram (b) in 
Fig.~\ref{fig:nonplanar}. We suspect that the BRS anomalies in our SFT 
occur this way only when the external states $\Psi_r$ contain no $\alpha_n^-$ 
modes. If so, then the relevant diagrams are always tight ones as in the 
light-cone gauge SFT and the anomalies for BRS and Lorentz invariance in
both theories will come from the same type of diagrams.

\section*{Acknowledgements}
The authors would like to express their sincere thanks to 
H.\ Hata, H.\ Itoyama, M.\ Kato, K.\ Kikkawa, N.\ Ohta, M.\ Maeno, 
Y.\ Matsuo, S.\ Sawada, K.\ Suehiro, Y.\ Watabiki and T.\ Yoneya for
valuable and helpful discussions. They also acknowledge hospitality at
Summer Institute Kyoto '97 and '98. T.~K.\ and T.~T.\ are
supported in part by the Grant-in-Aid for Scientific Research
(\#10640261) and the Grant-in-Aid (\#6844), respectively, from the
Ministry of Education, Science, Sports and Culture.

\appendix

\section{$\psi\,(\rho,w)$ and $\psi\,(z,q^2)$}
%%%%%%%%%%%%%%%%%%%%%%%%%%%%%%%%%%%%%%%%%%%%%%%%%%%%%%%%%%%%%%%%%%%%%%%

In the text, we have used the same variables as defined in GSW:\cite{rf:GSW}
\begin{equation}
\cases{\rho=e^{2\pi i\ttau\nu}\cr w=e^{2\pi i\ttau}\cr}, \qquad 
\cases{z=e^{2\pi i\nu}\cr q^2=e^{2\pi i\tau}\cr}, 
\end{equation}
with $\tau=-1/\ttau$. Then, clearly, $(z,q^2)$ corresponds to $(\nu,\tau)$ by
the same relation as $(\rho,w)$ to $(\tilde\nu\equiv\ttau\nu,\ttau)$. We have the 
following correspondences in the same manner:
\begin{equation}
\left\{
\begin{array}{ccc}
(\rho,w)   & \leftrightarrow & (\ttau\nu,\ttau) \\
(z,q^2)  & \leftrightarrow & (\nu,\tau) \\
(qz,q^2) & \leftrightarrow &(\nu+{\tau\over2},\tau) \\
(z^{-1/2},q^2) &\leftrightarrow & (-{\nu\over2},\tau) 
\end{array}
\right.
\label{eq:corresp}
\end{equation}

The functions $\psi, \ \psi^T$ and $\psi^N$ defined in the text are also 
the same as those in GSW, and are rewritten as follows in terms of the 
Jacobi theta functions:
\begin{equation}
\cases{ 
\displaystyle
\psi(\rho,w)
     = -2\pi ie^{\pi i(\ttau\nu)^2/\ttau}
          {\vartheta_1\left(\ttau\nu|\ttau\right)\over 
                   \vartheta'_1\left(0|\ttau\right)} 
     = {2\pi i\over\tau}
          {\vartheta_1\left(\nu|\tau\right)\over\vartheta'_1\left(0|\tau\right)}\,, 
\cr
\displaystyle
\psi^T(\rho,w)
    = 2\pi e^{\pi i(\ttau\nu)^2/\ttau}
       {\vartheta_2\left(\ttau\nu|\ttau\right)\over 
                  \vartheta'_1\left(0|\ttau\right)} 
     = {2\pi\over\tau}e^{\pi i(\nu+\tau/4)}
          {\vartheta_1\left(\nu+{\tau\over2}|\tau\right)\over 
                         \vartheta'_1\left(0|\tau\right)} \,,
\cr
\displaystyle
\psi^N(\rho,w)
    = -2\pi ie^{\pi i(\ttau\nu)^2/\ttau}
      {\vartheta_1\left(\ttau\nu|\ttau+\half\right)\over 
                  \vartheta'_1\left(0|\ttau+\half\right)} 
     = -{4\pi i\over\tau}
          {\vartheta_1\left(-{\nu\over2}|{\tau\over4}+\half\right)
                  \over\vartheta'_1\left(0|{\tau\over4}+\half\right)} \,,
\cr
}
\end{equation}
where the second equalities follow from the modular transformation 
properties of the theta functions. Now in view of the 
correspondences in Eq.~(\ref{eq:corresp}), the comparison of the first 
and second expressions of $\psi(\rho,w)$ immediately leads to the relation
\begin{equation}
\psi(\rho,w) = -{1\over\tau}e^{-\pi i\nu^2/\tau}\psi(z,q^2).
\label{eq:psirel}
\end{equation}
Comparison of the second expression of $\psi^T(\rho,w)$ with 
the first one of $\psi(\rho,w)$ gives
\begin{eqnarray}
\psi^T(\rho,w)&=&{i\over\tau}e^{\pi i(\nu+\tau/4)}e^{-\pi i(\nu+{\tau/2})^2/\tau}\psi(qz,q^2) %\cr
={i\over\tau}e^{-\pi i\nu^2/\tau}\psi(qz,q^2).
\hspace{2em}
\label{eq:psiTrel}
\end{eqnarray}
Further, comparing the first and second expressions of $\psi^N(\rho,w)$, we 
obtain
\begin{eqnarray}
\psi^N(\rho,w)&=&{2\over\tau}e^{-\pi i(-\nu/2)^2\over\tau/4}\psi^N(z^{-1/2},q^{1/2}) %\cr
={2\over\tau}e^{-\pi i\nu^2/\tau}\psi^N(z^{-1/2},q^{1/2}).
\hspace{2em}
\label{eq:psiNrel}
\end{eqnarray}
Moreover, using 
$\vartheta_1(-\nu|\tau)=-\vartheta_1(\nu|\tau)$ and 
$e^{\pi i(\nu+\tau)^2/\tau}\vartheta_1(\nu+\tau|\tau)=-e^{\pi i\nu^2/\tau}\vartheta_1(\nu|\tau)$, we immediately find
\begin{equation}
\cases{
   \psi(z^{-1},q^2)=-\psi(z,q^2) \cr %,\qquad \psi(\rho^{-1},w)=-\psi(\rho,w) \cr
   \psi(q^2 z,q^2)=-\psi(z,q^2) \cr %,\qquad \psi(w\rho,w)=-\psi(\rho,w) \cr
}.
\label{eq:psiproperty}
\end{equation}


\begin{thebibliography}{99}
%%%%%%%%%%%%%%%%%%%%%%%%%%%%%%%%%%%%%%%%%%%%%%%%%%%%%%%%%%%%%
% Some macros are available for the biblieography:
%   o for general use
%      \JL : general journals          \andvol : Vol (Year) Page
%   o for individual journal 
%      \PR  : Phys. Rev.               \PRL : Phys. Rev. Lett.
%      \NP  : Nucl. Phys.              \PL  : Phys. Lett.
%      \JMP : J. Math. Phys.           \CMP : Commun. Math. Phys.
%      \PTP : Prog. Theor. Phys.       \JPSJ: J. Phys. Soc. Jpn.
%      \JP  : J. of Phys.              \NC  : Nouvo Cim.
%      \IJMP: Int. J. Mod. Phys.       \ANN : Ann. of Phys.
% Usage:
%   \PR{D45,1990,345}            ==> Phys.~Rev.\ {\bf D45} (1990), 345
%   \JL{Phys.~Lett.,A30,1981,56} ==> Phys.~Lett.\ {\bf A30} (1981), 56
%   \andvol{B123,1995,1020}      ==> {\bf B123} (1995), 1020
%%%%%%%%%%%%%%%%%%%%%%%%%%%%%%%%%%%%%%%%%%%%%%%%%%%%%%%%%%%%%
\bibitem{rf:KugoTaka}
T.~Kugo and T.~Takahashi,
  \PTP{99,1998,649}.
\bibitem{rf:AKT2}
T.~Asakawa, T.~Kugo and T.~Takahashi, 
  \PTP{100,1998,831}.
\bibitem{rf:LPP}
A.~LeClair, M.~E.~Peskin and C.~R.~Preitschopf,
  \NP{B317,1989,411}.
\bibitem{rf:AGMV}
L.~Alvarez-Gaum\'e, C.~Gomez,
G.~Moore and C.~Vafa,
  \NP{B303,1988,411}.
\bibitem{rf:GM}
S.~B.~Giddings and E.~Martinec,
  \NP{B278,1986,91}.
\bibitem{rf:KugoSuehiro}
T.~Kugo and K.~Suehiro,
  \NP{B337,1990,434}.
\bibitem{rf:kugozwie}
T.~Kugo and B.~Zwiebach,
  \PTP{87,1992,801}.
\bibitem{rf:LPP2}
A.~LeClair, M.~E.~Peskin and C.~R.~Preitschopf,
  \NP{B317,1989,464}.
\bibitem{rf:ST1}
Y.~Saitoh and Y.~Tanii,
  \NP{B325,1989,161}.
\bibitem{rf:ST2}
Y.~Saitoh and Y.~Tanii,
  \NP{B331,1990,744}.
\bibitem{rf:KikkawaSawada}
K.~Kikkawa and S.~Sawada,
  \NP{B335,1990,677}.
\bibitem{rf:GSW}
M.~B.~Green, J.~H.~Schwarz and E.~Witten,
  {\it Superstring Theory}, (Cambridge Univ.\ Press, Cambridge, 1987).
\bibitem{rf:AKT1}
T.~Asakawa, T.~Kugo and T.~Takahashi, 
  \PTP{100,1998,437}.
\bibitem{rf:KK} 
M.~Kaku and K.~Kikkawa, 
  \PR{D10,1974,1823}.
\bibitem{rf:Mandelstam}
S.~Mandelstam,
  in {\it Unified String Theories}, ed.~by M.~Green and D.~Gross 
  (World Scientific, Singapore, 1986), p46.
\bibitem{rf:FGST}
D.~Z.~Freedman, S.~B.~Giddings, J.~A.~Shapiro and C.~B.~Thorn,
  \NP{B298,1988,253}.
\bibitem{rf:WhitWatson}
E.~T.~Whittaker and G.~N.~Watson,
  {\it A Course of Modern Aanlysis}, (Cambridge Univ.\ Press, London, 1927),
  Chapters XX and XXI.
\bibitem{rf:DougGrin}
M.~R.~Dougras and B.~Grinstein,
  \PL{183B,1987,52}.
\bibitem{rf:Weinberg}
S.~Weinberg,
  \PL{187B,1987,278}.
\bibitem{rf:ItoyamaMoxhay}
H.~Itoyama and P.~Moxhay,
  \NP{B293,1987,685}.
\bibitem{rf:Ohta}
N.~Ohta, 
  \PRL{59,1987,176}.
\bibitem{rf:KK1} 
M.~Kaku and K.~Kikkawa, 
  \PR{D10,1974,1110}.
\bibitem{rf:GiddingsD'Hoker}
E.~D'Hoker and S.~B.~Giddings,
  \NP{B291,1987,90}.
\bibitem{rf:HIKKO1}
H.~Hata, K.~Itoh, T.~Kugo, H.~Kunitomo and K.~Ogawa,
  \PR{D34,1986,2360}.

\end{thebibliography}
\end{document}